\documentclass{aa}

\usepackage[english]{}
\usepackage{natbib}
\usepackage[utf8]{inputenc}
\usepackage[colorlinks,urlcolor=magenta,citecolor=blue,linktocpage,breaklinks]{hyperref}
\usepackage[varg]{txfonts}
\usepackage{adjustbox} 
\usepackage{xspace}
\usepackage[]{footnote}
\usepackage{booktabs}
\usepackage{orcidlink}
\usepackage{float}
\usepackage{placeins}

\DeclareUnicodeCharacter{2212}{-}

\begin{document}

   \title{UV LIGHTS. New tools for revealing the low surface brightness regime in the ultraviolet}

      \author{Ignacio Ruiz Cejudo\inst{1,2}\orcidlink{0009-0003-6502-7714}
      \and Ignacio Trujillo\inst{1,2}\orcidlink{0000-0001-8647-2874}
      \and Giulia Golini\inst{1,2}\orcidlink{0009-0001-2377-272X}
      \and Nafise Sedighi\inst{1,2}\orcidlink{0009-0001-9574-8585}
      \and Mireia Montes\inst{3,1,2}\orcidlink{0000-0001-7847-0393}
      \and Sergio Guerra Arencibia\inst{1,2}\orcidlink{0009-0001-7407-2491}
      \and Mauro D'Onofrio\inst{4}\orcidlink{0000-0001-6441-9044}
      \and Dennis Zaritsky\inst{5}\orcidlink{0000-0002-5177-727X}
      \and Samane Raji\orcidlink{0000-0001-9000-5507}\inst{6}
      \and Nushkia Chamba\inst{7}\orcidlink{0000-0002-1598-5995}
      \and Chen-Yu Chuang\inst{5}    
      \and Richard Donnerstein \inst{5}\orcidlink{0000-0001-7618-8212}
      \and Sepideh Eskandarlou\inst{8}\orcidlink{0000-0002-6672-1199} 
      \and S.Zahra Hosseini-ShahiSavandi\inst{4}\orcidlink{0000-0003-3449-2288}
      \and Ra\'ul Infante-Sainz\inst{8}\orcidlink{0000-0002-6220-7133}
      \and Ouldouz Kaboud\inst{4}
      \and Garreth Martin\inst{9}\orcidlink{0000-0003-2939-8668}
      \and Javier Román\inst{10, 11}\orcidlink{0000-0002-3849-3467}
      \and Zahra Sharbaf\inst{1,2}\orcidlink{0009-0004-5054-5946}
      }

   \institute{Instituto de Astrofísica de Canarias,c/ Vía Láctea s/n, E38205 - La Laguna, Tenerife, Spain 
   \and Departamento de Astrofísica, Universidad de La Laguna, E-38205 - La Laguna, Tenerife, Spain
   \and Institute of Space Sciences (ICE, CSIC), Campus UAB, Carrer de Can Magrans, s/n, 08193 08193, Cerdanyola del Valles, Spain
   \and Department of Physics and Astronomy, University of Padua, Vicolo Osservatorio 3, 35122 Padova, Italy
   \and Steward Observatory and Department of Astronomy, University of Arizona, 933 N. Cherry Ave., Tucson, AZ 85721, USA 
   \and Departamento de Física Teórica, Atómica y Óptica, Universidad de Valladolid, 47011, Valladolid, Spain 
   \and NASA Ames Research Center, Moffett Field, CA 94035, USA 
   \and Centro de Estudios de F\'isica del Cosmos de Arag\'on (CEFCA), Plaza San Juan, 1, E-44001, Teruel, Spain
   \and School of Physics and Astronomy, University of Nottingham, University Park, Nottingham NG7 2RD, UK   
   \and Kapteyn Astronomical Institute, University of Groningen, PO Box 800, 9700 AV Groningen, The Netherlands
   \and Departamento de Física de la Tierra y Astrofísica, Universidad Complutense de Madrid, E-28040 Madrid, Spain}

   \date{}

  \abstract
  {Ultra-deep optical imaging surveys have reached unprecedented depths ($\gtrsim$30$ \,\rm{mag\, arcsec^{-2}}$; 3$\sigma$, 10\arcsec$\times$10\arcsec), facilitating the study of very faint galactic structures. However, the ultraviolet bands, which are key to the study of stellar populations, remain essentially unexplored at these depths. In this paper we present a detailed surface brightness and color analysis of the outermost regions of 20 nearby galaxies in the LBT Imaging of Galactic Haloes and Tidal Structues (LIGHTS) fields observed by GALEX in the FUV and NUV. We adapt and apply a low surface brightness oriented methodology that has proven effective in ultra-deep optical surveys.  A novel approach to background subtraction is proposed for UV imaging. Instead of subtracting a constant value from the background, we subtract a Poisson distribution that transforms the background into a pseudo-Gaussian distribution centered at zero. Furthermore, the PSF deconvolution algorithms developed for optical data are applied to our sample, using a novel set of very extended (R=750\arcsec) PSFs for the GALEX bands. This methodology allows us to obtain depths ranging from 28.5 to 30$\,\rm{mag\, arcsec^{-2}}$ (3$\sigma$; 10\arcsec$\times$10\arcsec), with reliable surface brightness profiles up to 31$\,\rm{mag\, arcsec^{-2}}$. This is about 1 mag deeper than with standard UV techniques. We use the surface brightness and color profiles to show that the application of PSF deconvolution, especially in the FUV, effectively mitigates the excess of light present in the outer regions of certain galaxies compared to the standard GALEX pipeline. This finding is crucial for any accurate stellar population inference from the color profiles. Additionally, a qualitative analysis of the results is presented, with particular emphasis on the surface brightness and color properties of the galaxies beyond their optical edges. Our work highlights the importance of developing innovative low surface brightness methods for UV surveys.}


\titlerunning{UV LIGHTS. New tools for revealing the low surface brightness regime in the ultraviolet}

\authorrunning{Ruiz-Cejudo et al.}

  \maketitle

\section{Introduction}\label{sec:intro}

The understanding of the outermost part of galaxies has evolved significantly in recent years thanks to the acquisition of increasingly deeper data. In this context, works such as \citet{jablonka+10} or \citet{trujillo+16} have shown that with an optimal observing and analysis strategy it is possible to reach depths of $\mu_{V}\gtrsim30\,\rm{mag\, arcsec^{-2}}$ (3$\sigma$; 10\arcsec$\times$10\arcsec). Such depths have recently been achieved by wide surveys such as the Hyper Suprime Cam Survey \citep{hscssp}, and are expected for the 10-year Legacy Survey of Space and Time \citep[LSST; ][]{lsst} coadd images, and for the Euclid \citep{euclid} VIS DEEP surveys \citep{2022A&A...657A..92E}.  

The achievement of these new surface brightness limits allows, for the first time, to systematically explore the boundaries of galaxies \citep[see e.g.][]{Martinez+19,Trujillo+20,Chamba+22}. The in-situ star formation edge of the galaxies  is expected to be produced by the location of a rapid drop in the star formation rate associated with a sudden decrease in the surface density of cold \citep[$T<$$100\,\rm{K}$; ][]{2020MNRAS.497..146D} gas  \citep[see e.g.][]{roskar+08}. 
The location of this sudden drop in in-situ star formation allows the definition of a radius, $R_{edge}$, to characterize the extension of the galaxies \citep{Trujillo+20}. This drop will manifest itself as a truncation on the stellar mass profiles, which will allow to measure the position of the stellar boundaries of the galaxies \citep{Trujillo+20,Chamba+22}. Taking advantage of the location of this feature, \citet{Chamba+24a,Chamba+24b} have related the boundaries of the galaxies to physical properties such as environmental quenching and stellar feedback \citep{Chamba+24a,Chamba+24b}, while \citet{2024A&A...682A.110B} have measured the growth of galaxies with cosmic time. In this sense, the localization of the radial position of these edges, $R_{edge}$, and their study, would be particularly aided by deep ultraviolet (UV) imaging, which can trace ongoing \citep[up to 100 Myr; see e.g. ][]{kennicutt+12} star formation. It is therefore not surprising that data from the Galaxy Evolution Explorer \citep[GALEX, ][]{galex} have been used to probe the stellar populations of galaxies, even at low optical surface brightness regimes. An example of this is the discovery of extended ultraviolet disks \citep[XUV disks, ][]{armando+05,thilker+05,thilker+07}. These are star-forming regions at galactocentric distances beyond the traditional definition of the size of galaxies given by the isophote $\mu_B$=25 mag arcsec$^{-2}$ \citep{Redman36}.

The goal of this work is to reexamine the GALEX data using the new tools we have learned from analyzing ultra-deep optical imaging over the last decade. In particular, we are interested in exploring whether the reanalysis of the background characterization \citep[see e.g.][]{kelvin+23,watkins+24} plus the use of extended Point Spread Functions \citep[PSFs, see e.g.][]{dejong08,sandin14,gnuastro-psf} will allow us to extend the current surface brightness limits of GALEX imaging and thus gain more insight into the process that forms the periphery of the disk galaxies.

In this paper, we use a sample of nearby disk galaxies observed to unprecedented surface brightness limits in the  Sloan \textit{g} and \textit{r} optical bands ($\mu_{g}$=31.2 mag arcsec$^{-2}$ and $\mu_{r}$=30.5 mag arcsec$^{-2}$; 3$\sigma$; 10\arcsec$\times$10\arcsec). These galaxies are part of the LIGHTS survey \citep{trujillo21, zaritsky2024lights}. The depth of the optical data allows us to characterize the edge of these objects with high precision, and therefore to compare whether  a sudden drop in the UV surface brightness profiles is also present. In this way, we use the optical information to probe physical parameters such as the stellar mass surface density at the edge of the galaxy \citep[which has been shown to be connected to the environment and the epoch of star formation, see e.g. ][]{Chamba+24a,2024A&A...682A.110B}, and the UV information to better characterize the age of the stellar population in these peripheral regions.

The paper is structured as follows: in Sec.~\ref{sec:samsec} we present the sample selection, with a total of 19 LIGHTS galaxies (14 including both NUV and FUV deep imaging) and one additional galaxy in one of the fields of view. In Sec.~\ref{sec:met} we present the methodology used in our work for the background subtraction and PSF deconvolution. In Sec.~\ref{sec:results}, we show the results of the application of our methodology. These results are presented in the form of surface brightness and color profiles, in addition to the measured limits. We additionally compare our results with those obtained using the standard GALEX pipeline. This comparison allows us to see that our methodology improves on the results of previous work. Finally, we summarize the main conclusions in Sec.~\ref{sec:conc}. Unless explicitly stated, the magnitudes are given in the AB system \citep{oke}.

\section{Sample Selection and Dataset}\label{sec:samsec}

    The selection of galaxies studied in this paper is based on the LIGHTS survey \citep{trujillo21, zaritsky2024lights}. The survey includes a total of 28 nearby galaxies (so far), selected according to the criteria described in \citet{zaritsky2024lights}. The selection criteria were primarily determined by instrumental limitations (such as the field-of-view (FOV) of the Large Binocular Camera (LBC), which is 23$\times$28 arcmin) and the scientific goals of the survey (including the ability to distinguish between extended stellar disk and halo components, and the detection of low surface brightness satellite galaxies). Most of the galaxies in the sample are nearby (D$<$20 Mpc) spiral galaxies with morphological properties and stellar masses similar to or slightly less than those of the Milky Way. These types of galaxies are particularly relevant because they have been extensively studied in cosmological simulations, and we therefore have extensive predictions of how they should grow \citep[see e.g.][]{cooper+10}. Notable exceptions in the LIGHTS survey are NGC 4220 and NGC 5866, which have been classified as S0-a \citep[][Fig. 1]{zaritsky2024lights}.  

    The UV data for these galaxies were obtained from the GALEX satellite GR6 and GR7 data releases \citep{bianchi14}. GALEX was a NASA Explorer mission launched in April 2003 that performed a space-based survey of the sky in the far-UV (FUV, 1344-1786 $\rm{\AA}$, $m_{ \rm{ZP}}=18.82$)  and near-UV (NUV, 1771-2831 $\rm{\AA}$, $m_{ \rm{ZP}}=20.08$) bands. Both bands were surveyed with a plate scale of 1.5\arcsec\ per pixel. The GALEX field of view is 1.24$\times$1.28 degrees, with a spatial resolution (as characterized by the Full Width Half Maximum (FWHM)) of 4.2\arcsec\  and 5.3\arcsec\ for FUV/NUV, respectively \citep{morrissey2007}. A summary of the surveys performance and the scientific highlights achieved can be found in \citet{GALEXsumary}.

    The GALEX data were obtained from three different surveys: the \textit{Nearby Galaxy Survey} \citep[NGS, ][]{GildePaz07}, the \textit{Medium Imaging Survey} (MIS), and the \textit{Deep Imaging Survey} \citep[Deep-DIS, ][]{morrissey2007,bianchi2017}; in addition to public data from guest investigator programs (GII). All data are publicly available in the MAST archive\footnote{\url{https://galex.stsci.edu/GR6/}}. To ensure sufficient depth to explore the outermost regions of the galaxies (i.e. $\mu_{lim}\gtrsim28.5\,\rm{mag\, arcsec^{-2}}$; 3$\sigma$;\, 10\arcsec$\times$10\arcsec$\,$ in both FUV and NUV images), only images with exposure times greater than 1000 seconds are selected. This selection criterion leads to the identification of 19 galaxies (14 with FUV), which are summarized in Table~\ref{tab:data}. In addition, IC 3211, a spiral galaxy in the field of NGC 4307, is included in the analysis to test the methodology on a smaller (in apparent size) galaxy.  In all cases we use the count maps (\textit{-cnt}) and the high resolution relative response maps (\textit{-rrhr}). To be consistent with the methodology we will present in Sec.~\ref{subsec:bcksub}, we convert the counts to intensity with $int=\frac{cnt}{rrhr}$ instead of downloading the intensity maps. Furthermore, background subtracted intensity maps (\textit{-intbgsub}) are used for comparison with the methodology presented in this paper.

    \begin{table*}
    
        \caption{List of galaxies analyzed in this work.} 
        \centering
        \begin{adjustbox}   {width=\linewidth}
        \begin{tabular}{ccccccccccccc}
        \hline
        \hline\\
        
            Name & R.A. (J2000) & Dec (J2000) & D & PA\tablefootmark{a}  & b/a & Survey    
             & \multicolumn{2}{c}{$t_{\rm exp}$} & \multicolumn{2}{c}{$\mu_{bck}$\tablefootmark{b} } &  \multicolumn{2}{c}{$\mu_{lim}(3\sigma; 10\arcsec\times10\arcsec$)}\\
              & [deg] & [deg] & [Mpc] &  & & & \multicolumn{2}{c}{[s]} & \multicolumn{2}{c}{[counts]}& \multicolumn{2}{c}{[$\rm{mag\, arcsec^{-2}}$]}\\
               & & & & & & & FUV & NUV & FUV & NUV & FUV & NUV\\
               \hline
               NGC1042	&	40.09986	&	-8.43354	&	13.5	&	$75^{\circ}$	&	0.83	&	NGS	&	2949.5	&	3775.7	&	0.649	&	7.755	&	29.09	&	29.17	\\
NGC2712	&	134.87698	&	44.9139	&	30.2	&	$90^{\circ}$	&	0.54	&	MIS	&	-\tablefootmark{c}	&	1691.1	&	-	&	2.991	&	-	&	28.77	\\
NGC2903	&	143.04212	&	21.50083	&	10.0	&	$-70^{\circ}$	&	0.42	&	NGS	&	1910.2	&	1909.2	&	0.411	&	4.597	&	28.79	&	28.67	\\
NGC3049	&	148.70652	&	9.27109	&	19.3	&	$-31^{\circ}$	&	0.64	&	NGS	&	1482.1	&	3172.2	&	0.427	&	7.152	&	28.56	&	28.95	\\
NGC3198	&	154.97897	&	45.54962	&	12.9	&	$-54^{\circ}$	&	0.36	&	NGS	&	14904.9	&	16587.9	&	1.925	&	26.272	&	30.26	&	30.00	\\
NGC3351	&	160.99042	&	11.70381	&	10.0	&	$-80^{\circ}$	&	0.72	&	NGS	&	1692.2	&	1692.2	&	0.351	&	4.028	&	28.81	&	28.56	\\
NGC3368	&	161.69058	&	11.81994	&	11.2	&	$80^{\circ}$	&	0.6	&	NGS	&	4529.1	&	11450.7	&	0.776	&	24.784	&	29.32	&	29.69	\\
NGC3486	&	165.09945	&	28.97514	&	13.6	&	$-10^{\circ}$	&	0.77	&	GII	&	2439.6	&	4023.6	&	0.430	&	6.925	&	29.06	&	29.18	\\
NGC3596	&	168.77586	&	14.78702	&	11.3	&	$-31^{\circ}$	&	0.92	&	GII	&	-	&	1661.1	&	-	&	3.908	&	-	&	28.61	\\
NGC3938	&	178.20604	&	44.12072	&	12.7	&	$-54^{\circ}$	&	0.93	&	GII	&	-	&	7248.7	&	-	&	12.686	&	-	&	29.56	\\
NGC3972	&	178.93787	&	55.32074	&	20.8	&	$27^{\circ}$	&	0.27	&	GII	&	-	&	2803.1	&	-	&	5.045	&	-	&	29.02	\\
NGC4013	&	179.6309	&	43.94702	&	14.6	&	$-25^{\circ}$	&	0.37	&	GII	&	1640.5	&	1640.5	&	0.271	&	2.595	&	29.06	&	28.79	\\
NGC4220	&	184.0488	&	47.88326	&	20.3	&	$50^{\circ}$	&	0.33	&	DEEP-DIS	&	12101.1	&	12101.1	&	2.313	&	19.455	&	29.82	&	29.89	\\
NGC4307	&	185.52368	&	9.04363	&	20.0	&	$-67^{\circ}$	&	0.24	&	GII	&	6337.4	&	6337.4	&	1.073	&	12.824	&	29.69	&	29.42	\\
NGC4321	&	185.72846	&	15.82182	&	15.2	&	$70^{\circ}$	&	0.77	&	GII	&	4805.1	&	6466.1	&	0.914	&	12.880	&	29.46	&	29.45	\\
NGC4330	&	185.82083	&	11.36806	&	27.7	&	$-32^{\circ}$	&	0.17	&	GII	&	3862.1	&	3862.1	&	0.754	&	7.579	&	29.25	&	29.15	\\
NGC5248	&	204.38343	&	8.88518	&	14.9	&	$17^{\circ}$	&	0.67	&	MIS	&	-	&	1648.1	&	-	&	3.479	&	-	&	28.73	\\
NGC5866	&	226.62291	&	55.76321	&	14.1	&	$38^{\circ}$	&	0.43	&	NGS	&	1526.4	&	1526.4	&	0.452	&	2.831	&	28.69	&	28.65	\\
NGC5907	&	228.97404	&	56.32877	&	16.5	&	$64^{\circ}$	&	0.13	&	GII	&	1543.6	&	15118.4	&	0.328	&	23.737	&	28.70	&	29.97	\\
IC3211	&	185.53048	&	8.99056	&	85.0	&	-54$^{\circ}$	&	0.89	&	GII	&	6337.5	&	6337.5	&	1.073	&	12.824	&	29.69	&	29.42	\\

               \hline
               \hline
        \end{tabular}
      \end{adjustbox}
        \vspace{0.1cm}
        \tablefoot{The position angle (PA) and the axis ratio (b/a) is based on the LIGHTS  observations using the \textit{g}-band data \citep{zaritsky2024lights}. Coordinates and distances are taken from the same work, with the exception of IC 3211, whose data is taken from \citet{sloandr}. \\
         \tablefoottext{a} {Position angle from West anti-clockwise.}\\
        \tablefoottext{b} {Mean value of the background, see Sec.~\ref{subsec:bcksub}. }\\
\tablefoottext{c} {If -, there is no data for this band or it is not reliable.} 
}
    
    \label{tab:data}
    
    \end{table*}

\section{Methodology}\label{sec:met}

In this section, we  explore both background characterization and the effect of the PSF in deep UV imaging. Our goal is to test whether the low surface brightness tools developed for deep optical imaging can additionally enhance the low surface brightness structures in the UV.

    \subsection{Background subtraction and mask building}\label{subsec:bcksub}

 The characterization of the background of an image depends critically on the ability to detect and mask all sources present in the data. In our case, the sources of the images are identified using \texttt{Gnuastro}\footnote{\url{https://www.gnu.org/software/gnuastro/}}'s tools \texttt{NoiseChisel} and \texttt{Segment} \citep{gnuastro, noisechisel_segment_2019}. The detection process has been optimized for low surface brightness features. \texttt{NoiseChisel} works under the assumption that the background of the image follows a Gaussian distribution. The difference between the quantiles of the mean and the median is then used to detect sources in the images. This approximation is effective for Gaussian backgrounds, such as those observed by ground-based optical and near-infrared telescopes, where the brightness of the atmosphere is very high. However, as has been pointed out in several papers \citep[see e.g., ][]{GildePaz07}, the background in the UV is significantly low, resulting in highly Poissonian statistics, especially in the FUV. As a result, the use of \texttt{NoiseChisel} to detect large extended sources in the FUV is inaccurate, and depending on exposure times, it also has problems in detecting the faintest parts of galaxies in the NUV. 

\begin{figure}
        \centering
        \includegraphics[width=\linewidth,height=0.6\linewidth]{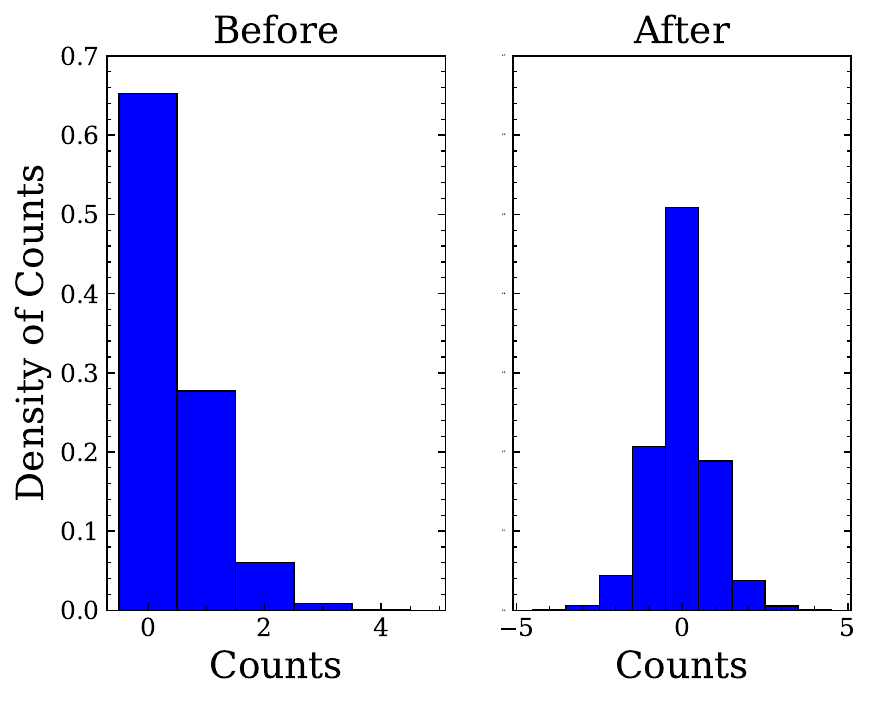}
        \includegraphics[width=\linewidth,height=0.6\linewidth]{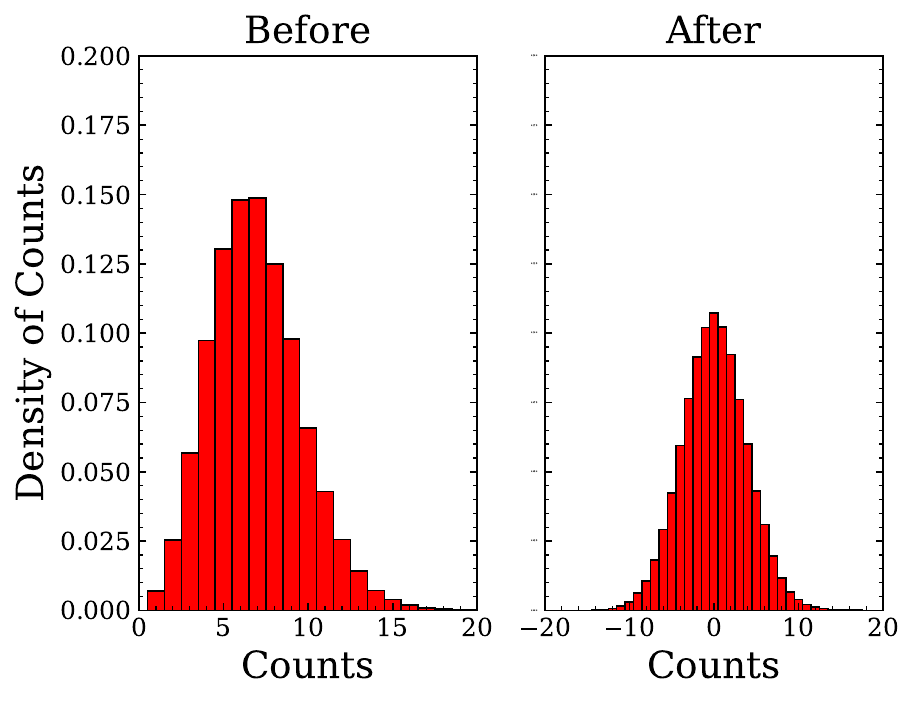}
        \caption{Histogram of background counts before and after statistical subtraction in NGC 3486 FUV (top) and NUV (bottom). The background is measured by using pixels within an elliptical ring between 8.75\arcmin\ and 11.25\arcmin\ from the center of the object.}
        \label{fig:background}
    \end{figure}

To work around this problem and allow the use of \texttt{NoiseChisel} to identify the faintest objects in both FUV and NUV, we have implemented the following strategy:

    \begin{enumerate}

      \item A preliminary mask is constructed using \texttt{NoiseChisel} and \texttt{Segment} on the NUV count map, since the NUV background is higher and nearly Gaussian, and the detection threshold works well on these images. This mask is then applied to both NUV and FUV, assuming that there are no FUV-only sources. However, to be sure that we are not leaving out any FUV-only sources, a visual inspection is performed to identify any FUV light sources that have not been masked in the NUV images.
 
\item After masking all sources, the mean ($\mu_{bck}$) of the Poissonian background in both FUV and NUV is characterized by measuring it at distances relatively far from the galaxy to avoid scattered light, but close enough to ensure that the measurements are representative of the local background beneath the galaxy. In practice, this is achieved by measuring the background at a distance equal to twice the apparent size of the galaxies (roughly characterized in these images by the isophote $29\,\rm{mag\, arcsec^{-2}}$ in NUV). The surface brightness profiles  in each band (see Sec.~\ref{subsec:sbcp}) are then examined to see if they behave in the outermost region as if they were dominated by the background (i.e. they remain flat). If this is the case, an annular region of width 150\arcsec\ is selected for the mean ($\mu_{bck}$) measurement. The mean background values found for each image are listed in the Table ~\ref{tab:data}.
         
     \item Once we have measured the mean of the background distribution, we begin the process of transforming a Poisson background into a pseudo-Gaussian one. This is done by generating a random 2D Poisson distribution map with the mean we measure for the background and subtracting this map from the actual image.  This subtraction results in an image where the background resembles a Gaussian distribution centered at 0. Fig.~\ref{fig:background} shows an example of the count distribution of the background in the galaxy NGC 3486 in the FUV (top) and NUV (bottom). The subtraction we perform is equivalent to subtracting the mean from a Gaussian background. We explain the mathematical details of this process in the Appendix~\ref{ap:skell}. It should be noted here that this is done on the counts map without taking into account the response of the CCD (\textit{-rrhr} images). This is because the Poisson distribution is defined in integer numbers (counts), while intensity maps have float values (counts per second).

     \item Having the counts map with a pseudo-Gaussian background centered at zero, we transform our images into intensity maps. This is done by dividing the counts map by the \textit{rrhr} map.
     We use the rrhr (i.e., high resolution relative response) maps because they are linearly interpolated to the same pixel scale as the counts map. The latter include the exposure time and the flat field correction. 

    \item  Now the images are ready to be reused by \texttt{NoiseChisel} and \texttt{Segment} to detect the faintest surface brightness sources and thus improve the masking of sources in the lowest surface brightness regime. After some exploration, we found the following \texttt{NoiseChisel} parameters to be appropriate for our purposes:  \texttt{-{}-tilesize=5,5 -{}-meanmedqdiff=0.02 -{}-outliersigma=3 -{}-qthresh=0.7 -{}-minskyfrac=0.7 -{}-dthresh=2 -{}-snminarea=3 -{}-detgrowquant=0.8 -{}-snthresh=3}; with small variations between each field. The final results are then visually inspected, and manually masked if any remaining sources were missed.

           \end{enumerate}

Motivated by the angular diameter of our galaxies \citep[see Table 1 in ][]{zaritsky2024lights}, we work in all fields with cropped images of $25\arcmin\times25\arcmin$ ($37.5\arcmin\times37.5\arcmin$ in the cases of NGC 3198 and NGC 5907), which are about 2 to 3 times the dimensions of our objects. This is done to avoid long \texttt{NoiseChisel} execution times. The use of GALEX cropped images also allows the avoidance of potential structural anomalies associated with vignetting or scattered light from external sources in the field when constructing the masks.
   
\subsection{Extended PSF characterization}\label{subsec:psf_sub}

Neglecting the effect of the PSF on the surface brightness distribution of galaxies can overestimate the amount of light in the outer parts of galaxies by nearly 1 mag or more \citep[see e.g.][]{trujillo+16}. It is therefore crucial to remove this effect by characterizing the extended PSF of the image. This implies knowing the shape of the PSF in extensions that are at least 1.5 times larger than the typical extension of the objects to be analyzed \citep{sandin14}. In our case, this implies knowing the PSFs up to distances greater than 10\arcmin.  In this study, we apply to the GALEX data the methodology for generating the extended PSFs of SDSS and LIGHTS data described in \citet{gnuastro-psf} and \citet{sedighi2024lightsextendedpointspread}. We do this by enlarging the currently available extended GALEX PSFs \citep[with a radial extension of 90\arcsec; ][]{morrissey2007} by nearly a factor of  9 up to a radius of $R=750\arcsec$ using PSF scripts from \texttt{Gnuastro} \citep{gnuastro-psf}. 

\begin{figure*}
    \centering
    \includegraphics[width=0.8\textwidth]{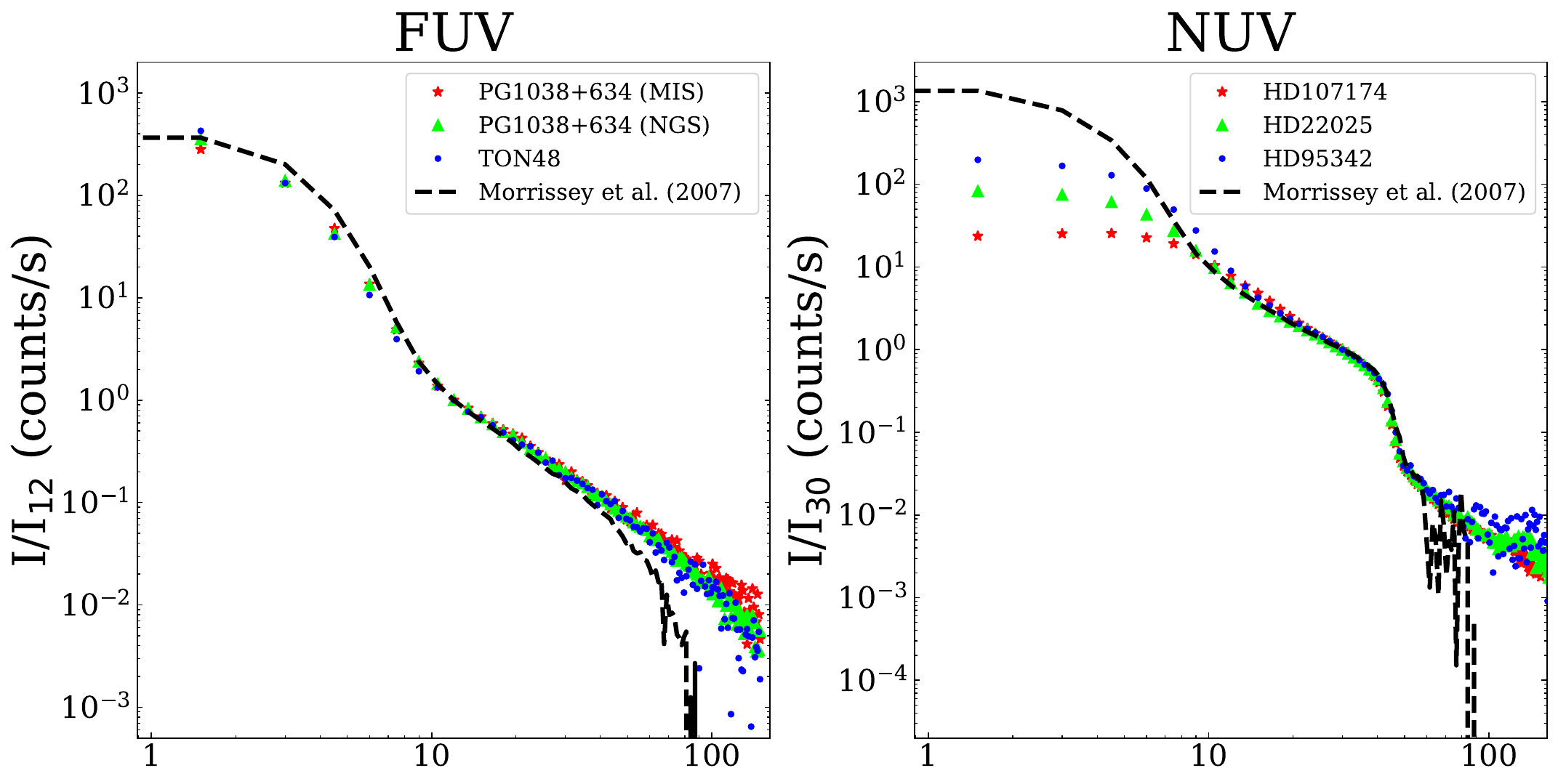}
    \includegraphics[width=0.8\textwidth]{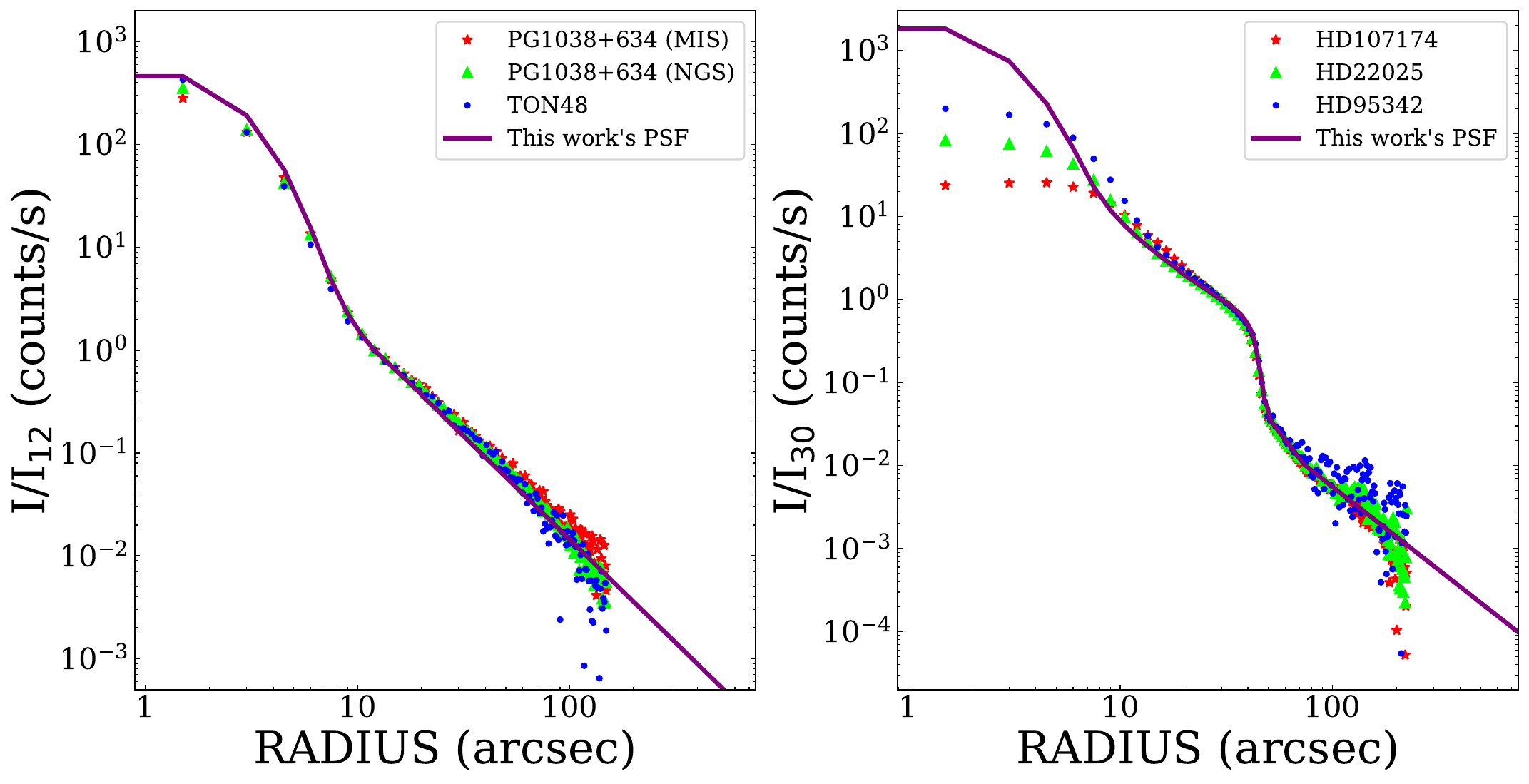}
    \caption{Profiles of several bright stars compared to the extended GALEX PSFs (FUV left, NUV right) by \citet{morrissey2007} (top) and to the PSFs built in this paper (bottom) using the cores from \citet{morrissey2007} and a power law extension based on real stars. The bright stars are from different fields and surveys. In NUV: HD107174 is in the GII survey ($t_{exp}=33700\rm{s}$). HD22025 is in the DEEP-DIS survey ($t_{exp}=29000\rm{s}$) and HD95342 is in the NGC 3486 field (see Table ~\ref{tab:data}). In the case of FUV: TON48 is in the NGC 3486 field, while PG1038+634 is in two different surveys: MIS ($t_{exp}=1655\rm{s}$) and NGS ($t_{exp}=1568\rm{s}$). The profiles are normalized to the value at $R=12^{\prime\prime}$ for FUV ($I_{12}$) and at $R=30^{\prime\prime}$ for NUV ($I_{30}$). The underestimation of the outer slopes by the extended PSFs provided by \citet{morrissey2007} is significant.}
    \label{fig:PSFs}
\end{figure*}

The extended GALEX PSFs of \citet{morrissey2007} are available at \href{http://www.galex.caltech.edu/researcher/techdoc-ch5.html}{GALEX technical documentation}\footnote{\url{http://www.galex.caltech.edu/researcher/techdoc-ch5.html}}. These PSFs are characterized up to R=90\arcsec$\,$ based on multiple observations of the star LDS749b. We note, however, that a comparison between the \citet{morrissey2007} PSFs and real stars from different fields and surveys suggests that these extended PSFs underestimate the outer slopes of the real stars (Fig.~\ref{fig:PSFs}, top). By reducing the data where these stars are located using both the standard GALEX pipeline and the background subtraction described earlier, we find that the underestimation of the outer slope of the \citet{morrissey2007} PSFs is not due to over-subtraction of the background.

As a consequence, we build new PSFs for the FUV and NUV bands based on different stars from different  GALEX surveys. In all cases, the same masking and background subtraction methodology described in Sec.~\ref{subsec:bcksub} is used. To avoid problems with saturation (see the central regions of the profiles in Fig.~\ref{fig:PSFs}), we keep the cores of the \citet{morrissey2007}  PSFs. The \citet{morrissey2007} PSF has been preserved up to a radius of 50\arcsec\, in NUV, and 25.5\arcsec\, in FUV. For the NUV, the profiles of the stars in Fig.~\ref{fig:PSFs} (right) suggest two different slopes. A power law fit to the region between 50\arcsec\, and 75\arcsec\, gives an exponent of $\beta_{1}=-3.30$, while $\beta_{2}=-2.01$ is found for radial distances greater than 75\arcsec. In the case of FUV (Fig.~\ref{fig:PSFs}, left), a single slope of $\beta=-2.01$ is a good description of the outer part of the PSF. For both bands, the PSFs are extended to a radius of 750\arcsec, following a power law with the same exponent as the one measured in the outer parts of real stars. The final PSFs are shown in the bottom panels of Fig.~\ref{fig:PSFs}, along with the stars used to construct them.

\subsection{Remove the PSF effect from galaxy images}

Following the work of Golini et al. (in prep.), we use the new extended GALEX PSFs to deconvolve the background-subtracted intensity maps by applying the Wiener deconvolution algorithm. This algorithm uses the   \citet{wiener1949extrapolation} filter to recover a signal that has been affected by noise and degraded by an impulse response. Consider that a given image can be expressed as $y$:
\begin{equation}
    y=Hx+n
    \label{eq:orim}
\end{equation}
where $H$ is the impulse response (in our case the PSF), and $n$ is the noise of the image; the Wiener deconvolution returns an estimation of the signal ($\hat{x}$) deconvolved from the impulse response. This estimation mitigates the effects of noise in a poor signal-to-noise ratio through the application of the Wiener filter in the frequency domain:
\begin{equation}
    \hat{x}=F^{\dagger}({\Lambda_{H}}^{2}+\lambda{\Lambda_{D}}^{2})^{-1}\Lambda_{H}^{\dagger}Fy
    \label{eq:wienfil}
\end{equation}
where $F,F^{\dagger}$ are the Fourier and inverse Fourier transforms, $\Lambda_{H}$ is the transfer function, and $\Lambda_{D}$ is a filter to penalize the recovered image frequencies, with $\lambda$ tuning the balance between data and regularization. The deconvolution was developed by \citet{hunt1971matrix}. 

The Wiener deconvolution algorithm is implemented in the \texttt{Python} package \texttt{Scikit-Image}, using the functions \texttt{restoration.wiener} and \texttt{restoration.unsupervised$\_$wiener}\footnote{\href{https://scikit-image.org/docs/stable/api/skimage.restoration.html}{\texttt{Skimage.restoration} manual}}. We mainly use unsupervised Wiener deconvolution, where $\lambda$ is automatically estimated. However, in the few cases where this algorithm produces particularly noisy results, we use the first one setting $\lambda=0.5$ after some exploration. 

Although the Wiener algorithm is introduced to reduce the effect of noise, the deconvolved images are noisier than the original ones. For this reason, in the present work, which aims to obtaining the lowest possible surface brightness in the GALEX data, Wiener deconvolution is used to characterize the slope of the galaxy surface brightness profile beyond its edge. This region of the galaxy immediately after the edge is the one most affected by the PSF \citep[see e.g., ][]{sandin14,trujillo+16}, and therefore having information about its shape is key to producing a deconvolved model of the galaxy.

\begin{figure*}
    \centering
    \includegraphics[width=\linewidth]{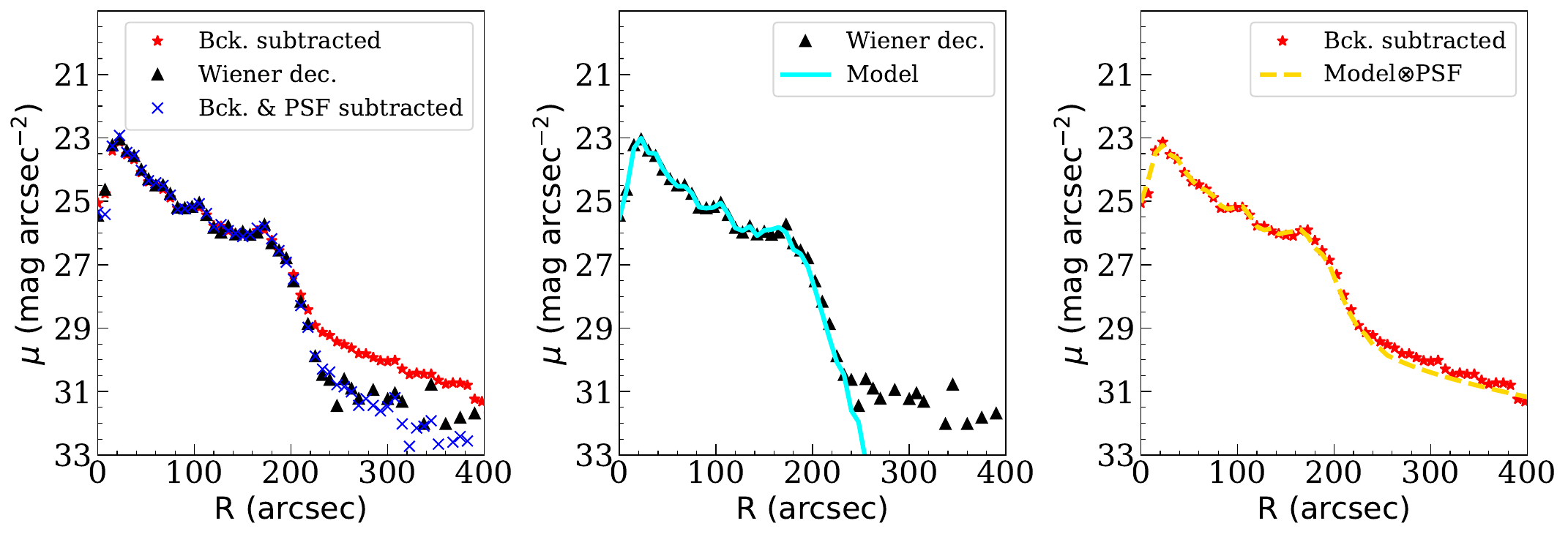}
    \caption{Example of the correction of the PSF effect applied to NGC 3486 in FUV. Left panel shows the profile after background subtraction (red), the profile from the Wiener deconvolved image (black), and the profile after applying all the PSF deconvolution procedures (blue). The middle panel shows the exponential fit (cyan) to the Wiener deconvolved image after the edge of the galaxy (in this case $R_{J}$=195\arcsec). The right panel shows the model (Wiener "core" plus exponential outer region) convolved with the extended PSF (yellow). This model is compared to the profile of the galaxy in the original with the background subtracted.}
    \label{fig:meth_NGC3486}
\end{figure*}

To deconvolve an image of the galaxy from the effect of the PSF while preserving the noise structure of the original GALEX images, we proceed as follows. First, the profiles of the Wiener deconvolved image of the galaxy and the original image of the galaxy are compared (see left panel of Fig. \ref{fig:meth_NGC3486}). The two surface brightness profiles are identical up to a radius where the effect of the PSF begins to be significant. For galaxies with well-defined borders, this radius coincides well with the edge of the galaxy \citep[see, e.g., ][]{Trujillo+20,Chamba+22}. Inside this radius we use the Wiener deconvolved image of the galaxy. Outside this radius we model the outer slope with an exponential model. We use the function \texttt{brokenexponential} from \texttt{Imfit} \citep{imfit} for that task. The inner part of our model of the galaxy is the Wiener deconvolved image up to a certain joining radius, $R_{J}$. In other words, we replace the inner exponential region of the \texttt{brokenexponential} model by the deconvolved image of the galaxy, while beyond $R_{J}$ the model follows a purely exponential decay characterized by a scale length $h_{out}$. The parameter $\alpha$ of the \texttt{brokenexponential} model is set to 1. The values of the free parameters ($R_{J}$ and $h_{out}$) that we have estimated for each galaxy are given in 
Table~\ref{tab:models}. An example can be seen in the middle panel of Fig.~\ref{fig:meth_NGC3486} (cyan line). We use this model to convolve it with the PSF, to compare with the original (see the right panel of Fig. \ref{fig:meth_NGC3486}). In most cases, this convolved model is a very accurate description of the original image of the galaxy, but some differences may be present, especially if the outermost part of the galaxy has some asymmetries such as stellar streams. For this reason, we subtract the convolved model from the original image of the galaxy, and the residuals are added to the deconvolved model of the galaxy.  The profile resulting of the entire process is shown in the left panel of Fig.~\ref{fig:meth_NGC3486} in blue. The equations describing the complete process are as follows:
\begin{equation}    
    G_{o}=(M\otimes \rm{PSF})+R\label{eq:gal_or}
\end{equation}
\begin{equation}
    G_{d}=M+R=M+[G_{o}-(M\otimes \rm{PSF})] \label{eq:gal_deconv}
\end{equation}
where $G_{o}$ is the original image, background subtracted, of the galaxy, $G_{d}$ is the final PSF deconvolved image of the galaxy, $M$ is the model obtained from the Wiener deconvolved image, and $R$ are the residuals of subtracting the PSF convolved model from $G_{o}$.

\begin{table}

    \caption{Joining radius $R_{J}$ and scale length $h_{out}$ used to model the outer part of the galaxies.}
    \centering
    \vspace{0.2cm}
    
        \begin{tabular}{cccc}
        \hline
        \hline\\
        Name &  \multicolumn{2}{c}{$h_{out}$}  & $R_{J}$ \\
         &  \multicolumn{2}{c}{[arcsec]}& [arcsec] \\
          &  FUV & NUV  & \\
          \hline
          NGC1042	&	16.50	&	17.39	&	156.0	\\
NGC2712	&	-	&	11.88	&	85.5	\\
NGC2903	&	44.30	&	28.68	&	352.5	\\
NGC3049	&	5.03	&	6.31	&	82.5	\\
NGC3198	&	70.67	&	30.00	&	217.5	\\
NGC3351	&	19.71	&	23.07	&	210.0	\\
NGC3368	&	13.80	&	31.94	&	255.0	\\
NGC3486	&	10.81	&	14.04	&	195.0	\\
NGC3596	&	-	&	13.63	&	111.0	\\
NGC3938	&	-	&	11.86	&	166.5	\\
NGC3972	&	-	&	15.64	&	69.0	\\
NGC4013	&	41.86	&	44.87	&	90.0 \tablefootmark{a}	\\
NGC4220	&	12.86	&	30.16	&	54.0	\\
NGC4307	&	15.97	&	26.11	&	40.5	\\
NGC4321	&	13.09	&	11.60	&	234.0	\\
NGC4330	&	27.17	&	29.90	&	121.5	\\
NGC5248	&	-	&	11.25	&	243.0	\\
NGC5866	&	66.83	&	59.81	&	108.0	\\
NGC5907	&	46.40	&	42.45	&	345.0	\\
IC3211	&	3.00	&	4.76	&	37.5	\\
          \hline
          \hline
        \end{tabular}
        \label{tab:models}
        
        \vspace{0.1cm}
        \tablefoot{
        \tablefoottext{a} {For the galaxy NGC 4013, the NUV the model is not joined with the Wiener image due to the  contamination of a bright star in the image. }
        
        }
        
\end{table}

\begin{figure*}
    \centering
    \includegraphics[width=0.9\textwidth]{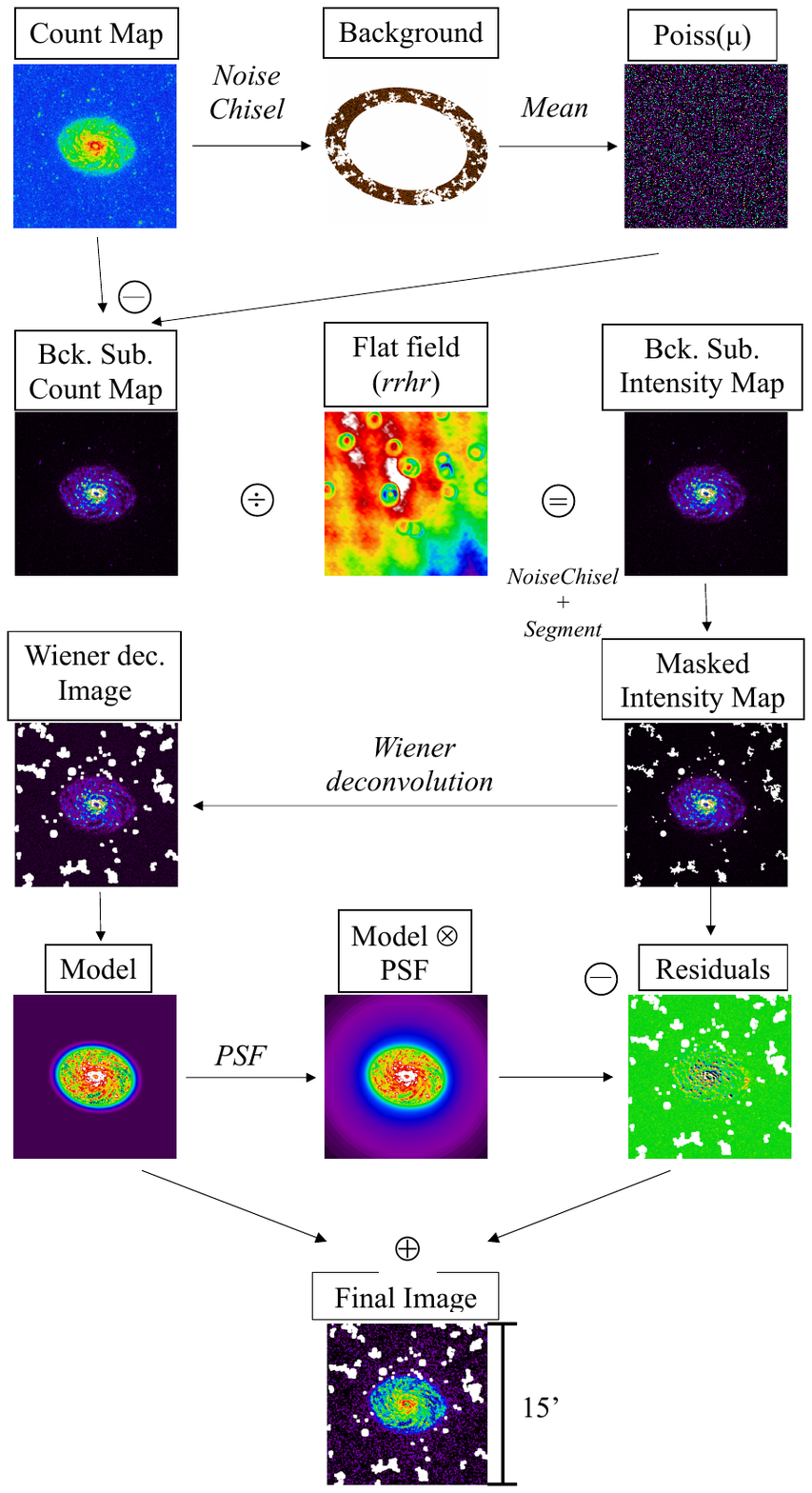}
    \caption{Flowchart of the methodology used for each galaxy, from the count map to the final background and PSF subtracted intensity map. NGC 3486 in FUV is used as an example. Wiener deconvolved image, model, residuals, and final image are smoothed for better visualization.}
    \label{fig:meth_summary}
\end{figure*}

We end this section by summarizing the process of both background subtraction (Sec.~\ref{subsec:bcksub}) and PSF deconvolution (Sec.~\ref{subsec:psf_sub}) of the galaxy in a single flowchart in Fig.~\ref{fig:meth_summary}.

\section{Results}\label{sec:results}

\subsection{Limiting surface brightness and magnitudes of the reprocessed GALEX dataset}\label{subsec:limits}

\begin{figure*}
        \centering

    \includegraphics[width=\textwidth]{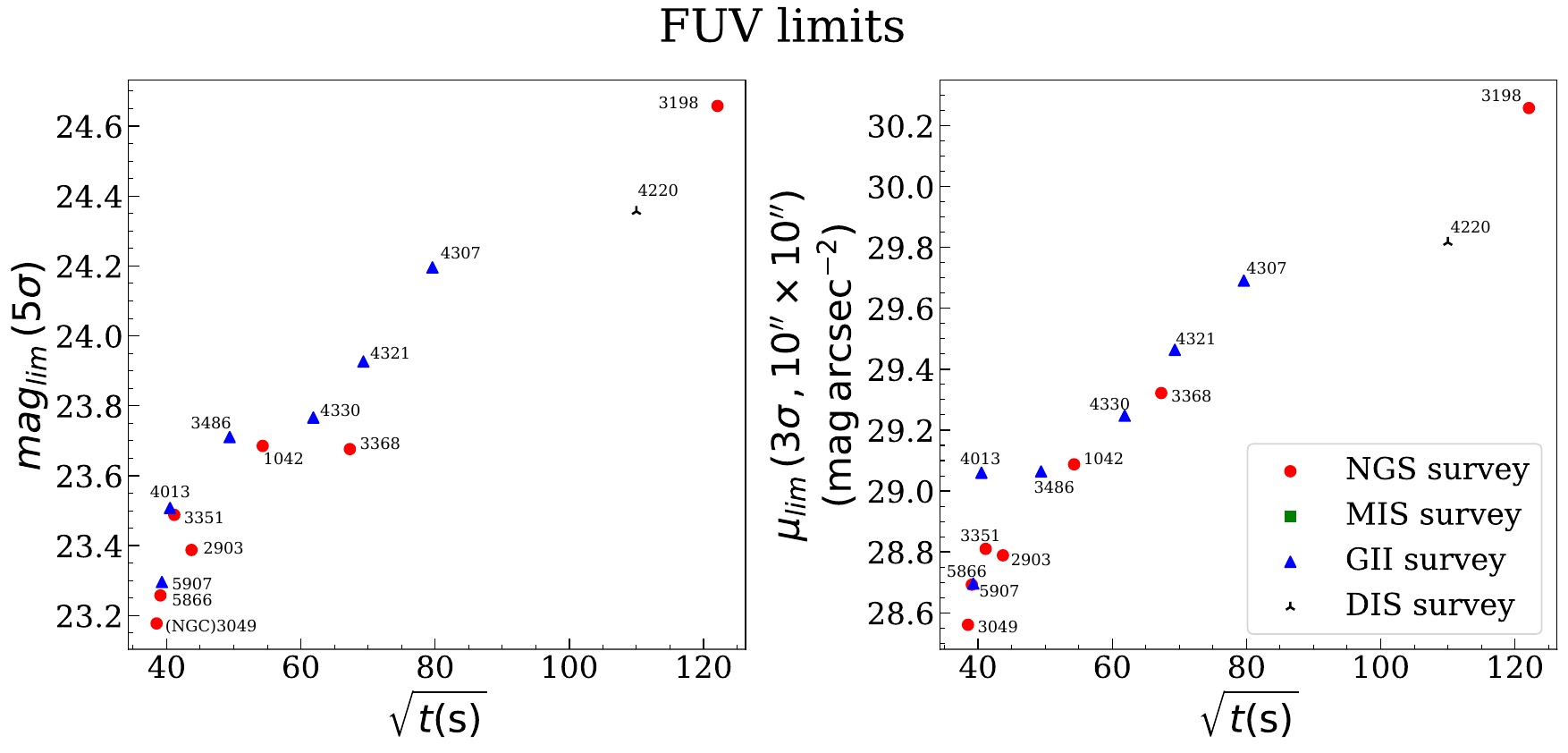}
    \includegraphics[width=\textwidth]{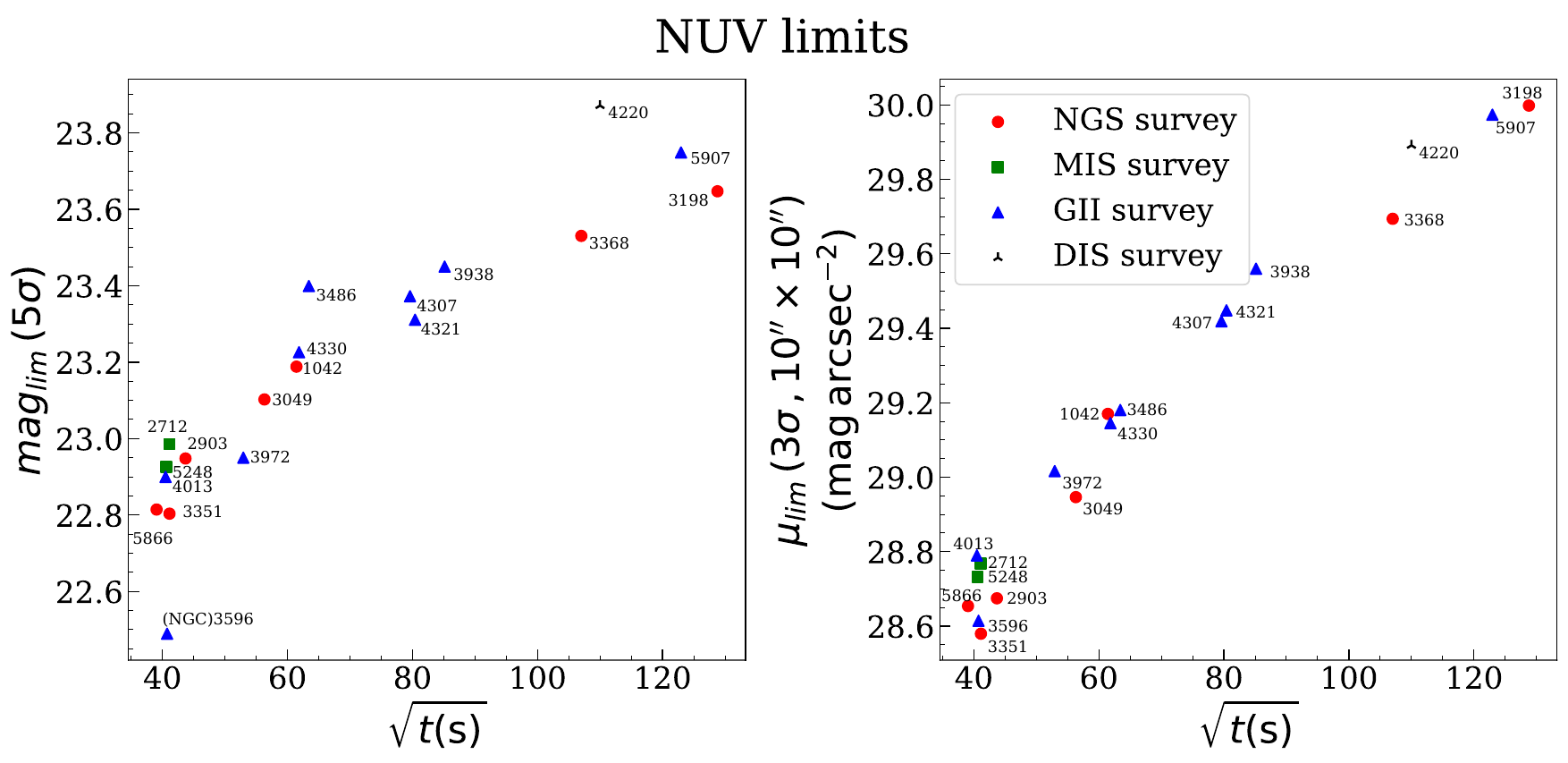}
    \caption{$5\sigma$ limiting magnitude (diameter of the aperture=$2\times \rm{FWHM}$) and $3\sigma$ surface brightness limit ($10\arcsec\times10\arcsec$ area) as a function of the square root of the exposure time in the FUV (top) and NUV (bottom) images. With some scatter, the depth of the data follows nicely the theoretical expectation ($\mu_{lim}\propto t^{1/2}_{exp}$). The numbers associated with the symbols indicate the NGC number of the galaxy observed by LIGHTS. }
    \label{fig:lims}
\end{figure*}

In addition to characterizing the depth of the data by quantifying the limiting magnitudes for point-like sources, in this low surface brightness oriented paper we measure the limiting surface brightness. We compute the surface brightness limits using \texttt{Gnuastro}'s program \texttt{MakeCatalog} \citep{MakeCatalog} with the standard low surface brightness metric of $3\sigma$ fluctuations in equivalent areas of $10\arcsec\times10\arcsec$ \citep[see e.g. ][]{trujillo+16,2020A&A...644A..42R}. In contrast, magnitude limits are obtained by measuring the equivalent $5\sigma$ background fluctuations in apertures (diameters) of $\varnothing=2\times \rm{FWHM}$ \citep[with FWHM being $4\arcsec.2$ in FUV, $5\arcsec.3$ in NUV, ][]{morrissey2007}. As shown in Fig.~\ref{fig:lims} for FUV (top) and NUV (bottom) , the surface brightness limits reached in the dataset after our background subtraction and source detection show the expected dependence on $t_{exp}^{1/2}$ (a Pearson test yields p-values $<1\times10^{-6}$), with values ranging from $\sim28. 5\rm{\,mag\, arcsec^{-2}}$ to $\sim30.5\rm{\,mag\, arcsec^{-2}}$. 

\subsection{Surface brightness and color profiles}\label{subsec:sbcp}

As mentioned in Sec.~\ref{sec:intro}, one of the main goals of this work is to connect the edges of galaxies located in ultra-deep optical images with sudden drops in the UV surface brightness profiles, and study the stellar populations around these locations using the UV bands. For that, we estimate the surface brightness profiles, using elliptical apertures. The ellipticity and position angles are obtained from the LIGHTS in the optical \textit{g}-band images (see Table~\ref{tab:data}). If FUV data are available, we also obtain the $(FUV-NUV)$ color profiles.  Depending on the apparent size of the galaxies, the width of the elliptical annuli varies. We use widths of $1.5\arcsec$, $3\arcsec$, $4.5\arcsec$, or $7.5\arcsec$. The profiles are obtained using the \texttt{Gnuastro} script \texttt{astscript-radial-profile} \citep{astscript-radial-profile}. In the cases of NGC 3972, NGC 4013, NGC 4220, NGC 4307, NGC 4330, NGC 5866, and NGC 5907, where the observed inclination is particularly high, the profiles are computed using wedge-shaped sectors centered on the major axis with apertures of $\pm5^{\circ}$ ($\pm3^{\circ}$ in NGC 4013, NGC 4330, and NGC 5907). We limit the surface brightness profiles when the data is fainter than $3\times\sigma_{bkg}$, i.e. when the surface brightness is below the limiting 3$\sigma$ surface brightness provided by the area of the annuli. The error bars are calculated by taking the square sum of each annulus standard deviation, where $\sigma_{bkg}$ is divided by the square root of the area of the annulus. 

The color profiles are computed only where the FUV surface brightness is above the surface brightness limit (since the FUV is generally the shallower band). The colors are corrected for Galactic extinction using \citet{cardelli+89} extinction curves for the Milky Way, as $E(FUV-NUV)=0.11E(B-V)$ \citep[see ][]{bianchi2017}. The optical extinctions ($E(B-V)$) are obtained from \citet{zaritsky2024lights}. In this work we do not consider the extinction associated with the internal dust of each galaxy. Although we do not expect a significant effect on the shape of the color profiles \citep[at least on the outer regions, where the star formation is expected to be very low, see e.g.][where they found only a $\sim$0.1 mag bluer color on LSB galaxies when taking internal dust extinction into consideration]{boissier+08}, it will be necessary to take it into account when measuring physical parameters such as star formation rates \citep[see e.g.][]{hao+11}.

\begin{figure*}
    \centering

        \includegraphics[width=0.9\textwidth]{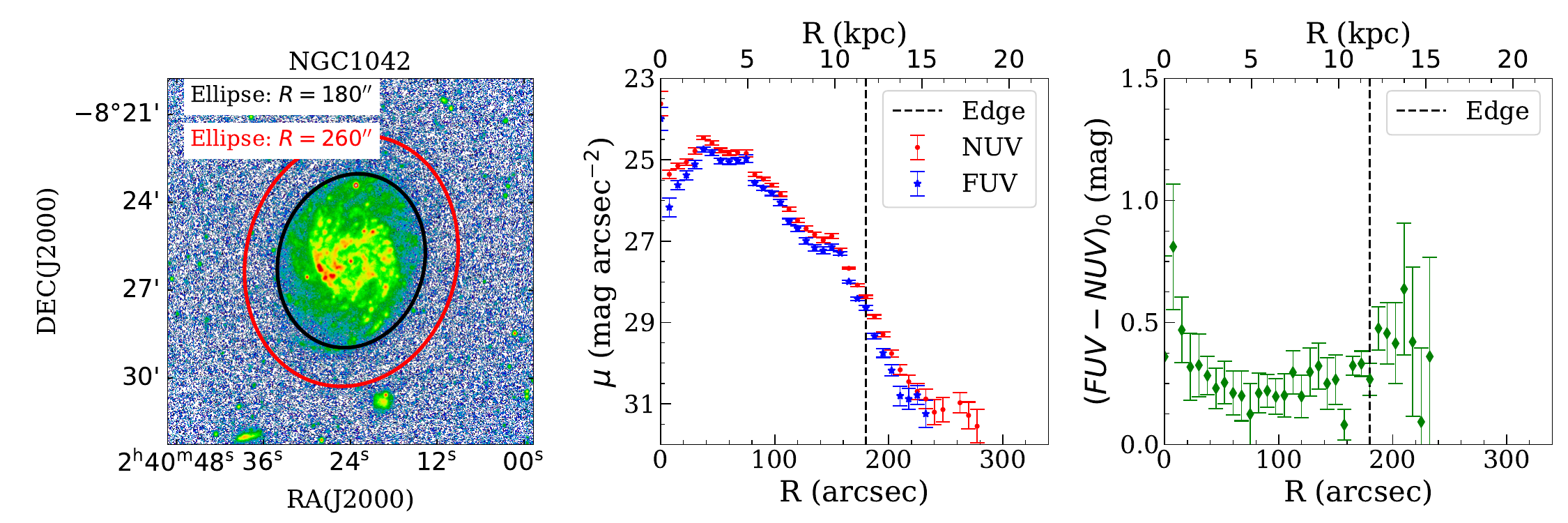}
        \includegraphics[width=0.9\textwidth]{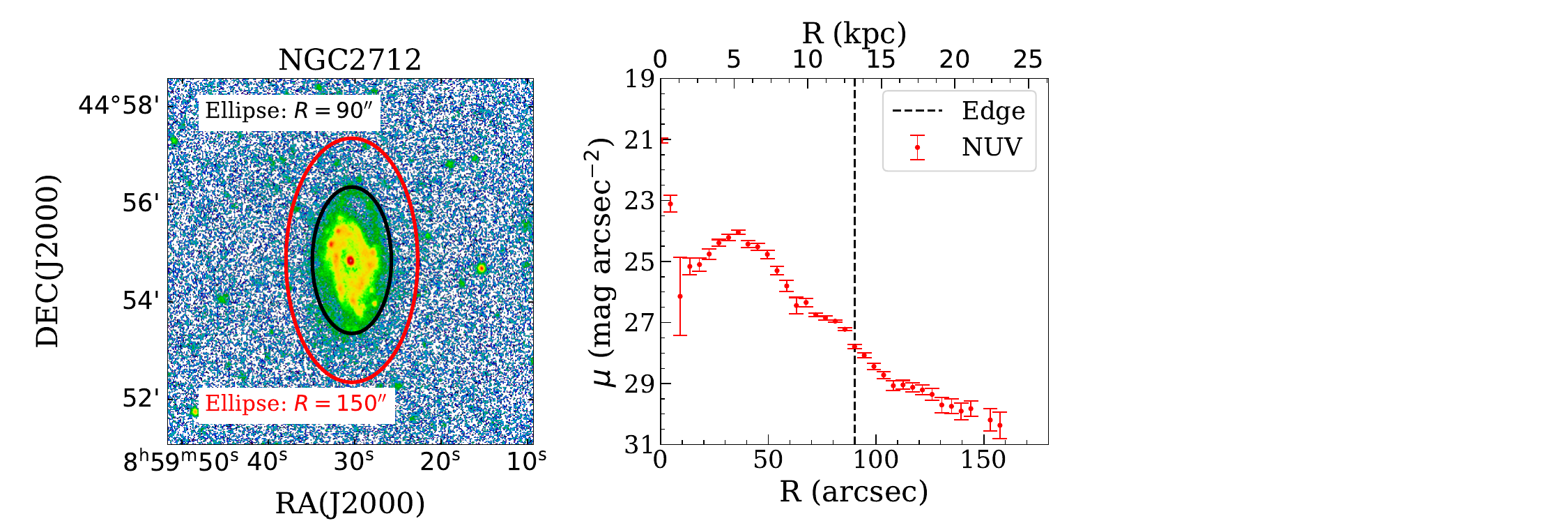}
        \includegraphics[width=0.9\textwidth]{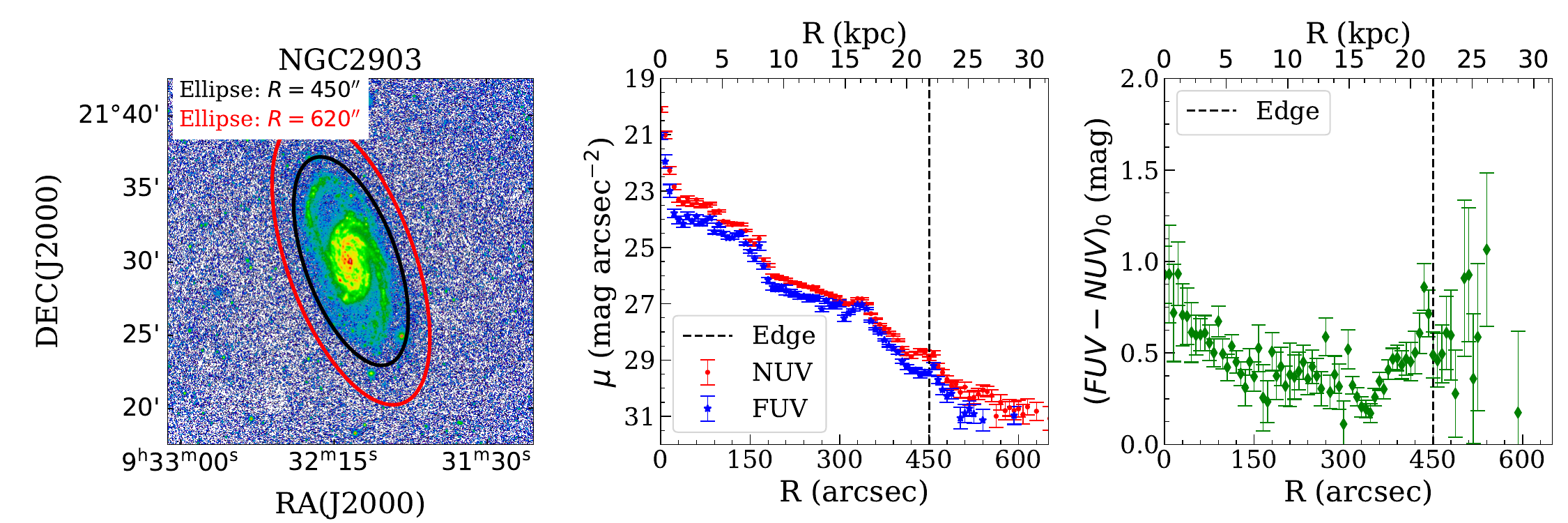}
        \includegraphics[width=0.9\textwidth]{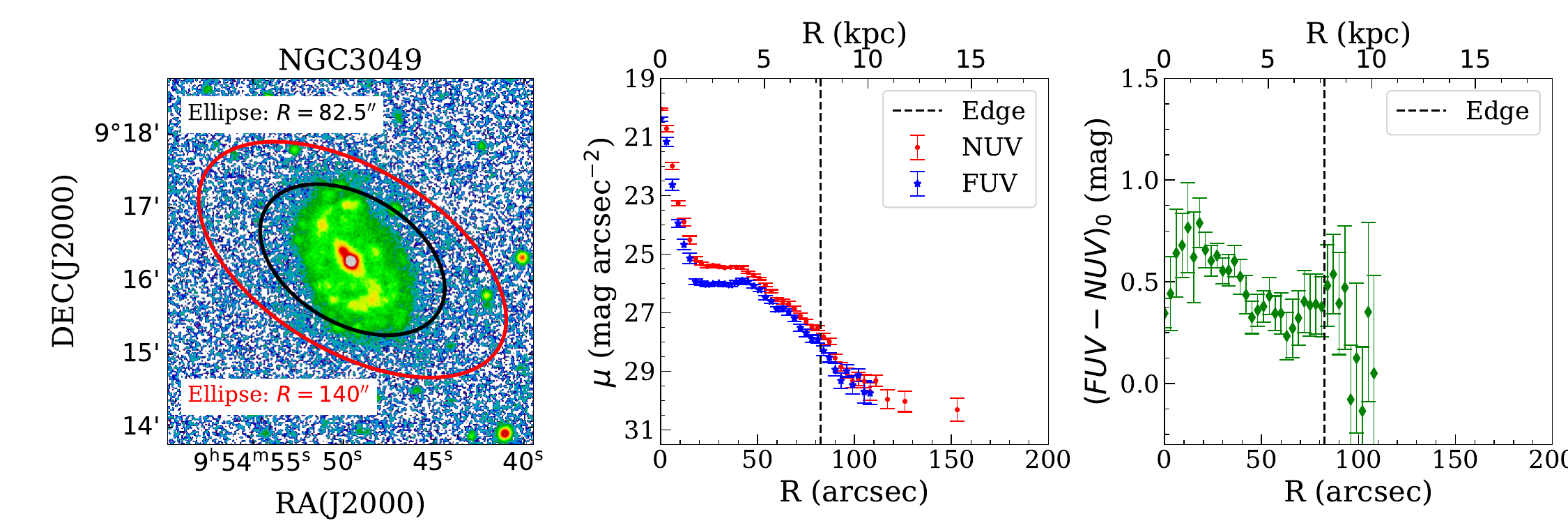}
     \caption{Left panel: NUV background subtracted images of the galaxies in our sample. Middle and right panels (where applicable): surface brightness and color profiles (corrected for Galactic extinction). These profiles have been corrected for the effect of the PSF, but not for galaxy inclination. The red ellipses represent the radii of the last measured point in NUV surface brightness profiles. The black ellipses represent the location of the galaxy edges \citep{Trujillo+20} according to the criteria described in \citet{Chamba+22} and \citet{2024A&A...682A.110B} (see Sec.~\ref{subsec:sbcp}). The semi-major radius of these ellipses are also shown as dashed vertical lines in the profiles. Black lines in the highly inclined galaxies represent the wedge shape used for the radial profile. The rest of the sample can be found in Appendix ~\ref{ap:rest_prof}. }
    \label{fig:profiles}
\end{figure*}

Fig.~\ref{fig:profiles} shows the surface brightness and color (if available) profiles for the galaxies in our sample. The edge radius  \citep[$R_{edge}$; ][]{Chamba+22} is represented by a black ellipse in the leftmost panels, while in red we represent the radius at which NUV surface brightness profiles are stopped. We identify the edge following the procedure outlined by \citet{Chamba+22} and \citet{2024A&A...682A.110B}: a sharp change in the slope of the surface brightness profiles in its outermost region. This edge should ideally coincide with the star formation threshold (past or present) of the disk \citep{Trujillo+20}. This should produce a reddening at the end of the $FUV-NUV$ profile for those galaxies where both FUV and NUV are available (rightmost panels in Fig.~\ref{fig:profiles}). If both criteria apply, the location of the edge is constrained by the color profile, and then the surface brightness profiles are examined to see if this reddening is at the change of the outermost slope. The NUV (and FUV) images are then examined in the UV images to see if this edge correlates with the visual edge of the galaxy. Using these criteria, an edge can be identified in 15 galaxies ($75\%$). Depending on the number of features that help to identify the edge, the total sample of 20 galaxies can be divided into 5 different cases:

\textbf{Case I. Spatial coincidence of the visual edge of the galaxy, sharp change in the surface brightness profiles, and reddening of the color profile.} For the galaxies NGC1042, NGC3351, NGC3486 and IC3211 the identification of the edge is the clearest one. It is based on both the observed break in the surface brightness profiles and the reddening in the color profiles. In these four cases, the identified radial position of the edge also agrees very well with the visual edge in the NUV images (see Fig. \ref{fig:profiles}). Note that in the surface brightness profiles the edge of the galaxy appears to be slightly shifted towards larger radii, as the definition of the edge here is the location of the intermediate point between two separate exponential behaviours (the inner and outer discs).

\textbf{Case II. Spatial coincidence of the visual edge of the galaxy and a sharp change in the NUV surface brightness profiles.} In the cases of NGC 2712, NGC 3596, NGC 3938, NGC 3972, and NGC 5248 no FUV and color information are available. However, in these cases we are also able to detect the edge, but only using the break in the NUV surface brightness profile as a selection criterion. In the first galaxy, the break at $\sim90\arcsec$ after a flat part of the profile is particularly clear (this is well above the surface brightness limit at this radius, see Table~\ref{tab:edges}). This flat part is caused by the outermost arm at the north of the galaxy. For NGC3596, in addition to the first break at $\sim110\arcsec$, we detect a second break at $\sim140\arcsec$. This further break seems to be related to an arm in the south- western part of the galaxy, which is particularly asymmetric along the major axis (note that opposite to this arm, i.e. in the north-eastern part, there is no light from the galaxy around the located edge; this could be related to the presence of a first break in the profile). In the cases of NGC3972 and NGC5248, there is a single break in the outermost part, close to the surface brightness limit of the images, but still observable. In these 5 cases, the detected breaks also correlate very well with the apparent sizes of the galaxies in the NUV images.

\textbf{Case III. No clear spatial coincidence of the visual edge of the galaxy and a sharp change in the surface brightness profiles with the reddening of the color profile.} For NGC 2903, NGC 3049, NGC 3368, NGC 4013, NGC4307 and NGC 4330, the reddening of the color profiles and the break in the surface brightness profiles are not exactly at the same radial position. In the cases of NGC2903 and NGC3049, the break is more clearly visible in the NUV profiles, and the reddening at this point is noticeable, but highly affected by noise (especially in NGC3049). In fact, in NGC2903 the break at $R_{edge}=450\arcsec$ is further away in NUV than the reddening by $10\arcsec$. This seems to be related to an arm in the north side of the galaxy (in Fig.~\ref{fig:profiles} the black ellipse is on top of this arm). In NGC4013 and NGC4330 there is a break in the NUV surface brightness profile coinciding with the visual extension of the galaxy, but the noise in the FUV prevents detection of this break in the surface brightness profile. This washes out a possible reddening at this point in the color profile. In NGC3368 the spatial coincidence between the NUV break and the reddening is less obvious, while the FUV break is clearer and present at the reddening point. Considering that the stellar population of this galaxy is particularly old according to the color profile ($(FUV-NUV)_{0}>0.5$), it could be that the galaxy beyond the edge is populated by older stars that emigrated there, still contributing to the NUV brightness of the galaxy. An extreme case where there is no spatial correlation between the break and the reddening is NGC 4307. In this galaxy the reddening is close to the center, while a break in the NUV is observed at $\sim85\arcsec$. While this break correlates with the visual boundary of the galaxy, the depth of the FUV data is not sufficient to detect this break in the FUV profile and thus in the color profile. Indeed, if we consider FUV as a tracer of star formation, with NUV more affected by older stars \citep[see ][]{hao+11}, one interpretation of this profile could be that the recent star formation activity of this galaxy is concentrated in the inner $50\arcsec$, and beyond this point the NUV light is associated with older stellar populations. However, in both of the 6 cases mentioned here, the location of the edge is in good agreement with the apparent visual edge in the NUV images.  

\textbf{Case IV. No clear spatial coincidence of the visual edge of the galaxy and a sharp change in the surface brightness profiles.} There are a few galaxies where, for various reasons, it is difficult to correlate a rim with an edge in the profiles. These are the cases of NGC 3198, NGC 4220, and NGC 5907. The scenarios are different for each galaxy. In the case of NGC4220, for example, the surface brightness and color profiles suggest that the galaxy is particularly compact. The scenario for this galaxy is similar to that of NGC4307 discussed above, except that in this case the surface brightness limit does not allow us to detect a break in the NUV profile where the edge appears on the image (about $110\arcsec$). In addition, as noted in Sec.~\ref{sec:samsec}, this galaxy is morphologically classified as S0-a, so the stellar population is expected to be older in this type of galaxy. More interesting are the cases of NGC3198 and NGC5907. In these cases we detect a first inner break in the color profiles (and a truncation in the surface brightness profiles in NGC 5907), but then we are able to detect a second further feature in the surface brightness profiles. In the case of NGC 3198, this second edge is associated with a low surface brightness spiral arm at the top of the galaxy, which we will analyze in more detail in Section~\ref{subsec:lsbfeatures}. In NGC 5907, the second feature is not observed in the FUV because the image is not very deep, but it is clearly visible in the NUV image. According to the NUV image (see Fig.~\ref{fig:profiles}), between the first truncation, determined here as $R_{edge}$ (which is in agreement with previous works such as \citealt{Martinez+19}), and the second feature there is an extension of the disk but lower in brightness, ending in a warped shape caused by its interaction with nearby or satellite dwarf galaxies \citep{1998ApJ...504L..23S, martinez+08}. Since truncations are associated with in-situ star formation \citep{Trujillo+20}, the detection of a second distant one could be related to the very recent ongoing star formation episode in the outermost region of the galaxy, triggered by a minor merger.

\begin{table*}[t]
    \caption{$R_{edge}$ location for the galaxies in our sample.} 
    \centering
    \vspace{0.4cm}

    \begin{tabular}{cccccccccc}
    \hline
    \hline
    \\
      Name  & \multicolumn{2}{c}{$R_{edge}$} & \multicolumn{2}{c}{$\mu\left(R=R_{edge}\right)$} & \multicolumn{2}{c}{$\mu_{lim}(3\sigma$; area annulus at $R_{edge}$)} &Criteria & \multicolumn{2}{c}{$R_{last}$} \\
         & [$^{\prime\prime}$] & [kpc] &\multicolumn{2}{c}{[mag arcsec$^{-2}$]} & \multicolumn{2}{c}{[mag arcsec$^{-2}$]} & & [\arcsec] & [kpc]\\
          & & & FUV & NUV & FUV & NUV & & &  \\
 \\
         \hline
\\
     NGC1042	&	$180.0\pm3.8$	&	$11.78\pm0.25$	&	28.95 & 28.50 & 31.40	&	31.48	&	1,2 & 260.0 & 17.02	\\
NGC2712	&	$90.0\pm2.3$	&	$13.18\pm0.33$	& - & 27.84 &	-	&	30.21	&	1 & 150.0 & 21.96	\\
NGC2903	&	$450.0\pm3.8$	&	$21.82\pm0.18$	& 29.71 & 29.00 &	31.20	&	31.01	&	1,2	  & 620.0 & 30.06 \\
NGC3049	&	$82.5\pm1.5$	&	$7.72\pm0.14$	& 28.58 & 27.76 &	29.79	&	30.24	&	1	 & 140.0 & 13.01 \\
NGC3198	&	$400.0\pm3.8$	&	$25.02\pm0.24$	& 29.26 & 29.21 &	32.61	&	32.35	&	1,2	 & 580.0 & 36.27 \\
NGC3351	&	$220.0\pm3.8$	&	$10.67\pm0.18$	& 28.02 & 27.68 &	31.15	&	30.92	&	1,2	 & 280.0 & 13.57 \\
NGC3368	&	$255.0\pm3.8$	&	$13.85\pm0.21$	& 29.27 & 28.33 &	31.58	&	31.99	&	2 & 400.0 & 21.72	\\
NGC3486	&	$210.0\pm3.8$	&	$13.85\pm0.25$	& 28.44 & 28.06 &	31.42	&	31.53	&	1,2 & 315.0 & 20.77	\\
NGC3596	&	$140.0\pm1.5$	&	$7.67\pm0.08$	& - & 28.63 &	-	&	30.33	&	1	& 160.0 & 8.77 \\
NGC3938	&	$170.0\pm2.3$	&	$10.38\pm0.14$	& - & 28.02 &	-	&	31.60	&	1 & 319.5 & 19.67	\\
NGC3972	&	$105.0\pm1.5$	&	$10.59\pm0.15$	& - & 27.43 &	-	&	29.05	&	1	& 130.0 & 13.11 \\
NGC4013	&	$150.0\pm2.3$	&	$10.62\pm0.16$	& 28.67 & 27.86 &	29.21	&	28.94	&	1 & 180.0 & 12.74	\\
NGC4220	&	$40.0\pm1.5$	&	$3.94\pm0.15$	& 26.25 & 25.33 &	29.30	&	29.37	&	1,2	 & 110.0 & 10.83\\
NGC4307	&	$85.0\pm1.5$	&	$8.24\pm0.15$	& 29.41 & 27.77 &	29.55	&	29.36	&	1	 & 110.0 & 10.67 \\
NGC4321	&	$280.0\pm3.8$	&	$20.63\pm0.28$	& 30.17 & 28.81 &	31.94	&	31.85	&	1	 & 420.0 & 30.95 \\
NGC4330	&	$95.0\pm2.3$	&	$12.76\pm0.30$	& 25.95 & 25.00 &	28.25	&	28.15	&	1	 & 150.0 & 20.14 \\
NGC5248	&	$280.0\pm3.8$	&	$20.23\pm0.27$	& - & 29.40 &	-	&	31.14	&	1	& 330.0 & 23.84 \\
NGC5866	&	$35.0\pm1.5$	&	$2.39\pm0.11$	& 25.50 & 23.86 & 	28.10	&	28.06	&	1,2	 & 220.0 & 15.04 \\
NGC5907	&	$330.0\pm3.8$	&	$26.40\pm0.30$	& 26.71 & 26.38 &	29.58	&	30.87	&	1,2	 & 600.0 & 48.00 \\
IC3211	&	$37.5\pm0.8$	&	$15.45\pm0.31$	& 27.95 & 27.72 &	30.31	&	30.04	&	1,2	 & 45.0 & 18.54\\

\\
     
     \hline
     \hline
    \end{tabular}
    \tablefoot{The criteria used to select the location of the edge: sharp change on the outermost slope of the surface brightness profiles (1) and/or color reddening (2) are provided. The error bars on the location of the $R_{edge}$ correspond to half the width of the elliptical annuli used to derive the surface brightness profiles. The observed (neither corrected of inclination nor Galactic extinction) surface brightness  at the edge of the galaxies (i.e. $\mu_{FUV}$(R$_{edge})$ and $\mu_{NUV}$(R$_{edge})$) are included. Surface brightness limits in the equivalent area of the annuli where $R=R_{edge}$ are also provided. We also provide the last radii of the surface brightness profiles ($R_{last}$), marked as a red ellipse on rightmost panels of Fig.~\ref{fig:profiles}. This radii corresponds to the last NUV point.}
    \label{tab:edges}
\end{table*}

\begin{figure*}
    \centering
    \begin{minipage}[t]{0.45\linewidth}
        {\includegraphics[width=\linewidth,height=0.999\linewidth]{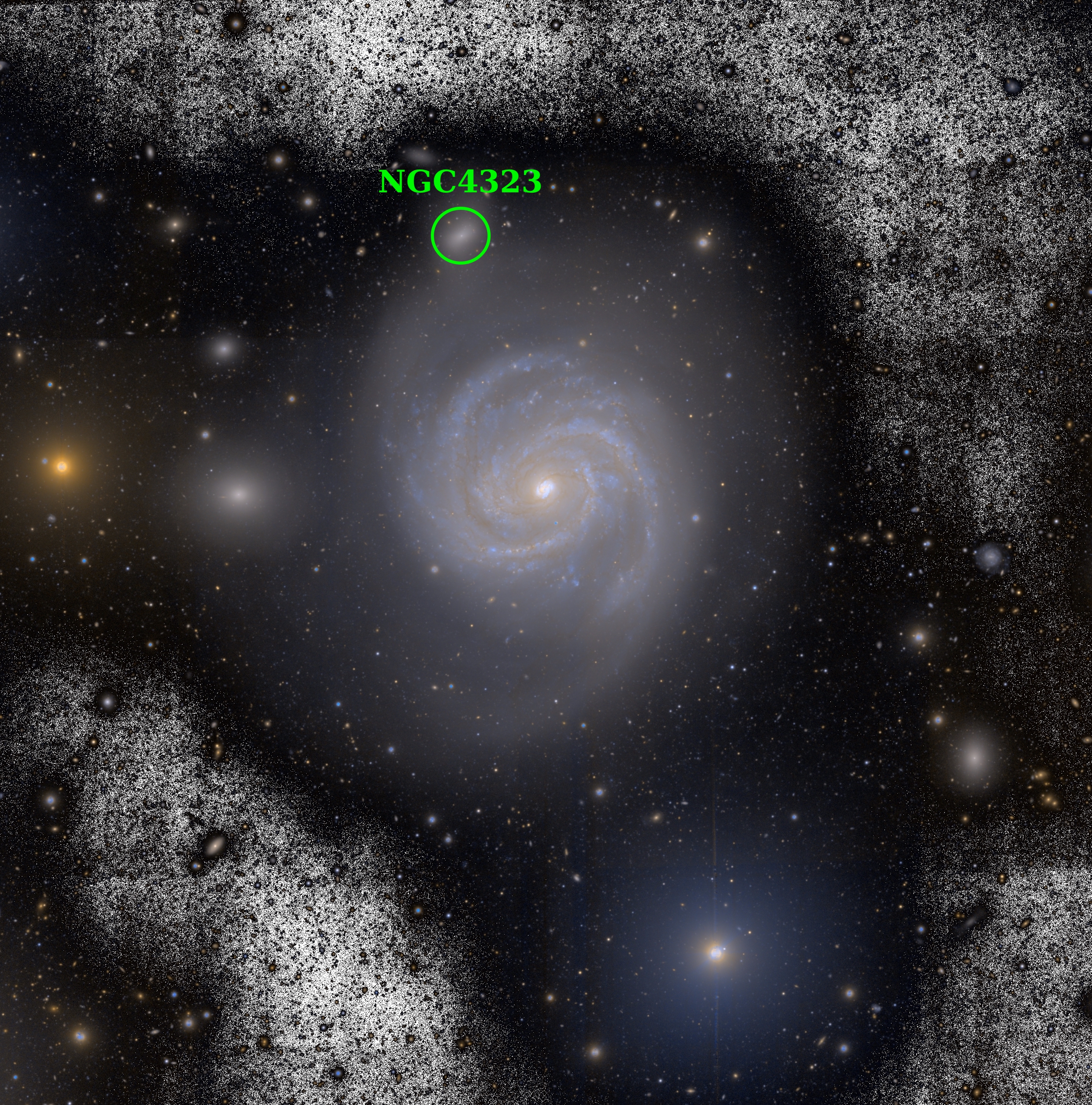}}
    \end{minipage}
    \begin{minipage}[t]{0.45\linewidth}
        {\includegraphics[width=\linewidth]{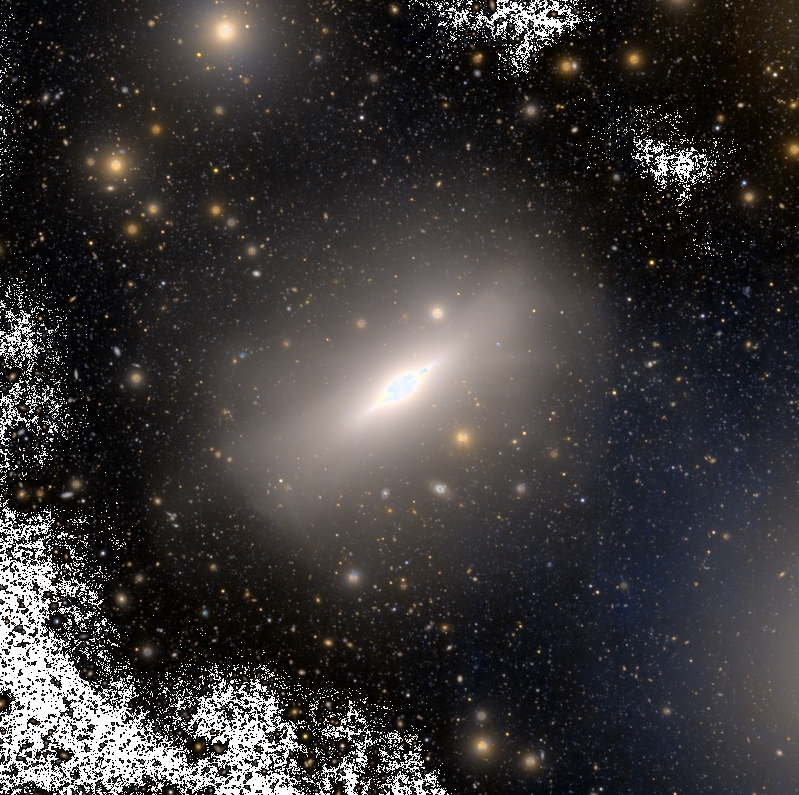}}
    \end{minipage}    
    \caption{LIGHTS color composed images of NGC 4321 ($21.5\arcmin\times21.5\arcmin$, left) and NGC 5866 ($15\arcmin\times15\arcmin$, right). NGC 4323 (in possible interaction with NGC4321) is marked with a green circle. The color composed images are constructed by combining \textit{g} (\textit{blue}), \textit{r} (\textit{red}) and a linear combination of the two ($(g+r)/2$, \textit{green}) using \texttt{Gnuastro}'s \texttt{astscript-color-faint-gray} \citep{astscript-color-faint-gray}.}
    \label{fig:opt_im}
\end{figure*}

\textbf{Case V. No clear edge identified}. Finally, in the cases of NGC 4321 and NGC 5866, there are no obvious edges in their surface brightness profiles. The explanation for this lack of edge comes from looking at the optical images. In Fig.~\ref{fig:opt_im} we show the LIGHTS color images in optical wavelengths for both galaxies. In both cases, the galaxies are undergoing a significant merger. In the case of NGC 4321, the uppermost dwarf galaxy (NGC 4323) appears to be interacting with the galaxy, causing the stream observed in Fig.~\ref{fig:opt_im}. This interaction washes out any potential edge of the galaxy. In addition, a deficit of neutral hydrogen has been observed \citep{cayatte+90} in this galaxy, which will also correlate with low star formation. In NGC 5866 the scenario is even more extreme. The presence of different tidal tails at different positions, together with the very low or null star formation activity apparently reflected in the very red UV colors beyond $R\sim40^{\prime\prime}$, can be linked to a history of multiple dry mergers for this galaxy. Despite the lack of clear edge, we set a $R_{edge}$ value for both galaxies according to the criteria described for the surface brightness and color profiles as in the rest of the sample.\\

\begin{figure*}
    \centering
    \includegraphics[width=\textwidth]{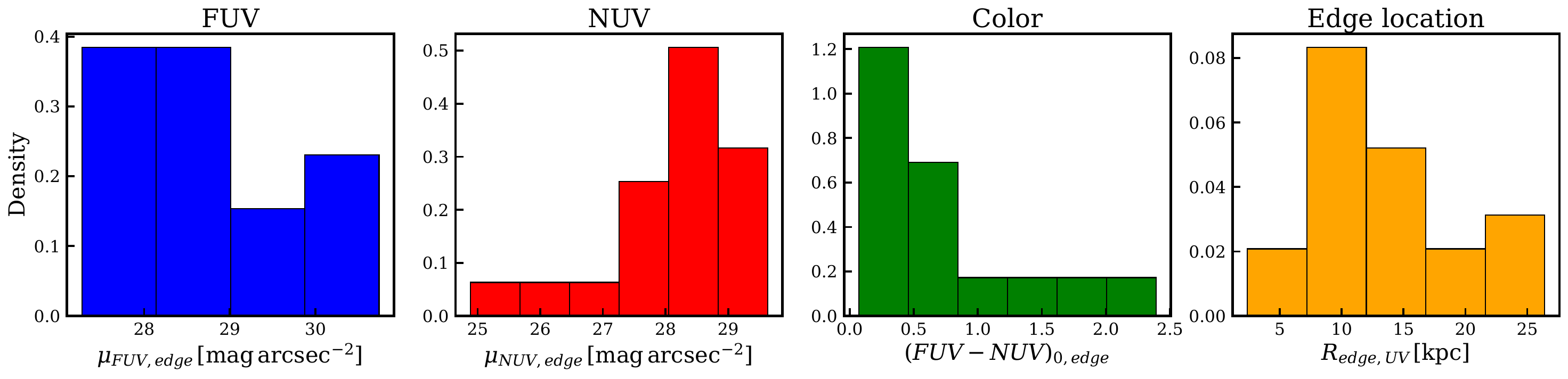}
    \caption{Distribution of surface brightness and color values of our galaxies at $R=R_{edge}$. Surface brightness values are corrected for Galactic extinction using $A_{NUV}=7.95E(B-V)$ and $A_{FUV}=8.06E(B-V)$ \citep[see ][]{bianchi2017}, and also for their respective inclinations using Eq. 2 from \citet{Trujillo+20} and assuming a ratio of scale height to scale length of $z_{0}/h=0.12$. The rightmost panel shows the distribution of the edge radius in the UV bands in kpc.}
    \label{fig:sb_redge}
\end{figure*}

The radial positions of the edges of the galaxies, together with the observed surface brightness values at these radial positions are given in the Table \ref{tab:edges}. The corrected surface brightness and color of the galaxies at their edge are given in the Fig. \ref{fig:sb_redge}. The values in this figure have been corrected for Galactic extinction and the inclination of the galaxies, so that they are all shown at the equivalent surface brightness they would have if they were face-on. The values of the edge location are also given in Fig. \ref{fig:sb_redge}.

\subsection{UV low surface brightness features in the outermost regions of the galaxies: the  extended UV disk of NGC 3198}\label{subsec:lsbfeatures}

\begin{figure}
    \centering
    \includegraphics[width=0.8\linewidth]{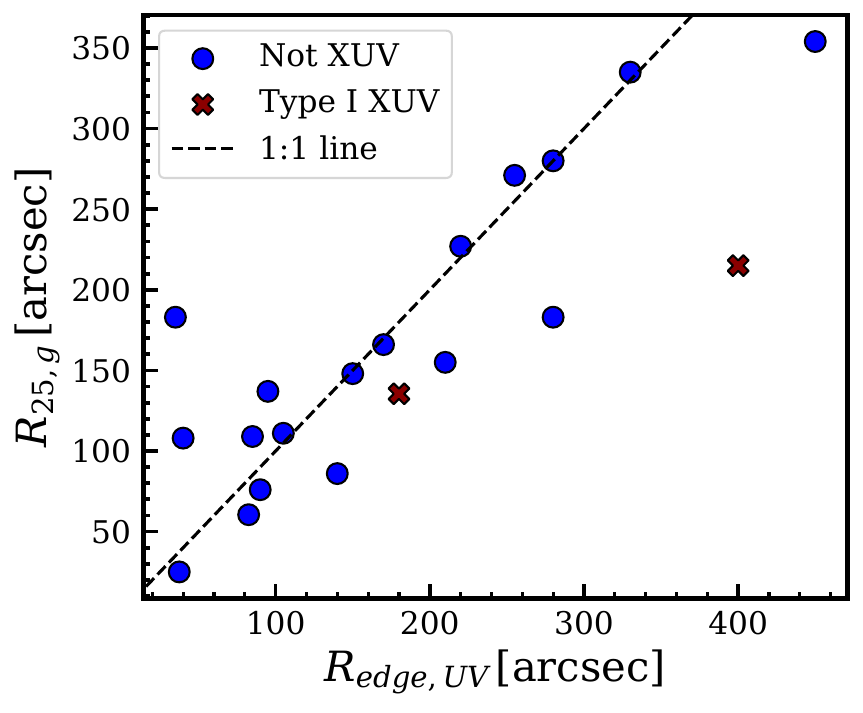}
    \caption{Comparison between the edge radius $R_{edge,UV}$ given in the Table ~\ref{tab:edges} and the optical radius $R_{25}$ measured on the LIGHTS \textit{g}-band images. The red cross corresponds to galaxies with an extended UV disk (i.e. Type I XUV) according to \citet{thilker+07}. The dashed line corresponds to the one-to-one line.  }
    \label{fig:comp_th07}
\end{figure}

One of the scientific highlights of GALEX was the detection of extended UV disks (XUV disks) in nearby spiral galaxies. \citet{armando+05} in NGC 4625 or \citet{thilker+05} in M83 presented the first results of star formation activity in the outskirts at larger distances than expected, i.e. far beyond the traditional optical extension given by the position of the isophote $26\,\rm{mag\, arcsec^{-2}}$ in B-band. Other works, such as \citet{zaritsky07} or \citet{herbert12} also explored the presence of star forming clumps beyond $R_{25}$ using GALEX data. One of the most complete studies of XUV disks can be found in \citet{thilker+07}, which analyzed 189 disk galaxies from the NGS survey and classified them into two types of extended disks, correlating these features with the imprints of galaxies growing in star formation activity at the outskirts. Two of the galaxies of our sample (NGC1042 and NGC3198) were  classified as Type I XUV \citep[see ][ Table 2]{thilker+07}. In Fig.~\ref{fig:comp_th07}, we compare the  $R_{25}$ values of our sample of galaxies with the edge radius $R_{edge,UV}$ we have found (Table ~\ref{tab:edges}). $R_{25}$ value is measured from the \textit{g}-band of LIGHTS. As observed in Fig. \ref{fig:comp_th07}, NGC1042 follows a similar trend as galaxies not classified by \citet{thilker+07} as XUV disks. However, NGC3198 appears to be an extreme case, with a significant discrepancy between $R_{edge}$ and $R_{25}$. Moreover, in the reanalysis of the UV data that we are doing in this paper, we are able to detect an extension of the disk up to at least up to $\sim520\arcsec$ ($\sim32.5$ kpc), $\gtrsim60^{\prime\prime}$ than the $D_{25}$ value \citep[][Table 1]{thilker+07}. Examining the NUV and FUV images, this extension seems to be associated with a low surface brightness ($\sim30-31\,\rm{mag\, arcsec^{-2}}$) spiral arm in the northern part of the major axis.

\begin{figure*}
    \centering
    \includegraphics[width=\textwidth]{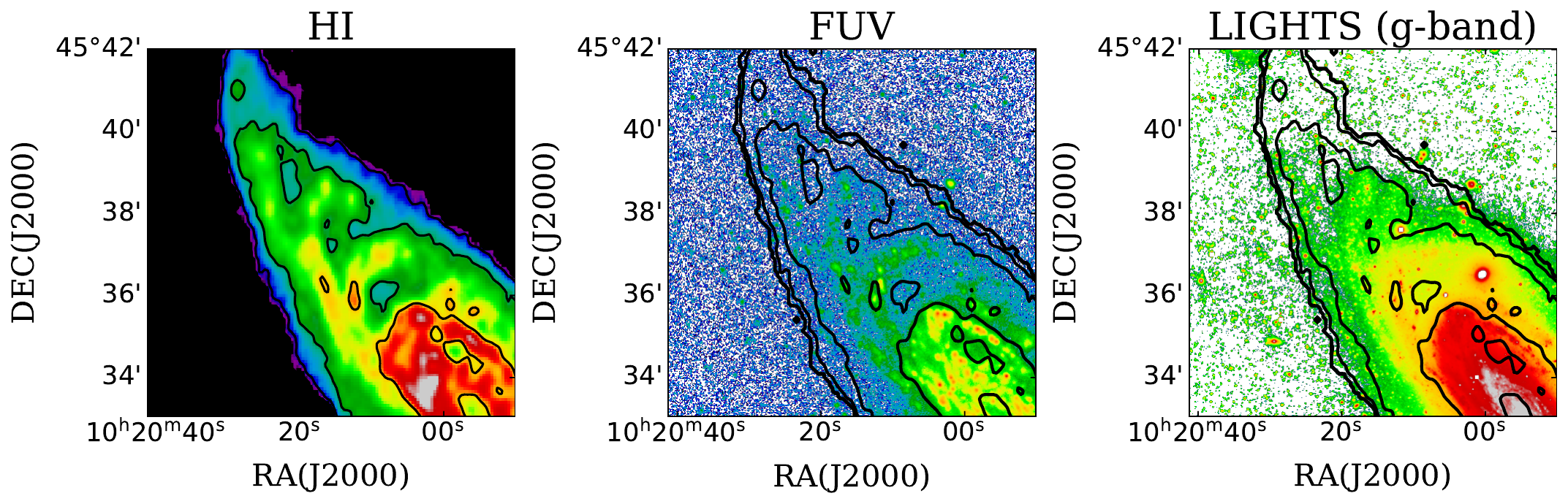}
    \caption{North-east part of NGC 3198 in different bands: H\textsc{i} map from \citet{gentile+13}, FUV band from this work, and LIGHTS g-band (scattered light subtracted) from LIGHTS survey. HI black contours are over plotted in the FUV and g-band imaging corresponding to the following HI densities: 0.1, 1, 5 and 15$\times10^{20}\rm{atoms\, cm^{-2}}$ (i.e. 0.08, 0.8, 4 and 12 $\rm{M_{\odot}pc^{-2}}$).}
    \label{fig:hi_fuv_g}
\end{figure*}

To better analyze this outer disk, we can compare our UV results with deep H\textsc{i} maps and ultra-deep optical images from LIGHTS. \citet{gentile+13} used data obtained with the Westerbork Synthesis Radio Telescope as part of the \textit{Hydrogen Accretion in LOcal GAlaxieS} (HALOGAS) survey \citep{heald+11} to show the extension of H$\textsc{i}$ in NGC 3198. The results of our work suggest that the distribution of H\textsc{i} in NGC 3198 is consistent with an extension of the stellar disk with the same position angle and axis ratio. In Fig.~\ref{fig:hi_fuv_g} we compare the H\textsc{i} map from \citet{gentile+13} with our FUV image and the LIGHTS g-band data after removing the scattered light from the bright top star \citep{sedighi2024lightsextendedpointspread}.

The H\textsc{i} map shows 3 spiral arms (green-yellow regions) whose boundaries are well defined by the H\textsc{i} contour $\Sigma_{HI}$=5$\times10^{20}$ $\rm{atoms\, cm^{-2}}$. This contour is equivalent to 4 $\rm{M_{\odot}\, pc^{-2}}$, where we use the following unit transformation: $1\,\rm{M_{\odot}\, pc^{-2}}=1.25\times10^{20}$ $\rm{atoms\, cm^{-2}}$ \citep[see Section 7.8 in ][]{condon2016essential}. In both the FUV and deep optical data, it is possible to see stellar emission within such a H\textsc{i} contour, but hardly anything beyond this gas density. This could suggest that there is a minimum gas density required for efficient star formation in this galaxy \citep[see e.g. ][]{schaye04,leroy+08,Trujillo+20}. Whether this is something we can generalize to other galaxies is beyond the scope of this paper and will be analyzed in a future paper.

\subsection{Comparison with standard GALEX pipeline data reduction}\label{sec:galcom}

In this paper we present an analysis of GALEX UV data based on a methodology optimized for low surface brightness studies. However, the GALEX archive also contains data products with a background subtraction and object detection pipeline that are in principle suitable for extracting the radial and color profiles of Sec.~\ref{subsec:sbcp}. The pipeline is described in detail in \citet{morrissey2007}. In the following sections we compare the surface brightness and color profiles obtained using the standard GALEX pipeline (hereafter referred to as GALPIP) with those using the low surface brightness methodology presented here. 

For the comparison, we use the background subtracted intensity maps (\textit{-intbgsub}) of the galaxies resulting from the GALEX pipeline. To facilitate the comparison, we use the same masking that we have derived in both datasets. In this section we present the cases where the differences are most pronounced, due to background (Sec.~\ref{subsec:disc_bck}) or due to PSF effects (Sec.~\ref{subsec:disc_psf}). The full set of differences for all galaxies between different methods is presented in Appendix~\ref{ap:comp}.

\subsubsection{Effects of masking and background}\label{subsec:disc_bck}

In Sec.~\ref{subsec:bcksub} we presented a methodology for background subtraction and source detection based on the \texttt{Gnuastro} tools and a global determination of the background as a Poisson distribution. This methodology differs from the one presented in \citet{morrissey2007}. In the GALEX pipeline, while a Poisson distribution is also considered for the background, it is determined locally in large bins across the image \citep[as described in ][section 3.3]{morrissey2007} and based on \texttt{SExtractor} source detection \citep{sextractor}. 

\begin{figure*}
    \centering
    \begin{minipage}[t]{0.49\linewidth}
        {\includegraphics[width=\linewidth]{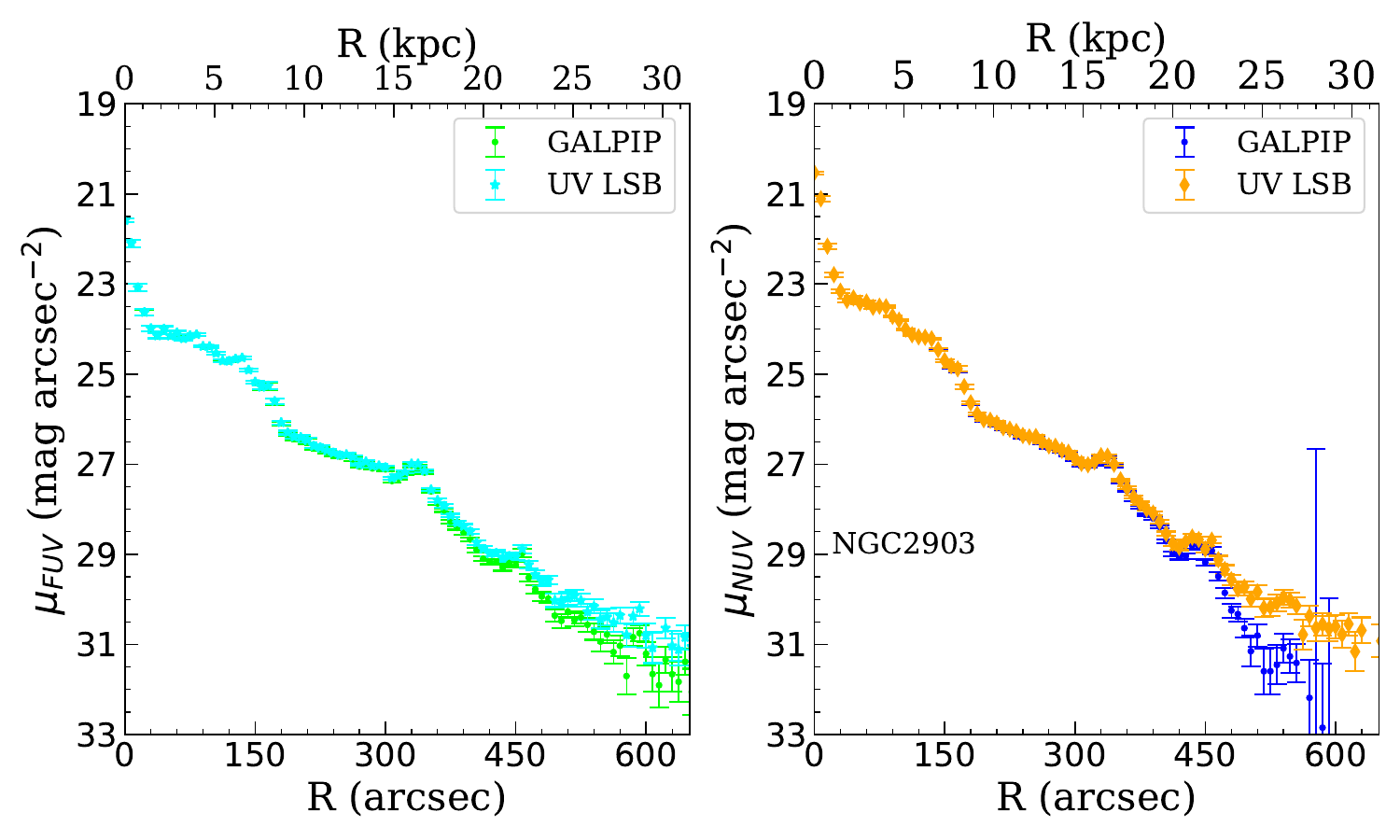}}
    \end{minipage}
    \begin{minipage}[t]{0.49\linewidth}
        {\includegraphics[width=\linewidth]{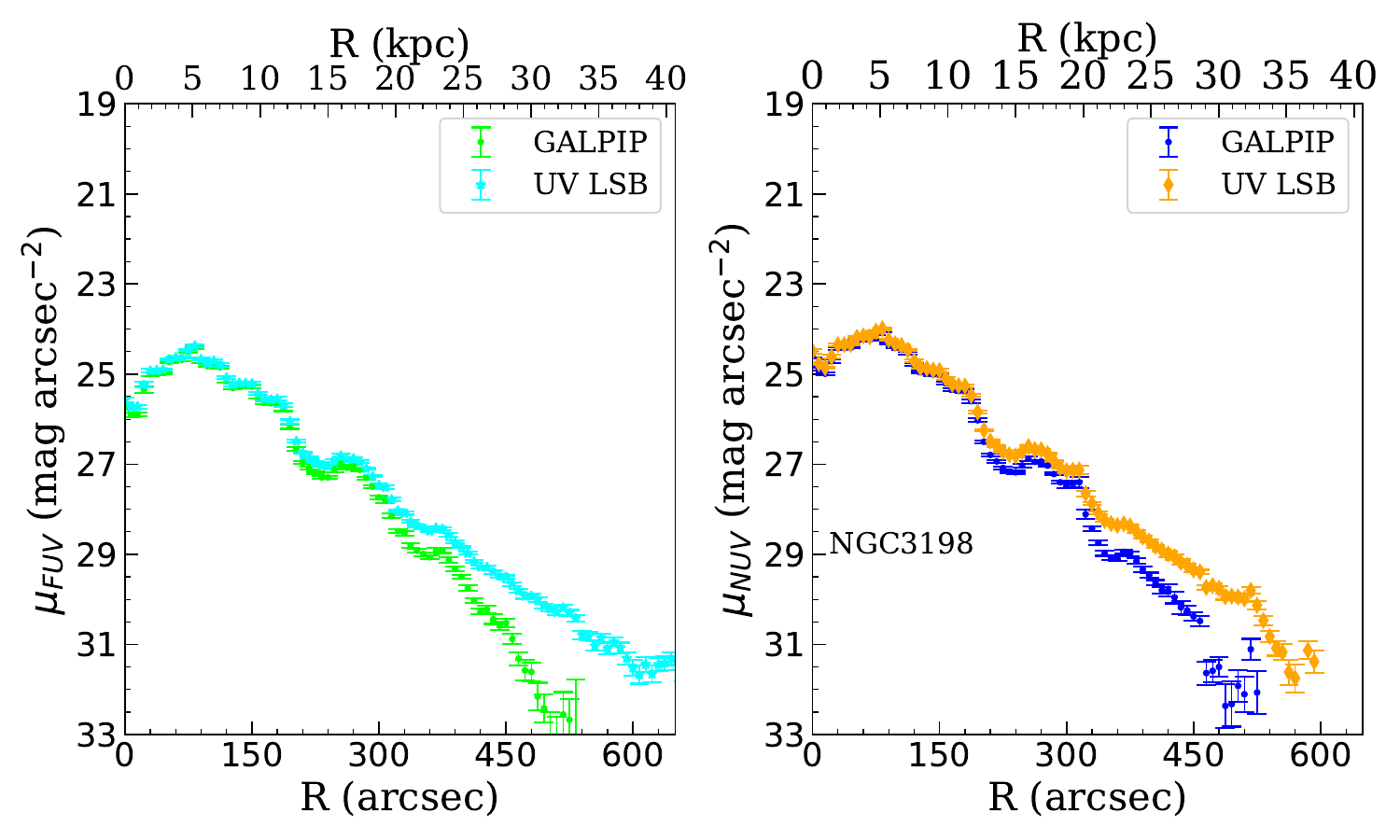}}
    \end{minipage}
    \begin{minipage}[t]{0.49\linewidth}
        {\includegraphics[width=\linewidth]{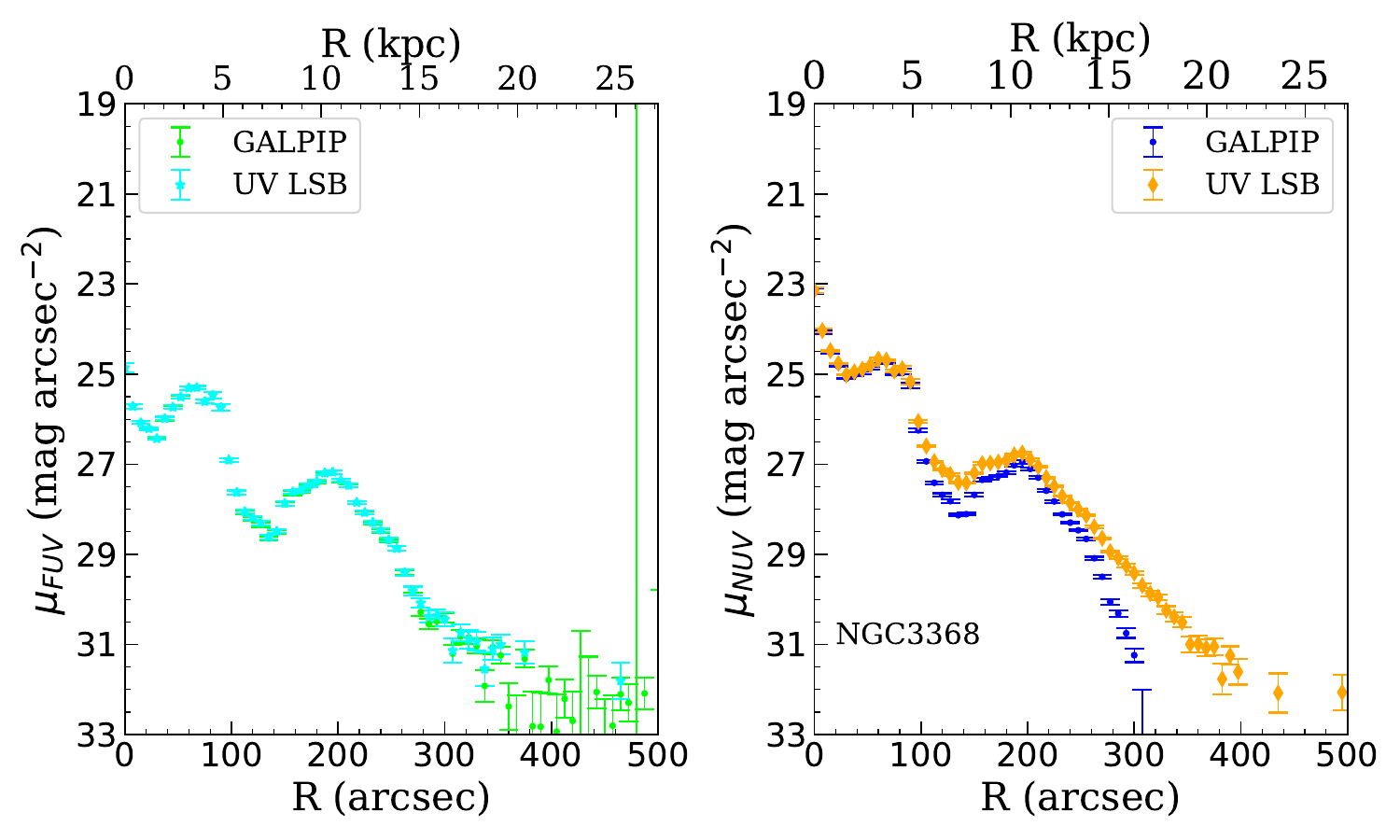}}
    \end{minipage}
    \begin{minipage}[t]{0.49\linewidth}
        {\includegraphics[width=\linewidth]{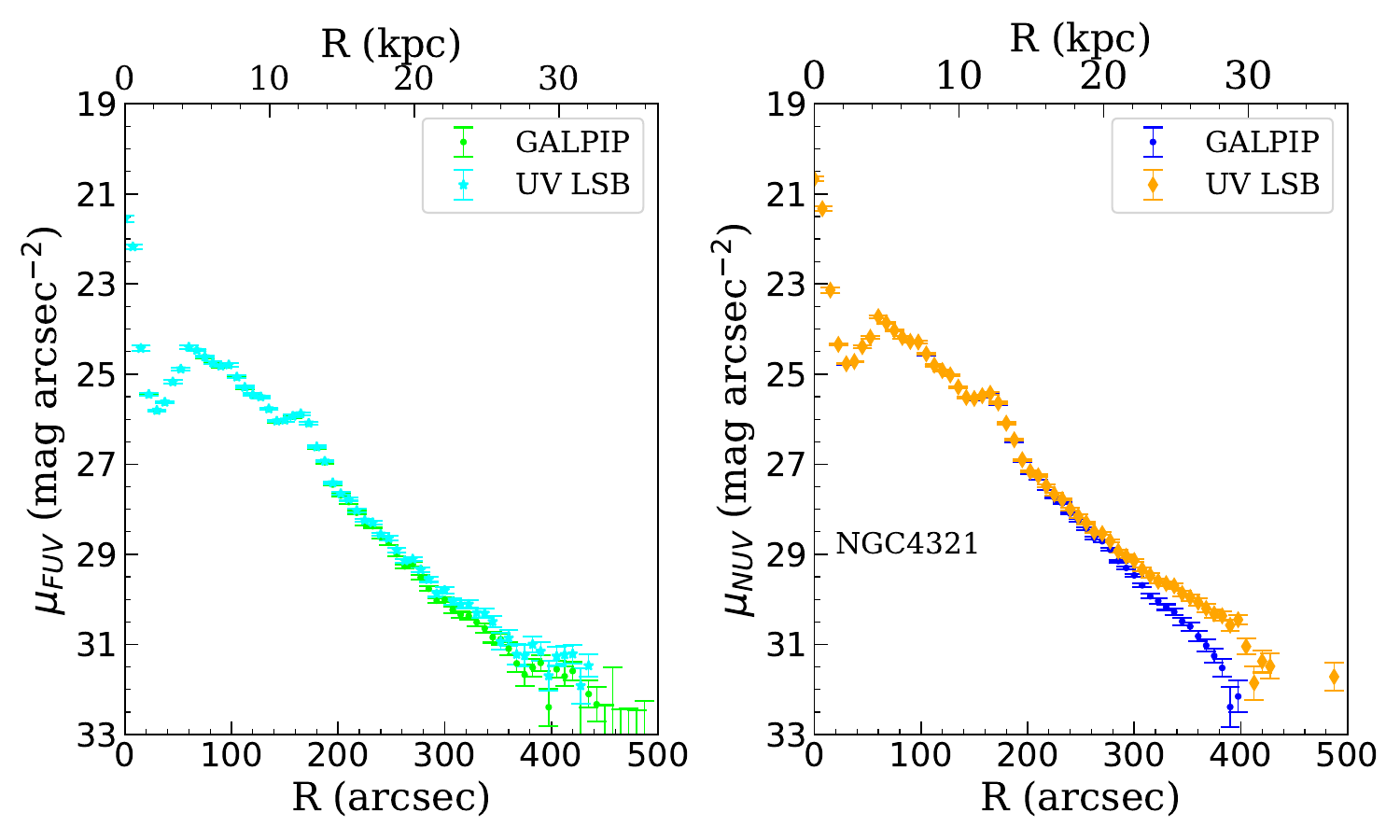}}
    \end{minipage}
    \caption{Comparison of the surface brightness profiles of four galaxies using the standard GALEX pipeline (named GALPIP) and those derived here using a different estimation of the background (\textit{UV LSB}). For each galaxy, the right panel is NUV and the left panel is FUV.}
    \label{fig:skycomp}
\end{figure*}

A different background estimation between the two methods (ultimately based on different areas and masking of contaminating sources) can lead to different surface brightness profiles. Here we make a detailed comparison in those cases where the surface brightness profiles differ more: NGC 2903, NGC 3198, NGC 3368, and NGC 4321. We have labeled the profiles obtained with our background estimation methodology as \textit{UV LSB}. Fig.~\ref{fig:skycomp} shows a comparison between the GALPIP and the \textit{UV LSB} profiles for the four galaxies mentioned above. We find two clear trends: a) the GALPIP profiles have systematically a lower intensity in the outermost regions of the galaxies and b) the difference in the surface brightness profiles are more important (in general) in the NUV than in the FUV profiles.

To test which of the two background estimates is more likely to produce a better representation of the outer part of the galaxies, we have analyzed the asymptotic values of the counts of the profiles in the NUV bands. For accurate background estimation and removal, we would expect the values in the outer part of the profiles to tend to zero. This is not the case for the GALPIP profiles, where NUV presents profiles with asymptotic values of $-0.19\pm0.04$, $-0.93\pm0.31$, $-0.25\pm0.20$, and $-0.36\pm0.13$ counts in NGC 2903, NGC 3198, NGC 3368, and NGC 4321, respectively. These results already indicate the presence of consistent negative backgrounds in GALPIP in the vicinity of these four galaxies. This does not happen by construction in our method, since the background estimation is done in the immediate vicinity of the galaxies. 

\begin{figure}
    \centering
    \includegraphics[width=\linewidth]{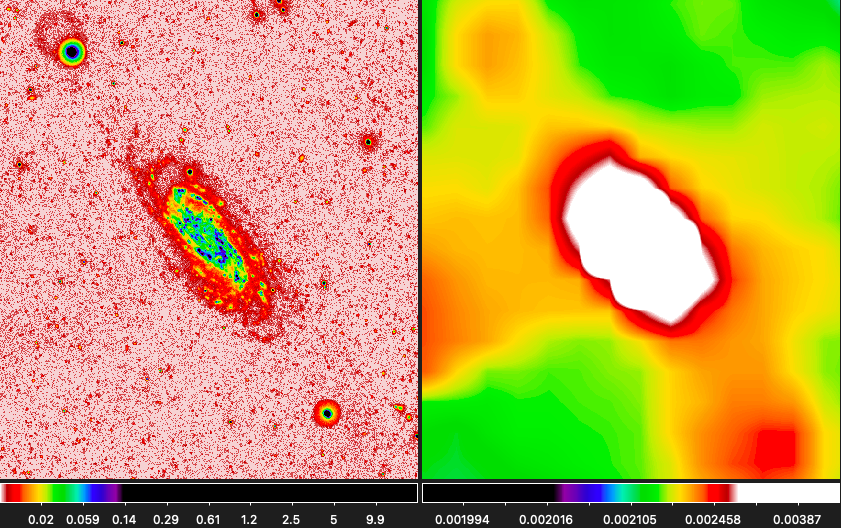}
    \includegraphics[width=\linewidth]{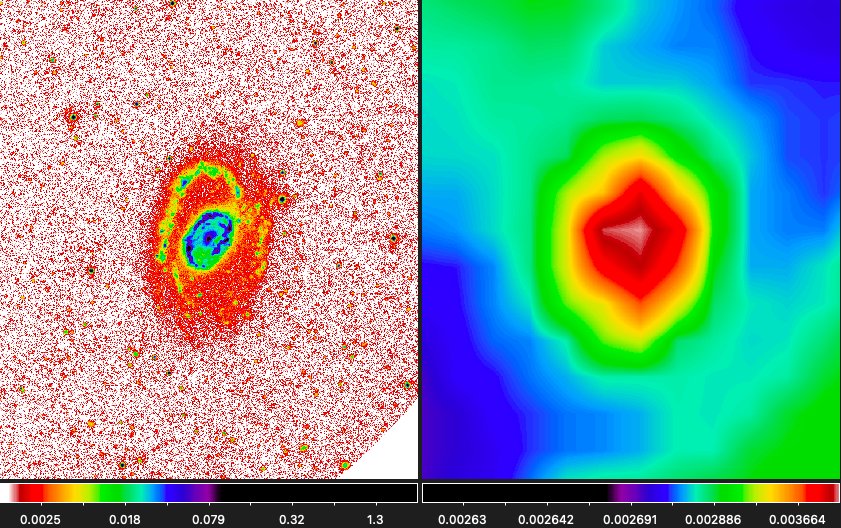}
    \caption{SAOImageDS9 image of NUV NGC 3198 (top) and NGC 3368 (bottom) intensity (left) and background (right) maps from GALPIP. The contrast is leveled to see the full galaxy and to appreciate the structure in the background.}
    \label{fig:3198_3368bck}

\end{figure}

\begin{figure}
    \centering
    \begin{minipage}[t]{0.49\linewidth}
        {\includegraphics[width=\linewidth]{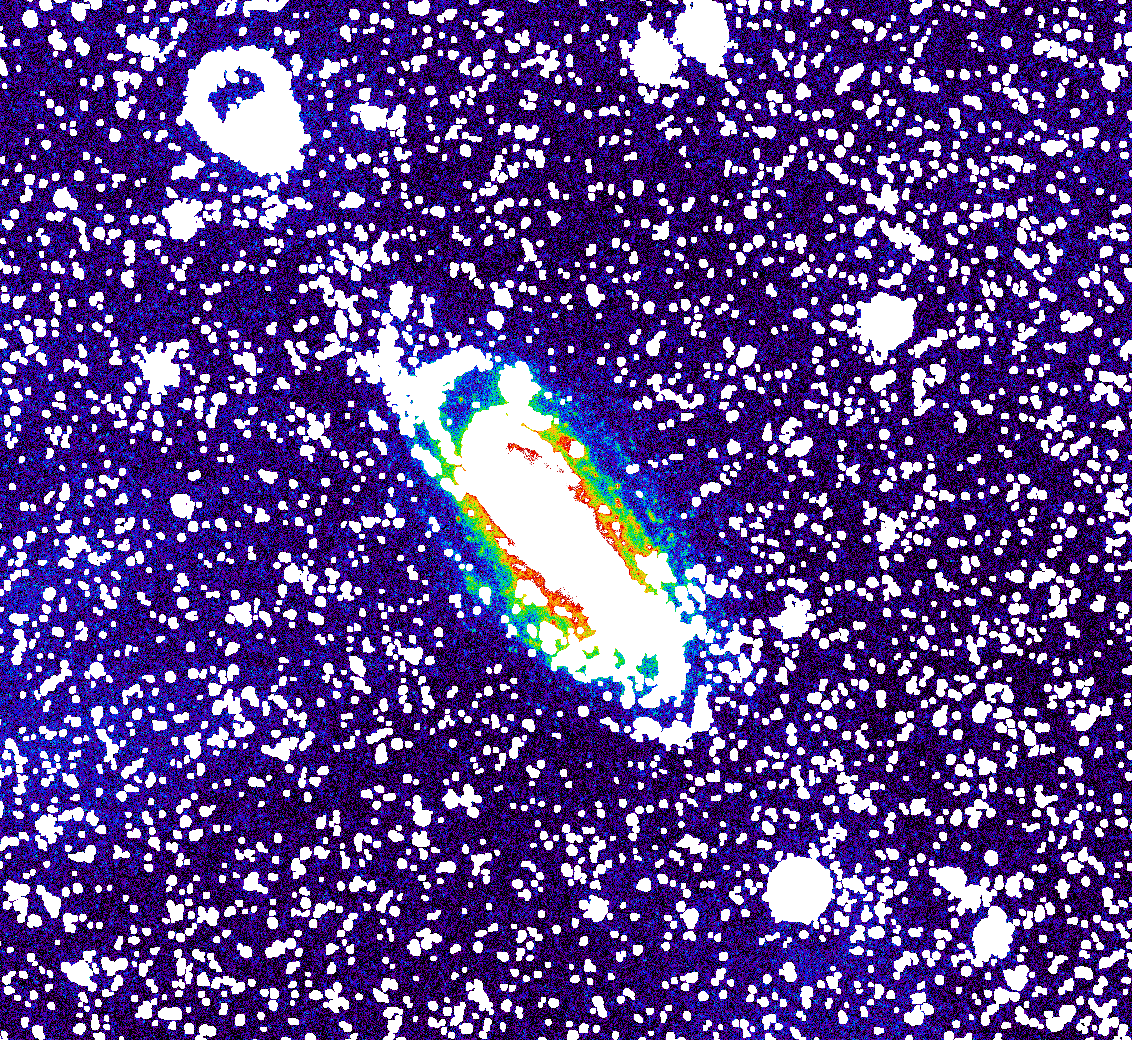}}
    \end{minipage}
    \begin{minipage}[t]{0.49\linewidth}
        \includegraphics[width=\linewidth]{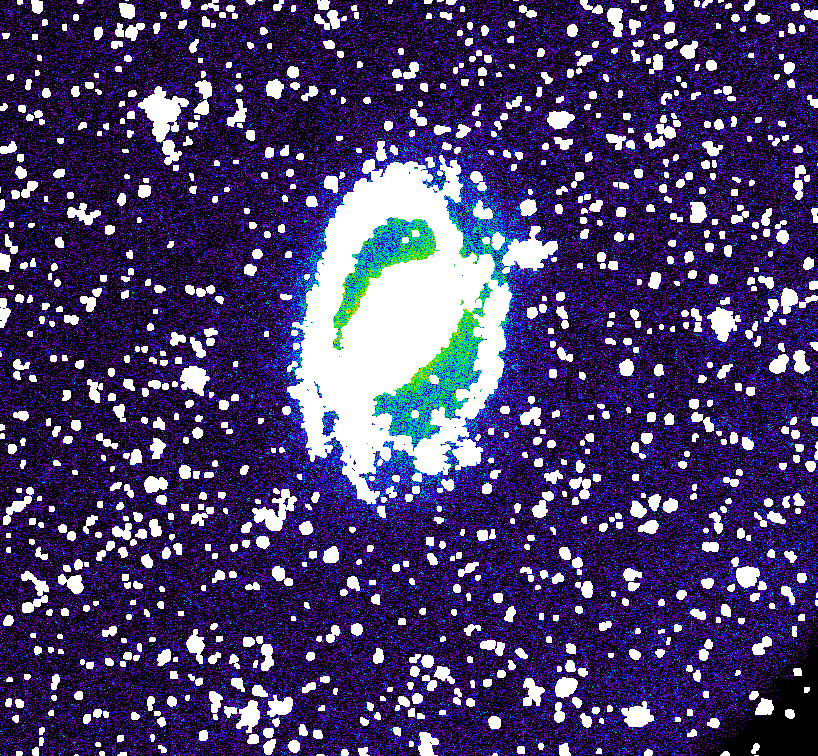}
    \end{minipage}
    \caption{NGC 3198 (left) and NGC 3368 (right) NUV count maps masked with the \textit{-objmask} files available in GALEX archives.}
    \label{fig:objmask}
\end{figure}

To determine the reasons for the differences in background estimates between GALPIP and our method, we examine the GALPIP background maps (\textit{-skybg} in the GALEX archive). In the cases of NGC 3198 and NGC 3368 (see Fig.~\ref{fig:3198_3368bck}), the GALPIP background maps show a complex structure that seems to reflect the effect of scattered light from the brightest sources (both galaxies and bright stars) on the images. It is likely that these structures are residuals left by inaccurate masking of the brightest objects. This is supported by the masks associated with these galaxies, which can be found in the GALEX archive (Fig.~\ref{fig:objmask}). In addition to having larger masks, our methodology involves measuring the background at a sufficient distance from the galaxy to minimize the potential for light scattered from the galaxy itself.

\begin{figure}
    \centering
    \includegraphics[width=\linewidth]{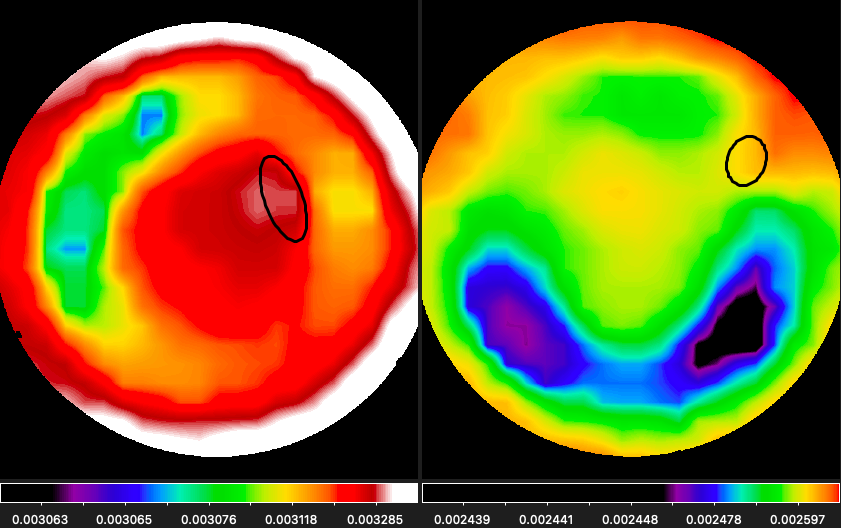}
    \caption{NGC 2903 (left) and NGC 4321 (right) NUV background maps from GALPIP. The positions of the galaxies are indicated by black ellipses with a semi-major length corresponding to $R=R_{edge}$ (see Table~\ref{tab:edges}).}
    \label{fig:29034321bck}
\end{figure}

The cause of the background oversubtraction in NGC 2903 and NGC 4321 by GALPIP is less clear. Examining the GALPIP background maps (see Fig.~\ref{fig:29034321bck}), a large ring-shaped structure is observed around the center of the image. NGC 2903 and NGC 4321 lie in a transition zone between lower (green-yellow) and higher (red-white) background levels. Accordingly, the GALPIP profiles could be affected by different backgrounds and therefore be prone to oversubtraction. The reason for the presence of this global ring in the NUV images is not clear. We note that similar structures are present in other NUV backgrounds, but they do not lead to different background estimates between GALPIP and our method. 

It is worth noting that since the background estimate is critical to obtain accurate surface brightness profiles in the outermost regions of galaxies, some works \citep[see e.g.][]{GildePaz07,bianchi14} have also followed their own background estimate instead of the GALPIP one. In the Appendix ~\ref{ap:comp_prev}, we present a comparison between our work and \citet{Bouquin18}, which also uses a different background subtraction strategy than the GALEX background maps. Except for three highly inclined galaxies (NGC4307, NGC4330, and NGC5907), the agreement between their work and ours is very good, with our profiles generally extending further out than theirs. Since the difference in the surface brightness profiles of the high-inclination galaxies is mostly in the central region, we speculate that the difference with these authors may be related to the strategy used to represent or not the dust lane in the profiles.

\subsubsection{Effects of the PSF in the surface brightness profiles}\label{subsec:disc_psf}

\begin{figure*}
    \centering
    \includegraphics[width=\textwidth]{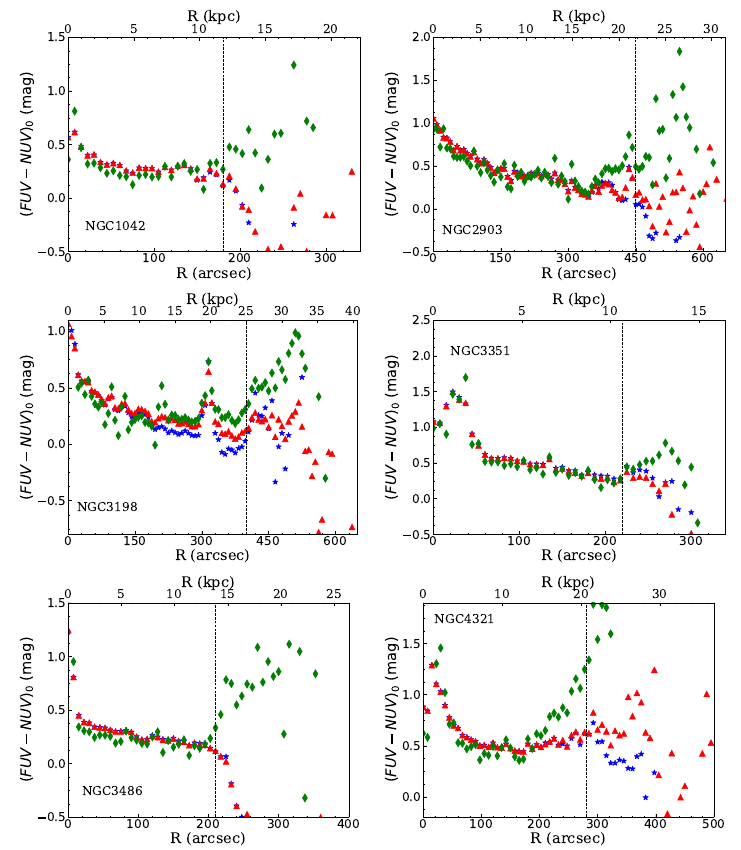}
    \caption{Radial UV color profiles for 6 different galaxies. The blue profiles correspond to the color profiles using the background estimate from the standard GALEX pipeline (here called GALPIP). The color profile resulting from our background estimate (\textit{UV LSB}) is shown in red. When the effect of the PSF is taken into account (green profiles), the shape of the color profiles changes dramatically in the outermost region beyond the galaxy edge. The edge position is marked as a vertical dashed line.}
    \label{fig:color_comp}    
\end{figure*}

As mentioned above, a different background estimate could lead to different surface brightness profiles in the outermost regions. Here we explore how taking into account the effect of the PSF of the galaxy itself can introduce an additional change in the outer part of the galaxies. This effect will be particularly evident in the color profiles of these objects.  In Fig.~\ref{fig:color_comp} we show 6 cases where the color profiles, after PSF deconvolution, show significant differences with both the GALPIP and our background subtracted images. For these galaxies, the difference is the same: after PSF deconvolution of our background subtracted images, the color profiles at the outermost parts of the galaxies become redder, with the transition located around the edge of the galaxy (see Sec.~\ref{subsec:sbcp}). In this section we will show how these differences can be explained by the effect of the PSF alone, and why it is crucial to take these effects into account for an accurate stellar population analysis of the outermost regions of the objects.

\begin{figure*}[h!]
    \centering
    \includegraphics[width=\linewidth]{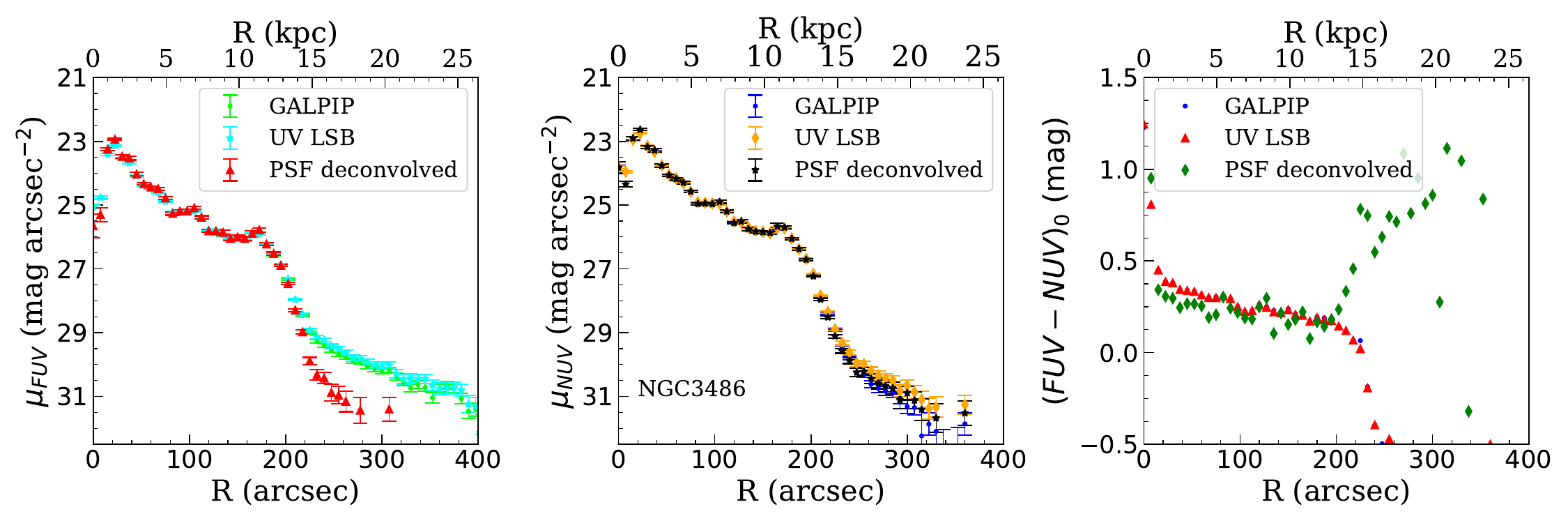}
    \caption{The effect of different background estimates (GALPIP vs. \textit{UV LSB}) and the deconvolution by the effect of the PSF (PSF deconvolved) on the surface brightness and color profiles of the galaxy NGC 3486. The color profiles have been corrected for Galactic extinction.}
    \label{fig:3486_diff}
\end{figure*}

To understand the origin of the reddening of the color profiles once the effect of the PSFs has been taken into account, we illustrate the process of deconvolution for a particular case: NGC 3486.  In Fig.~\ref{fig:3486_diff} we show the differences in the FUV, NUV, and color profiles depending on the background estimate (\textit{UV LSB} and GALPIP for reference) and when the effect of the PSF is considered (PSF deconvolved). None of the profiles show significant deviations up to the location of the edge of the galaxy (i.e. R$\sim$195\arcsec). Beyond this radius, the NUV profiles also remain quite similar, while in the case of the FUV, the profiles begin to diverge dramatically. This strong difference in the FUV surface brightness profile produces a strong reddening in the PSF deconvolved $(FUV-NUV)$ color profile beyond $220\arcsec$.

\begin{figure*}[h!]
    \centering
    \includegraphics[width=0.5\textwidth]{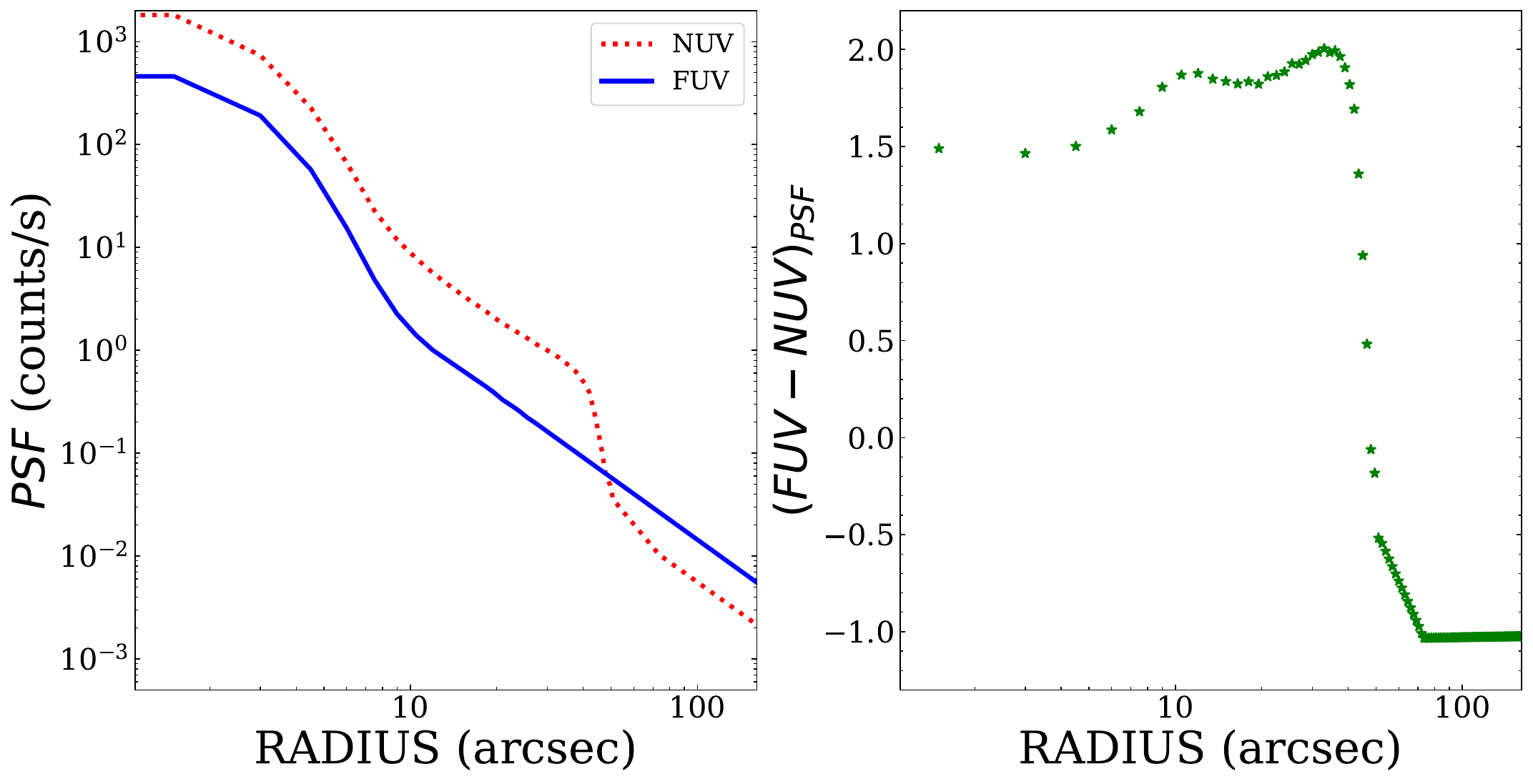}
    \caption{The left panel shows the FUV and NUV PSF profiles normalized as explained in Sec.~\ref{subsec:psf_sub} (to the value at $R=12\arcsec$ in FUV and at $R=30\arcsec$ in NUV). The right panel shows the $(FUV-NUV)$ radial color profile of the PSF.}
    \label{fig:colpsf}
\end{figure*}

To understand why the correction of the PSF effect is so different in the NUV and FUV images, we need to look at the PSF profiles in each band. The profiles of both PSFs are shown in the left panel of Figure \ref{fig:colpsf}. The FUV PSF profile shows the commonly observed PSF structure of a Gaussian-like core followed by power-law behavior \citep[see e.g.][]{sandin14}. However, the NUV is very different, with a sharp drop at $\sim40\arcsec-50\arcsec$. In the right panel of Fig. \ref{fig:colpsf} we show the radial $(FUV-NUV)$ color profile of both PSFs. Due to the sharp transition of the NUV PSF at $\sim40\arcsec-50\arcsec$, the PSF color transitions from red to blue. The color step is very significant (i.e. greater than 2 magnitudes).

\begin{figure*}[h!]
    \centering
    \includegraphics[width=\textwidth]{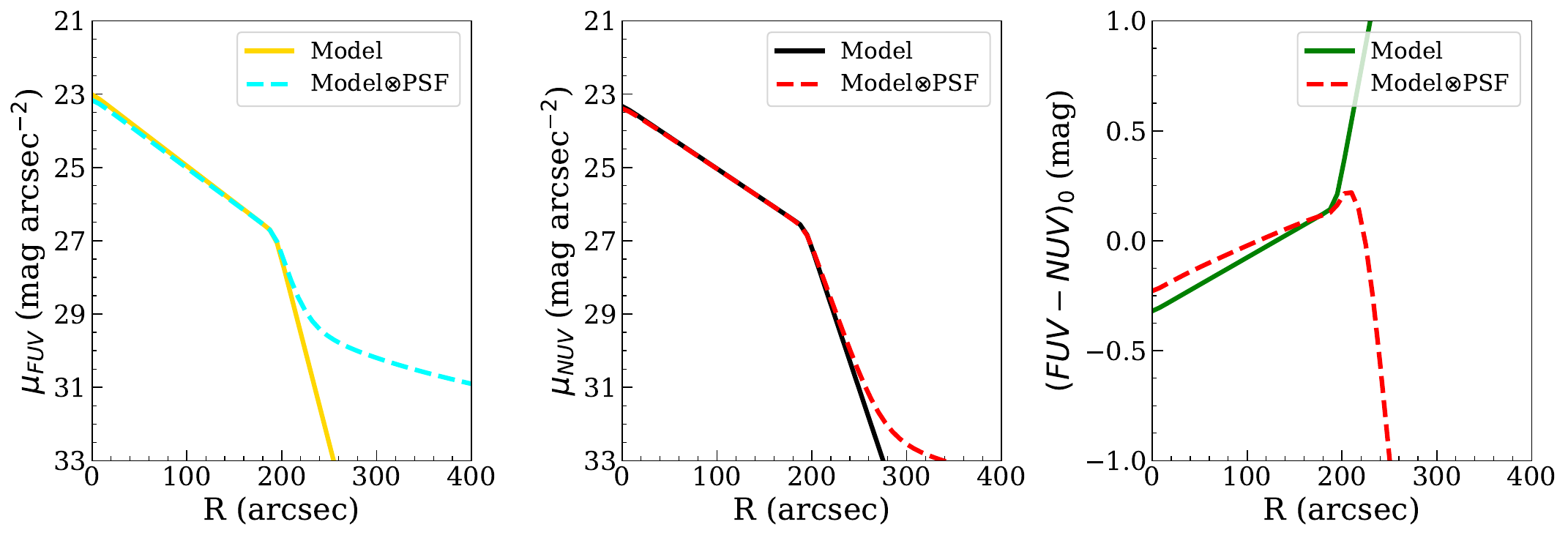}
    \caption{The effect of the FUV and NUV PSF on the surface brightness (left and middle) and color (right) profiles of a broken-exponential galaxy model simulating the galaxy NGC 3486. The larger effect of the FUV PSF in the outer parts of the galaxies cause an artificial bluing of the color profiles beyond the edge of the galaxies.}
    \label{fig:models_psf}
\end{figure*}

When applied to galaxies, the fact that the NUV PSF decays so dramatically at $\sim40\arcsec-50\arcsec$ would imply that there is less scattered light from the galaxies at large distances compared to the effect produced by the FUV PSF. The expectation then is that at large distances the different effects of the two PSFs should result in an artificial bluing of the galaxy color profile. To explore this, we make a practical illustration with a simple broken exponential model in Fig. \ref{fig:models_psf}, which represents the observed surface brightness profiles of NGC 3486. The models created with \texttt{Imfit} \citep{imfit} have the following parameters: a break at $195\arcsec$ and slopes of $h_{1}=54.9\arcsec$, $h_{2}=10.8\arcsec$ for FUV, and $h_{1}=62.7\arcsec$, $h_{2}=14.0\arcsec$ for NUV. These models are then convolved with their respective PSFs.  The larger effect of the FUV PSF produces an excess of FUV light beyond the edge of the galaxy, causing the color profile to be artificially blue.

The reddening of the color profiles beyond the edge of the galaxies is something that we find on a regular basis in the optical color profiles. These U-shaped profiles in the optical were originally predicted by \citet{roskar+08} and found almost immediately by \citet{azzollini+08} and \citet{bakos+08}. The reddening beyond the edge of the galaxy is interpreted as the aging of stellar populations that have formed in-situ within the edge of the galaxy and have migrated outwards. Alternatively, other works explain this reddening as a result of the accretion of older stars from minor mergers (see for example \citealt{ruizlara+16} for Milky Way-mass galaxies, or \citealt{mostoghiu+18} for an M33 analog).  Since the GALEX FUV-NUV color is extremely sensitive to young stellar populations \citep{bianchi2011_sf}, it would be reasonable to expect this reddening beyond the edge also in the UV color of our galaxies, as emission in the UV decreases with age, specially for stellar populations greater than 100 Myr \citep[see e.g. ][]{bruzual03}. The PSF subtraction applied in this paper thus reconciles the observed GALEX FUV-NUV color profile with that expected from theories of star formation in late-type galaxies and the optical color profiles. \\

To conclude this section, it is important to note that in this study, single observational runs are utilised. However, for a number of galaxies within our sample, multiple observing runs with sufficient depth have been conducted. The impact of employing multiple observing runs on the outcomes is examined in Appendix ~\ref{ap:coadd}.

\section{Summary and conclusions}\label{sec:conc}

The analysis of the outermost regions of galaxies has received a significant boost with the advent of very deep optical imaging. However, to carry out an in-depth analysis of the physical mechanism associated with galaxy growth, we need to extend the analysis of these regions to a larger wavelength coverage. In this work we perform the UV analysis of the outermost regions of a sample of 20 galaxies selected from the ultra-deep imaging survey LIGHTS \citep{trujillo21,zaritsky2024lights}. To obtain reliable UV information in the faintest regions of the galaxies, we have applied and adapted to deep GALEX data a number of techniques commonly used in deep optical surveys to characterize very faint extended structures. These are the following:

\begin{itemize}

    \item We characterize the background as a single mean value around the galaxy, following procedures similar to those used in other UV studies \citep[e.g. ][]{GildePaz07}. To remove the background, we construct a Poisson statistical background with this mean value. Removing a Poissonian random distribution creates a pseudo-Gaussian background distribution. This is an important step forward, as it opens the possibility of using low surface brightness optimized detection software, such as \texttt{NoiseChisel}, which assumes a Gaussian-like nature of the background. This greatly improves the detection and masking (if necessary) of extremely faint structures.

    \item We use a deconvolution algorithm to subtract the effect of the extended PSFs from the surface brightness distribution of the galaxies. This deconvolution uses a combination of Wiener filtering and modelization of the galaxies using \texttt{Imfit} routines. In order to apply such techniques, we generate very extended GALEX PSFs up to a radial range of 750\arcsec. This is a factor of $\sim$9 increase over previously available extended GALEX PSFs.
    
\end{itemize}

We present the results of applying both steps in the form of surface brightness and color (where FUV is available) profiles. We show that this methodology yields surface brightness depths of $\sim28.5-30.5\,\rm{mag\, arcsec^{-2}}$ ($3\sigma,\,10^{\prime\prime}\times10^{\prime\prime}$), with radial profiles reaching reliable surface brightness as faint as $\sim31\,\rm{mag\, arcsec^{-2}}$, about 1 magnitude deeper than in previous GALEX studies.

We also compare our results with those obtained using the standard GALEX pipeline. We find that our background subtraction strategy generally avoids over-subtracting the outermost parts of the galaxies compared to the standard GALEX pipeline. This is probably a combination of better masking of the sources and/or better determination of the background locally around the galaxies.

We also show that removing the effect of the GALEX PSFs is key to obtaining meaningful color information from the outermost part of the galaxy. This problem is particularly acute in this UV data, since the structure of the FUV and NUV PSFs is very different. Failure to account for this problem will result in artificially blue outer regions of the galaxies.
    
Our methodology is used to qualitatively analyze the profiles of the galaxies. In $75\%$ of the sample, a well-defined edge can be found using surface brightness and color profiles. In the majority of cases, the edge coincides with the visual edge of the galaxy.

In summary, this study presents the ultraviolet counterpart of a significant fraction of the galaxies in the LIGHTS survey. This is done with the ultimate goal of understanding how galaxies grow with time and how their edges move from the inside out. The combination of both data sets, which we will present in the future, will allow us to study the evolution of the stellar population properties before and beyond the star formation edge of these objects. Moreover, the methodology developed here yields some of the deepest results in UV studies of galaxies to date, and may prove beneficial in future ultraviolet analyses, not only with GALEX, but also with newer UV telescopes such as the \textit{UltraViolet Imaging Telescope}  \citep[UVIT/AstroSat, ][]{UVIT} or the \textit{UltraViolet EXplorer} \citep[UVEX, ][]{UVEX}.

\begin{acknowledgements}

 We thank the referee for his detailed reading of the manuscript, which helped to clarify and improve the presentation of the results of this work.
 IRC and SGA acknowledge support from grant PID2022-140869NB-I00 from the Spanish Ministry of Science and Innovation. IT acknowledges support from the ACIISI, Consejer\'{i}a de Econom\'{i}a, Conocimiento y Empleo del Gobierno de Canarias and the European Regional Development Fund (ERDF) under a grant with reference PROID2021010044 and from the State Research Agency (AEI-MCINN) of the Spanish Ministry of Science and Innovation under the grant PID2022-140869NB-I00 and IAC project P/302302, financed by the Ministry of Science and Innovation, through the State Budget and by the Canary Islands Department of Economy, Knowledge, and Employment, through the Regional Budget of the Autonomous Community. MM acknowledges support from grant RYC2022-036949-I financed by the MICIU/AEI/10.13039/501100011033 and by ESF+.
    SR acknowledges the support of the grant PID2023-150393NB-I00 from the Spanish Ministry of Science, Innovation and Universities. RIS acknowledges financial support from the Spanish Ministry of Science and Innovation through the project PID2022-138896NA-C54. JR acknowledges financial support from the Spanish Ministry of Science and Innovation through the project PID2022-138896NB-C55. This work has been done using the following software: \texttt{Astropy}     \citep{astropy1,astropy2},
\texttt{Gnuastro}    \citep{gnuastro,noisechisel_segment_2019},
\texttt{Imfit}       \citep{imfit},
\texttt{Matplotlib}  \citep{matplotlib},
\texttt{NumPy}       \citep{numpy} and 
\texttt{Scikit-Image} \citep{scikit-image}.

\end{acknowledgements}

\bibliographystyle{aa}
\bibliography{bibliografia.bib}

\appendix
\onecolumn
\section{Pseudo-Gaussian backgrounds resulting from the subtraction of two Poisson distributions}\label{ap:skell}

In this paper we use the subtraction of two Poisson distributions with the same mean to show that the resulting distribution is similar to a Gaussian centred at 0. It should be noted, however, that the theoretical distribution is not exactly Gaussian. In a previous study, \citet{skellam} showed that the probability distribution of the difference between two Poisson distributions, $P(K=X-Y)$, can be expressed as a function of the probability distributions of the individual variables, $P(X=\mu_{1})$, $P(Y=\mu_{2}$), as 

\begin{equation}
    p\left(k;\mu_{1},\mu_{2}\right)=Pr\{ K=k\} = e^{-\left(\mu_{1}+\mu_{2}\right)}\left(\frac{\mu_{1}}{\mu_{2}}\right)^{k/2}I_{k}\left(2\sqrt{\mu_{1}\mu_{2}}\right)
    \label{eq:skellam}
\end{equation}

where $I_{k}(x)$ is the modified Bessel function of the first kind. Then \citet{abramowitz1948handbook} showed that in the case where $\mu_{1}=\mu_{2}$ (as in this paper) the Skellam distribution can be approximated by an asymptotic expansion of the Bessel function:
\begin{equation}
    p(k;\mu,\mu)\sim\frac{1}{\sqrt{4\pi\mu}}\left[1+C\right]
    \label{eq:abramow}
\end{equation}
where $C$ is defined as
\begin{equation}
    C=\sum_{n=1}^{\infty}(-1)^{n}\frac{\{4k^{2}-1\}\{4k^{2}-3^{2}\}\cdots\{4k^{2}-(2n-1)^{2}\}}{n!2^{3n}(2\mu)^{n}}
\end{equation}

Eq.~\ref{eq:abramow}, for sufficiently large values of $k$, can be approach to a Gaussian distribution of $\sigma=\sqrt{2\mu}$:

\begin{equation}
    p(k;\mu,\mu)\sim\frac{e^{-k^{2}/4\mu}}{\sqrt{4\pi\mu}}
    \label{eq:normaprox}
\end{equation}

\begin{figure*}[h!]
    \centering
    \includegraphics[width=\textwidth]{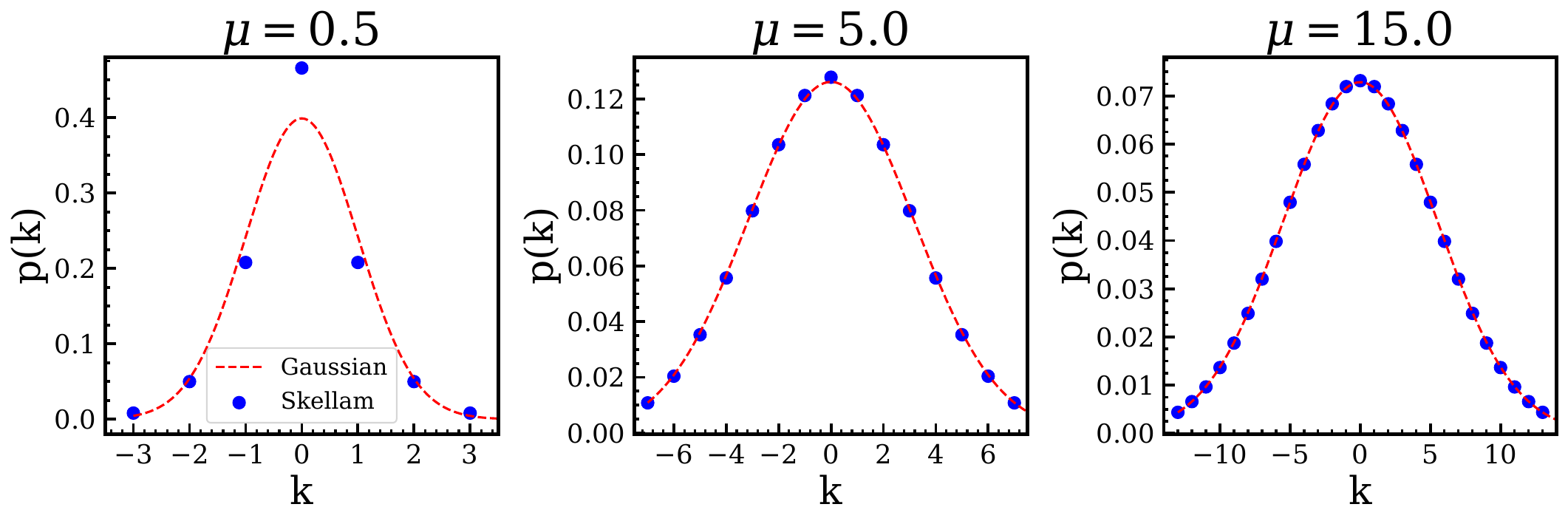}
    \caption{Skellam distribution for $\mu_{1}=\mu_{2}=\mu$ compared with a Gaussian distribution centered on 0 with $\sigma=\sqrt{2\mu}$.}
    \label{fig:eqsdist}
\end{figure*}

Fig.~\ref{fig:eqsdist} shows the Skellam distribution for typical backgrounds of our data. These are characterised by a mean of $\mu=0.5$ for typical FUV backgrounds, $\mu=5$ for low NUV backgrounds, and $\mu=15$ for high NUV backgrounds. Overlaid on this are the approximations to a Gaussian distribution as described by eq.~\ref{eq:normaprox}. The plots indicate that it is reasonable to claim that if the background is subtracted as a Poisson distribution, the resulting distribution is similar to a Gaussian distribution, with greater similarity observed at higher background levels. Therefore, these mathematical properties of the Poisson distributions we use in this paper are key to allowing software such as \texttt{NoiseChisel} to be applied to our dataset. This is because this type of software assumes a Gaussian statistic for the background in order to identify the sources of the images.

\section{Surface brightness and color profiles}\label{ap:rest_prof}
In this section, we include the full set of surface brightness and color profiles, with the exception of those present in Fig.~\ref{fig:profiles}.

\begin{figure*}[h!]

    \centering
\includegraphics[width=0.9\textwidth]{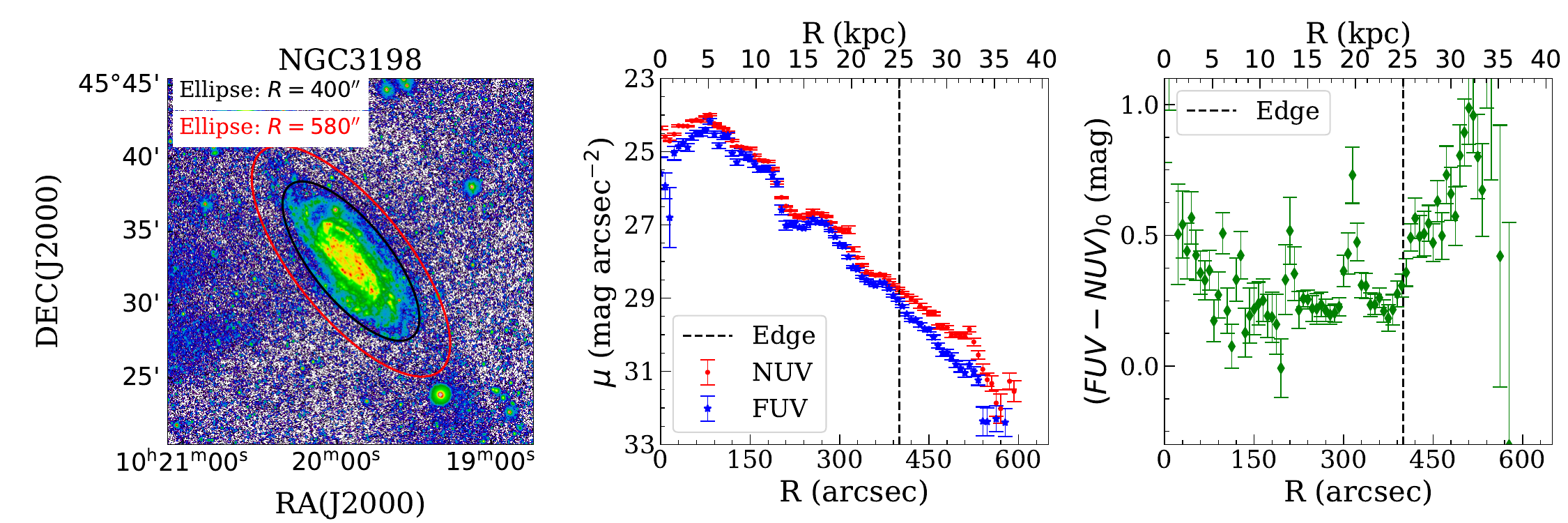}
        \includegraphics[width=0.9\textwidth]{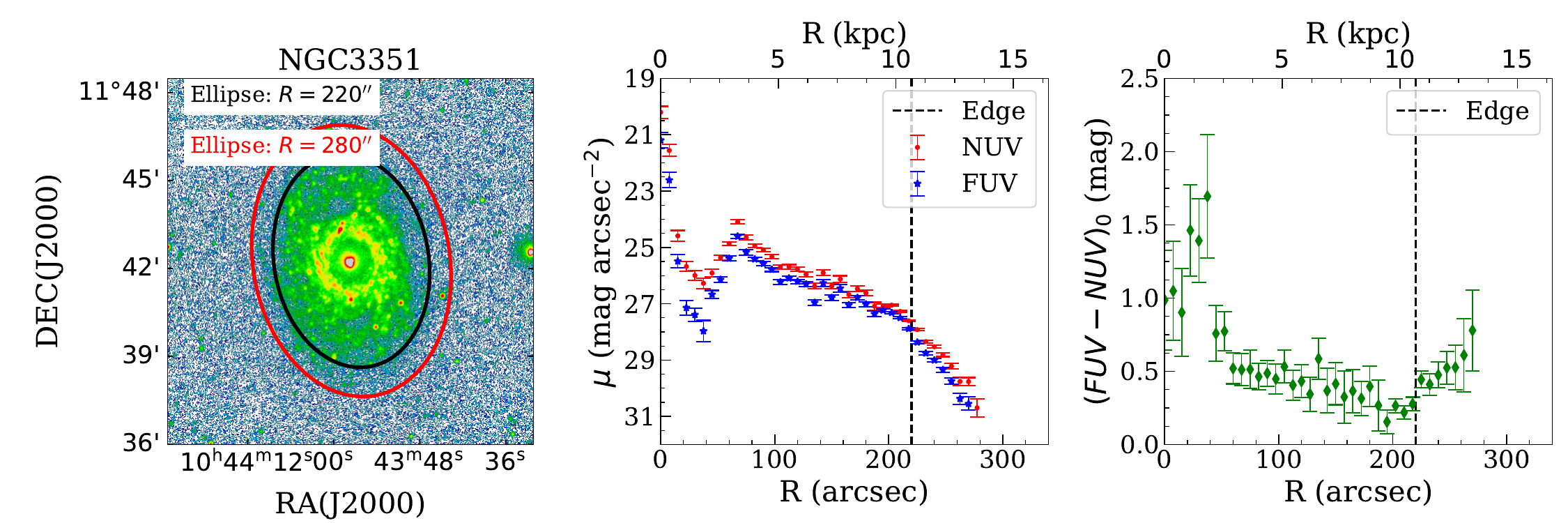}
        \includegraphics[width=0.9\textwidth]{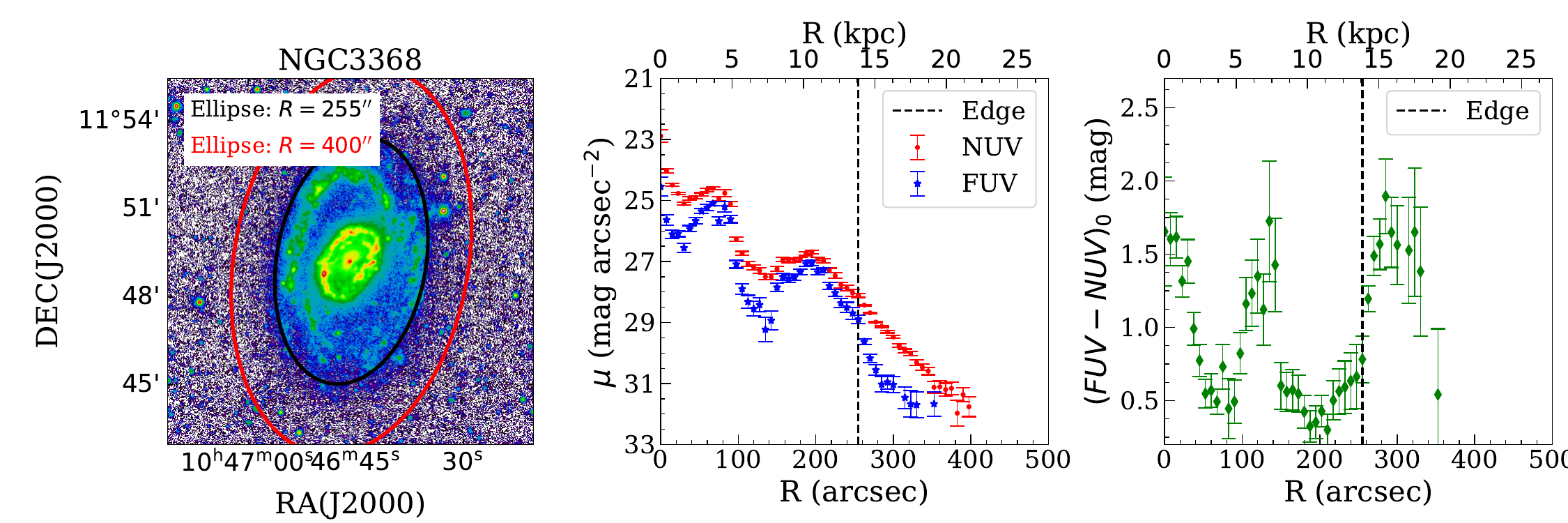}
        \includegraphics[width=0.9\textwidth]{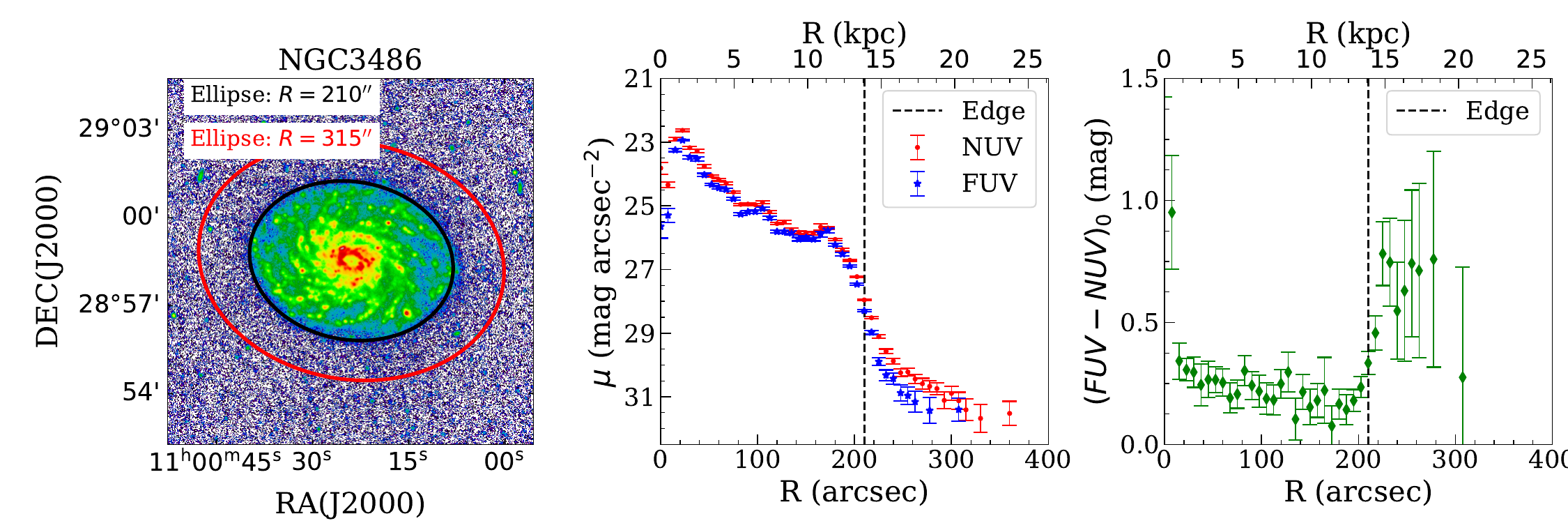}
        \caption{Surface brightness and color profiles of the full sample (see Fig.~\ref{fig:profiles} for the description).}
\end{figure*}
\begin{figure*}[h!]
    \setcounter{figure}{0}
    \centering
        \includegraphics[width=0.9\textwidth]{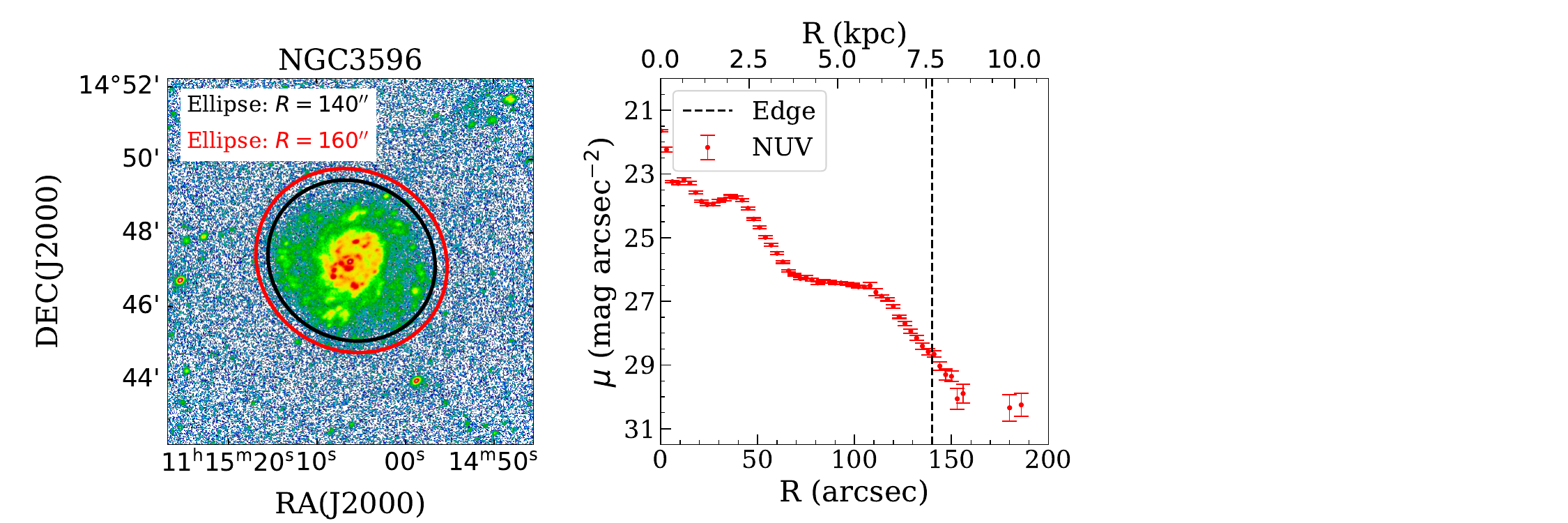}
        \includegraphics[width=0.9\textwidth]{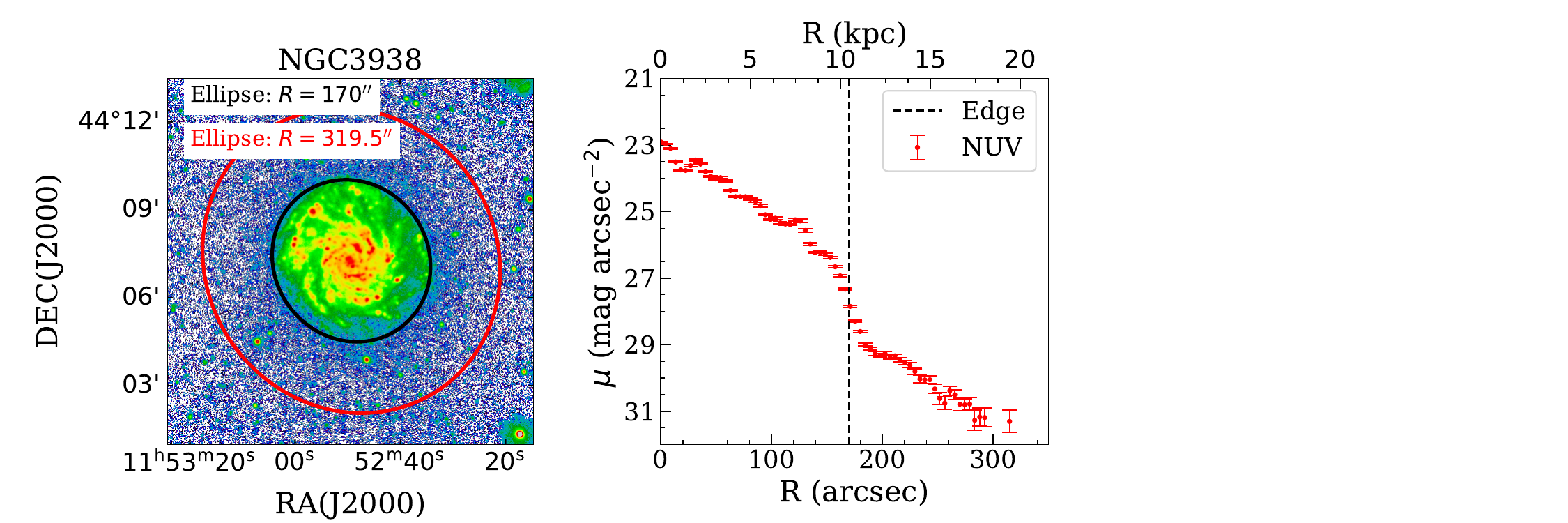}
        \includegraphics[width=0.9\textwidth]{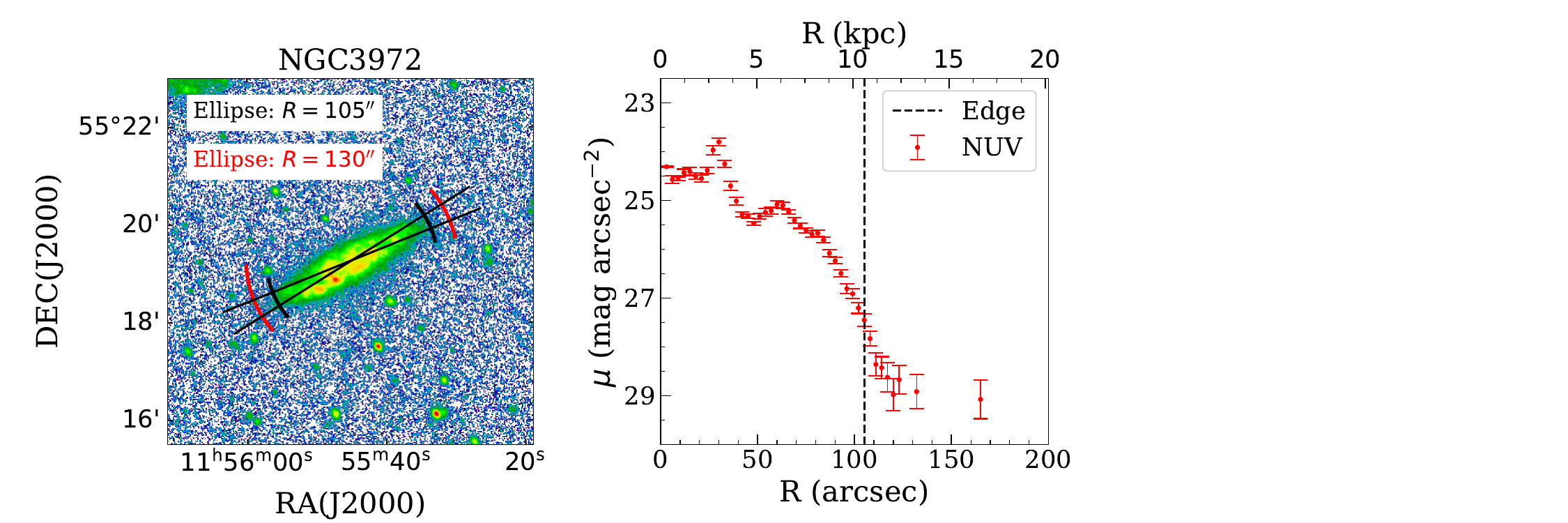}
        \includegraphics[width=0.9\textwidth]{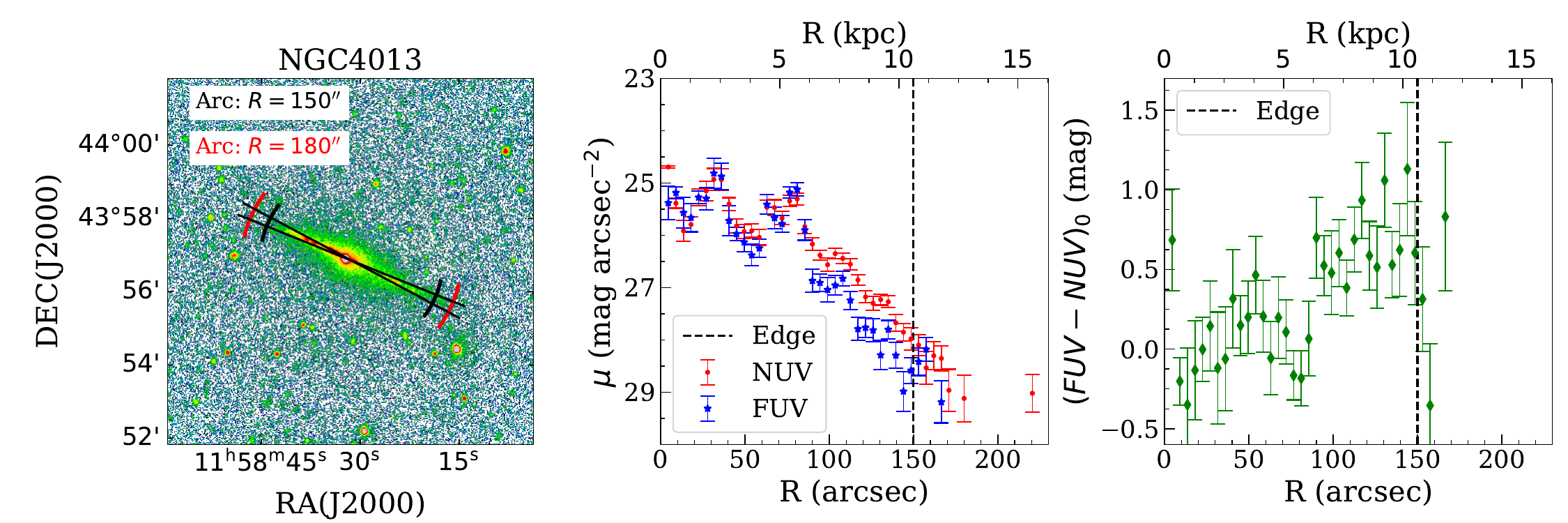}
    \caption{Profiles (cont.)}

\end{figure*}

\begin{figure*}[h!]
\centering
\setcounter{figure}{0}
        \includegraphics[width=0.9\textwidth]{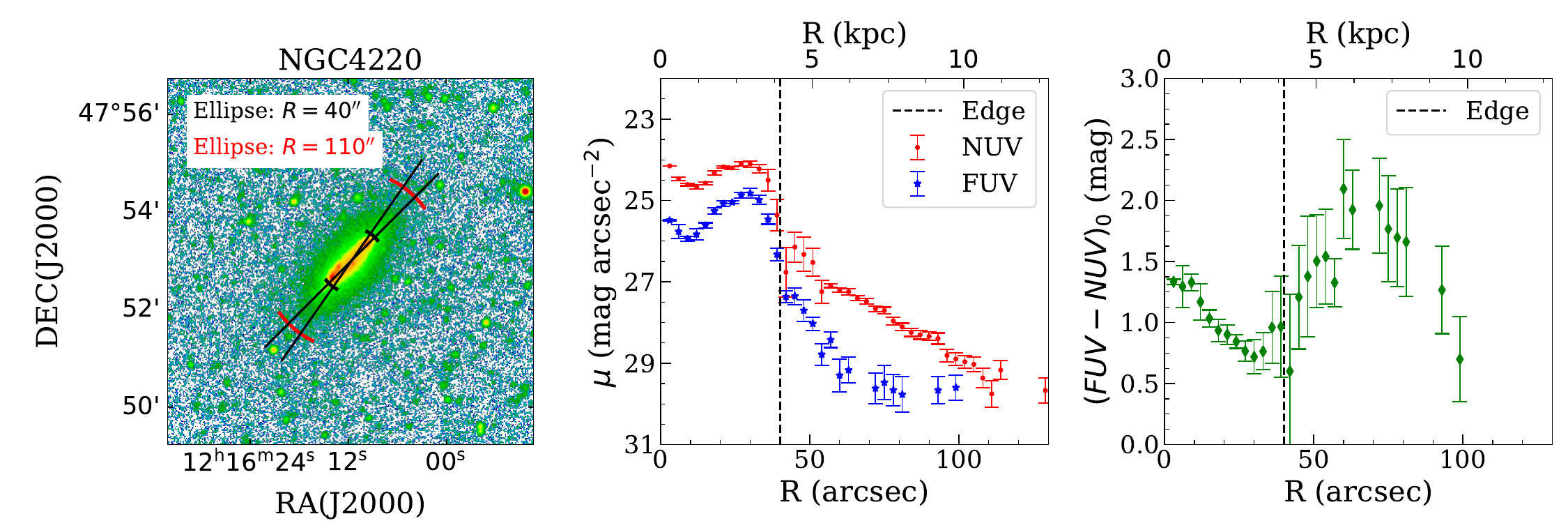}
        \includegraphics[width=0.9\textwidth]{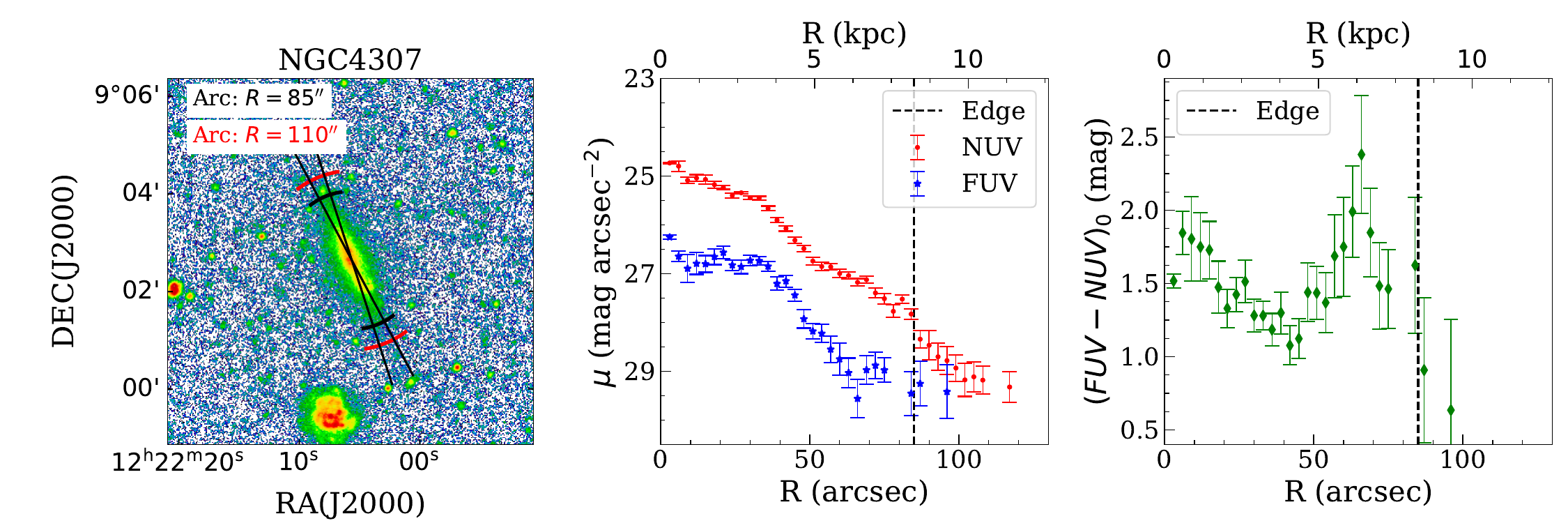}
        \includegraphics[width=0.9\textwidth]{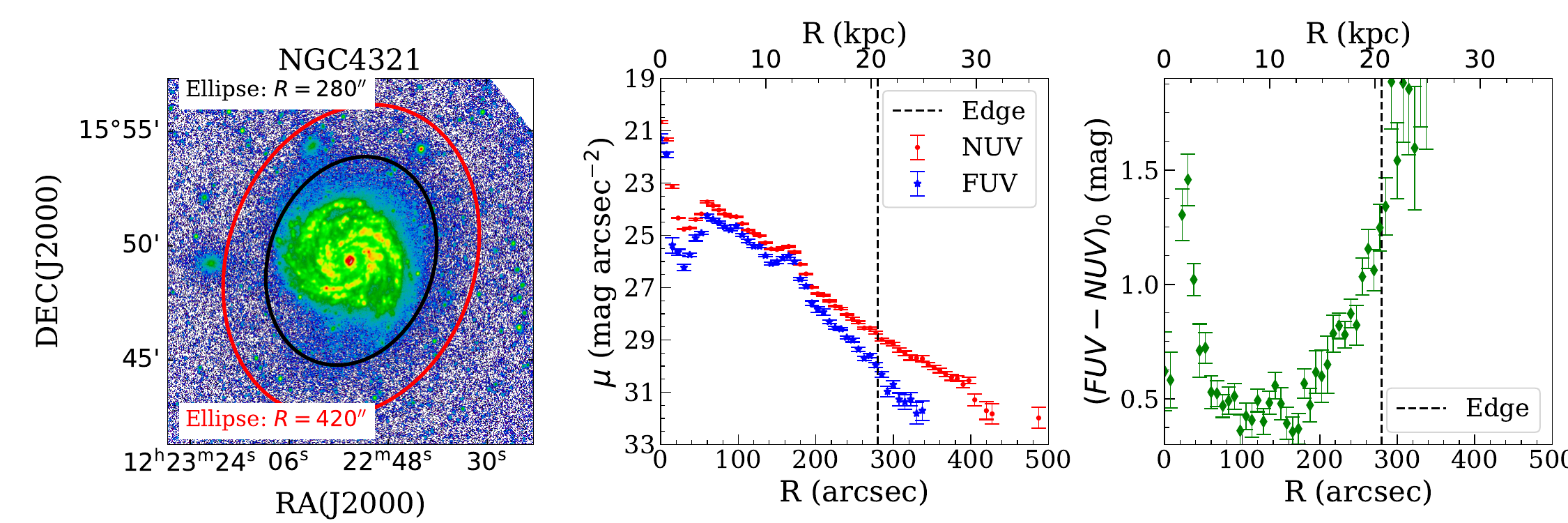}
        \includegraphics[width=0.9\textwidth]{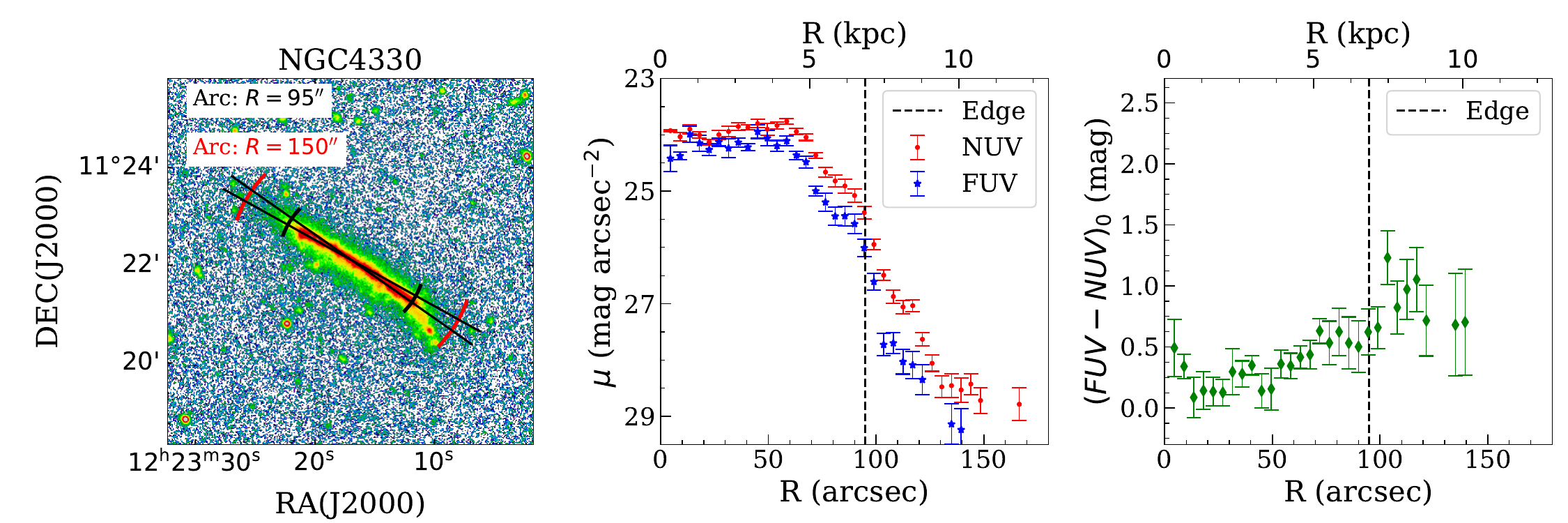}
    \caption{Profiles (cont.)}

\end{figure*}

\begin{figure*}[h!]
\centering
\setcounter{figure}{0}
        \includegraphics[width=0.9\textwidth]{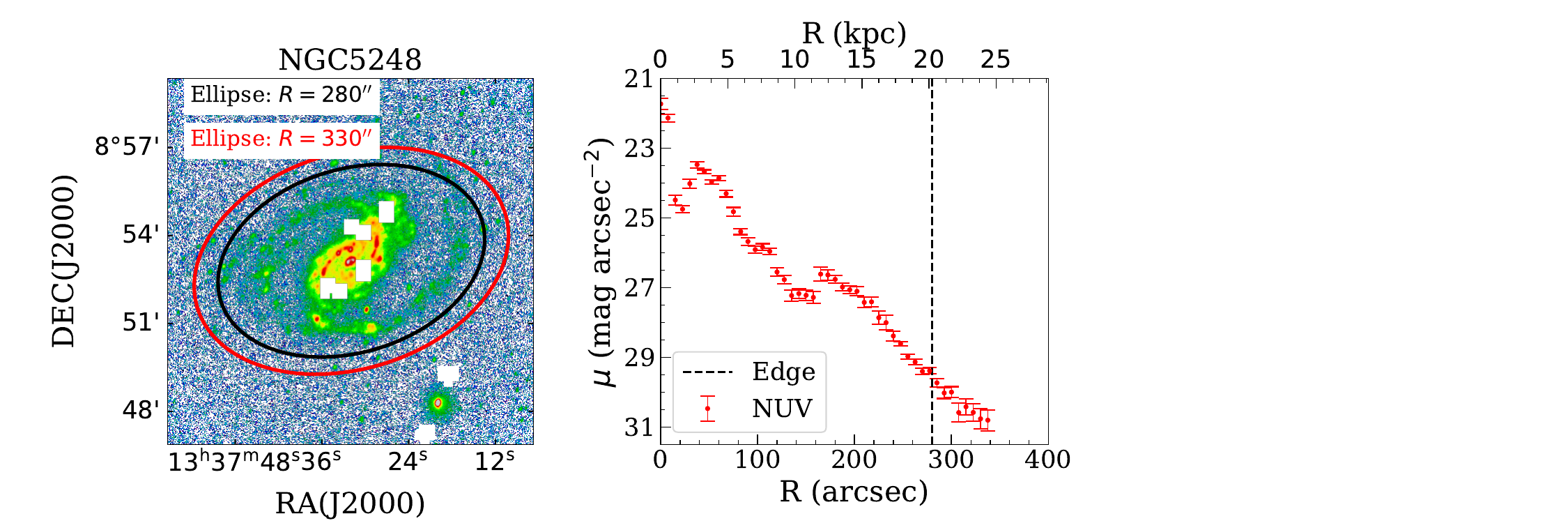}
        \includegraphics[width=0.9\textwidth]{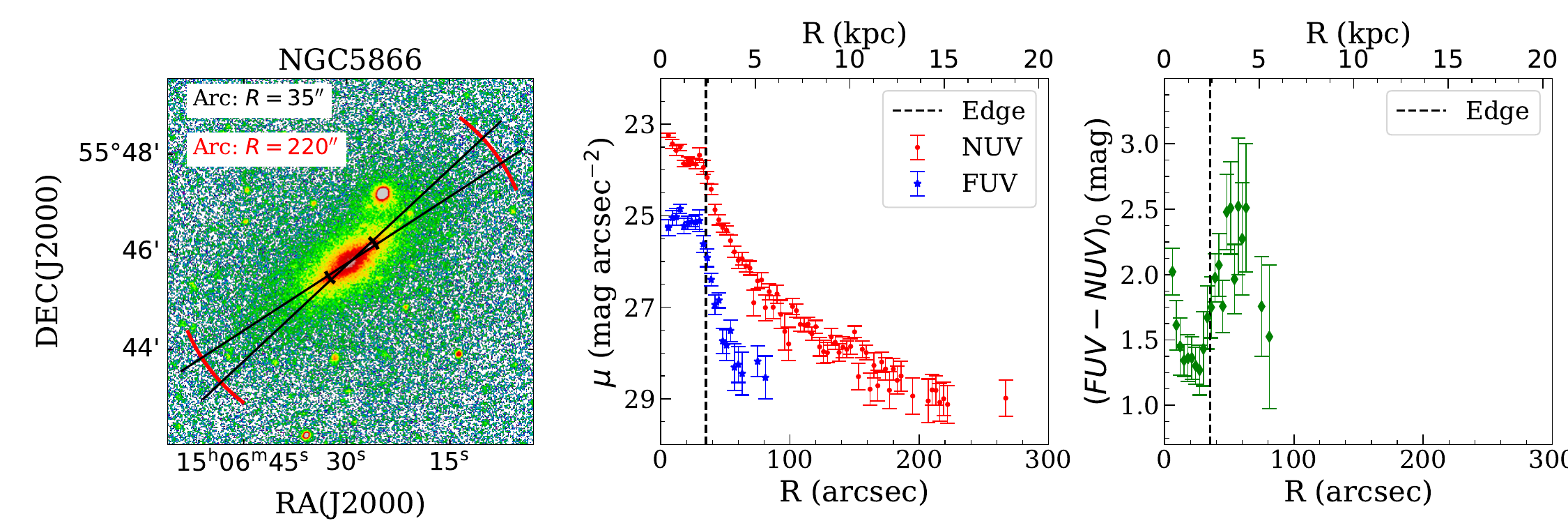}
        \includegraphics[width=0.9\textwidth]{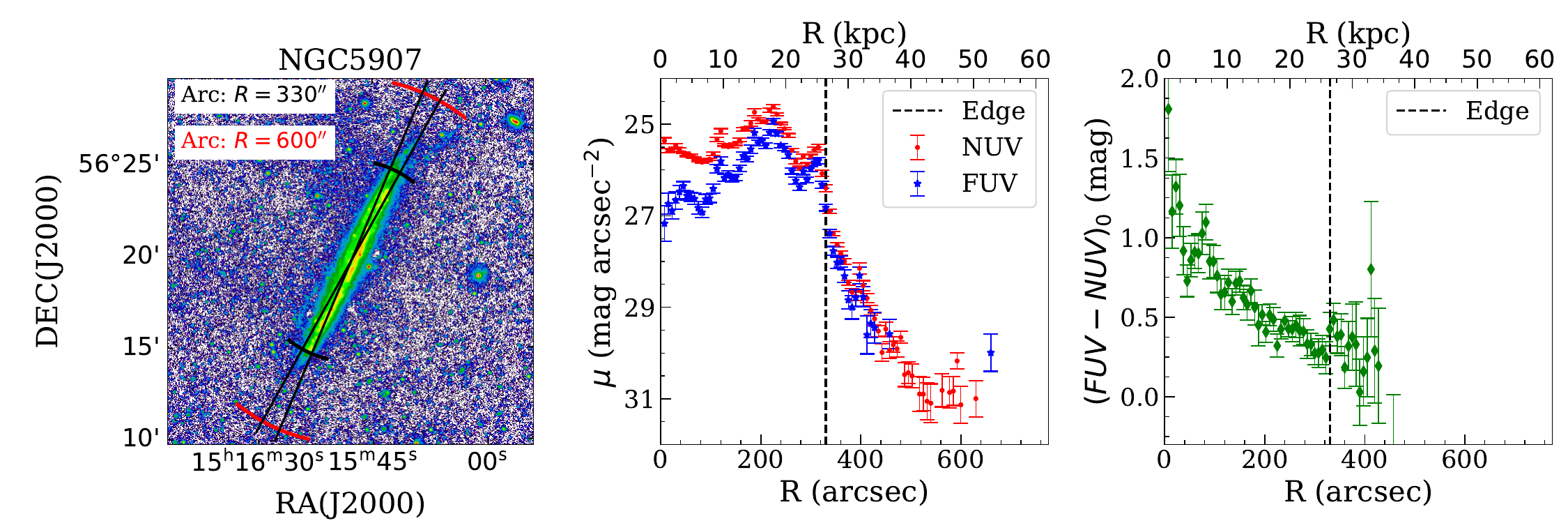}
        \includegraphics[width=0.9\textwidth]{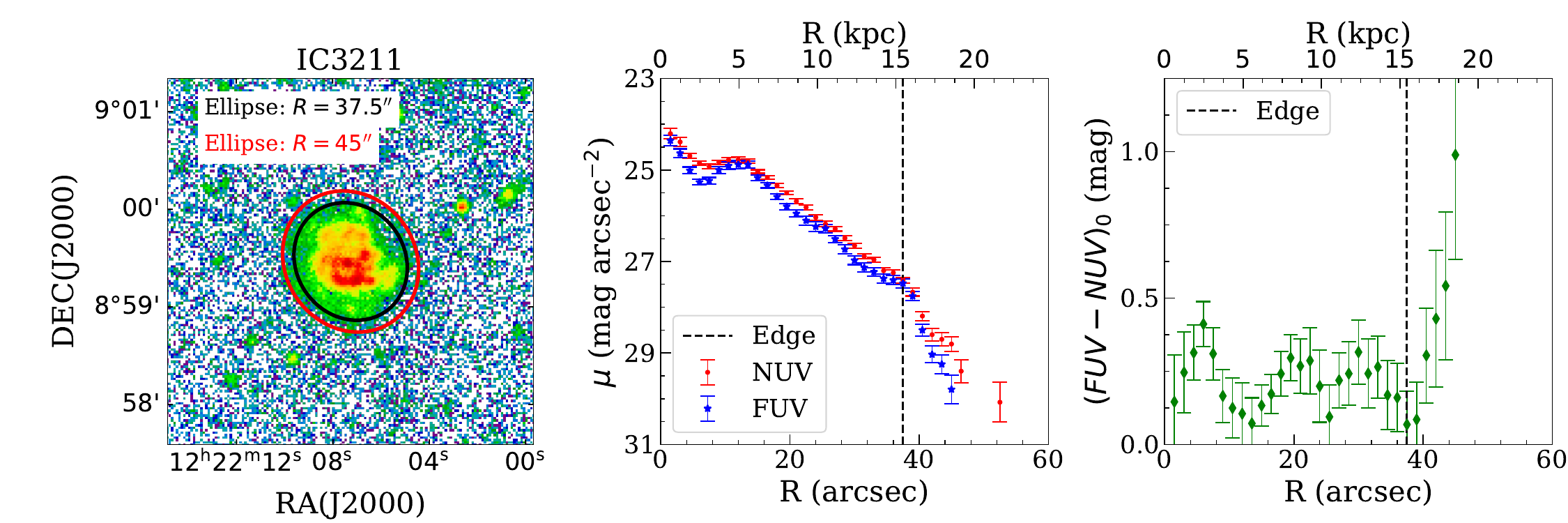}
    \caption{Profiles (cont.)}

\end{figure*}
\FloatBarrier
\section{Full comparison with GALEX pipeline}\label{ap:comp}

In this section we present the surface brightness profiles in the FUV and NUV, as well as the color profiles for all the galaxies in our sample (except NGC 3486, which has already been presented in section ~\ref{subsec:disc_psf}). We show the differences when using different background estimates (i.e. those using the standard GALEX pipeline \citep[][]{morrissey2007}, here called \textit{GALPIP}, and ours (\textit{UV LSB})). We also show how the profiles change when the effect of the PSF is taken into account (\textit{PSF deconvolved}).

\begin{figure*}[h!]
    \centering

        \includegraphics[width=0.8\textwidth]{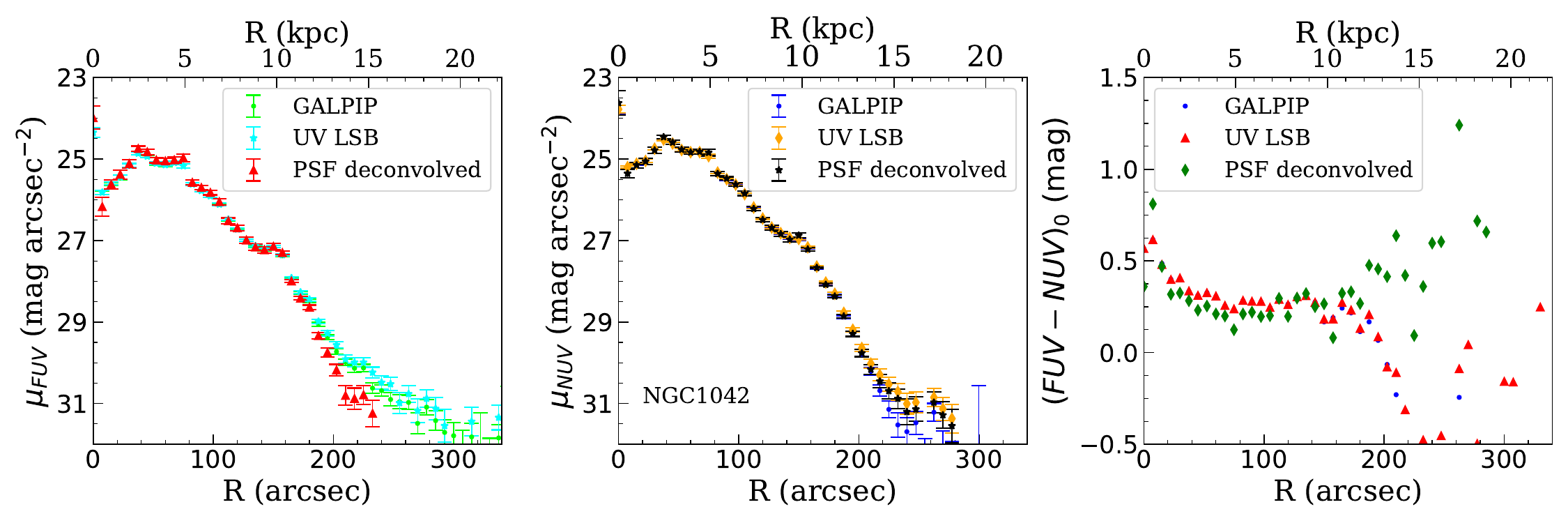}

        \includegraphics[width=0.8\textwidth]{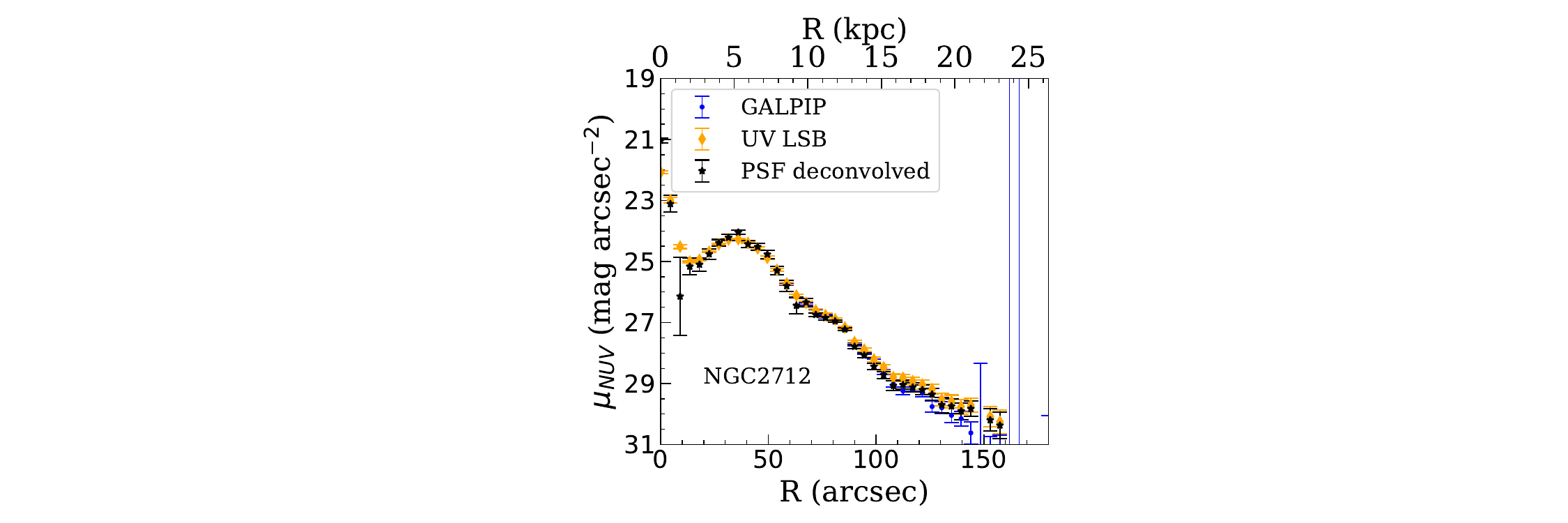}
        \includegraphics[width=0.8\textwidth]{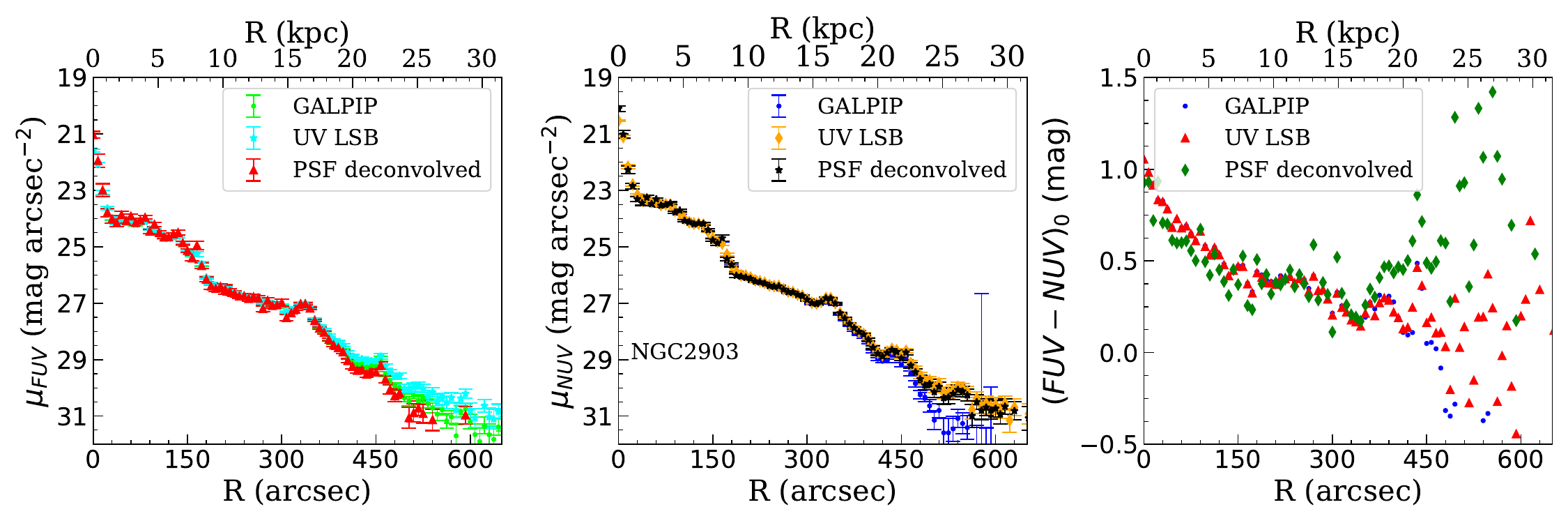}
        \includegraphics[width=0.8\textwidth]{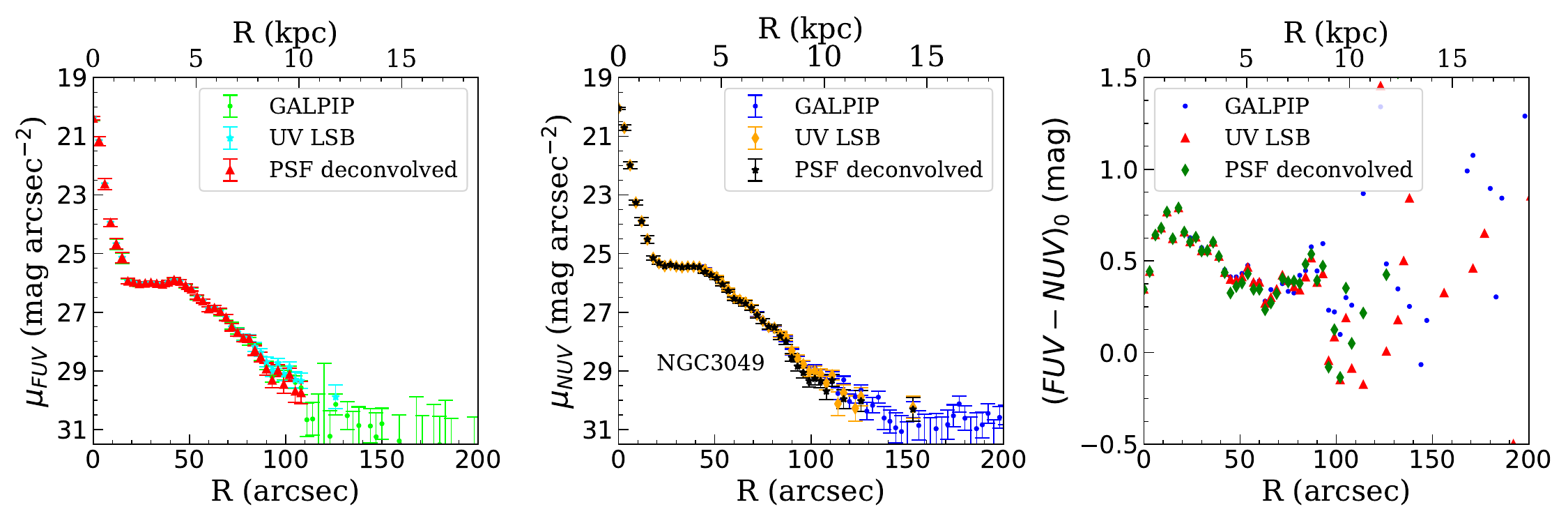}

    \caption{Comparison of the surface brightness and color profiles for the galaxies in our sample (except NGC 3486, which has already been presented in section ~\ref{subsec:disc_psf}) using the standard GALEX pipeline (called GALPIP) and those derived here using a different estimation of the background (\textit{UV LSB}). We also show how the profiles change when the effect of the PSF is taken into account (\textit{PSF deconvolved}).}
    \label{fig:fullcomp}
\end{figure*}
\begin{figure*}[h!]
    \centering
    \setcounter{figure}{0}
        \includegraphics[width=0.8\textwidth]{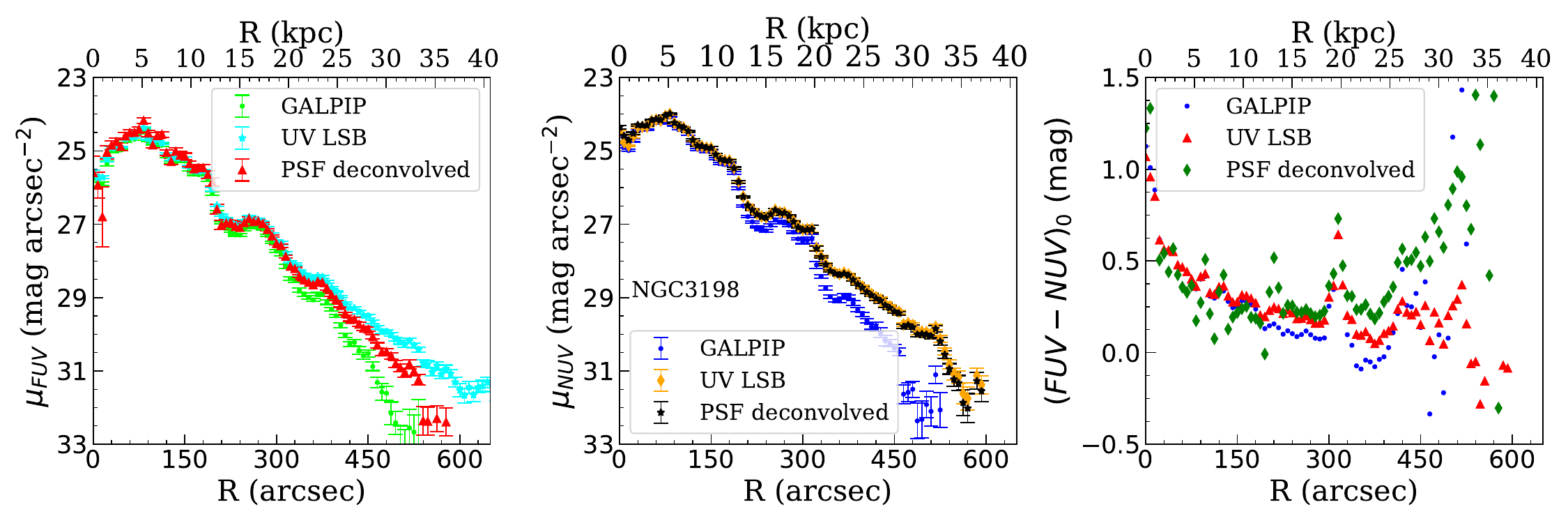}
        \includegraphics[width=0.8\textwidth]{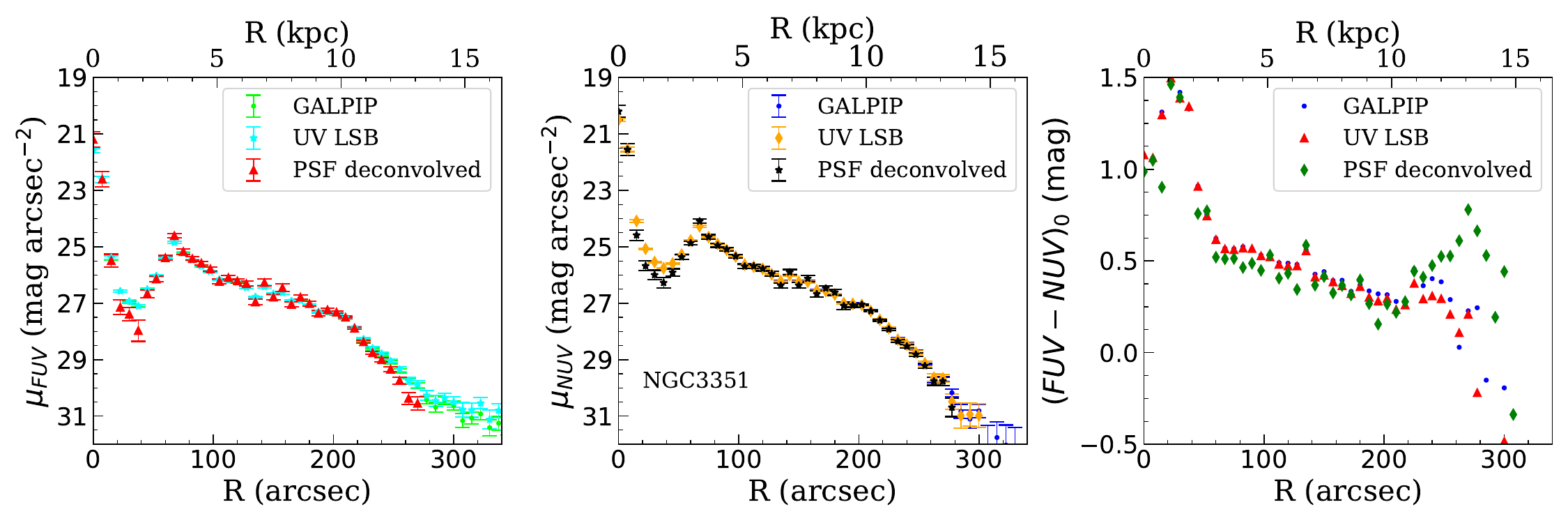}
        \includegraphics[width=0.8\textwidth]{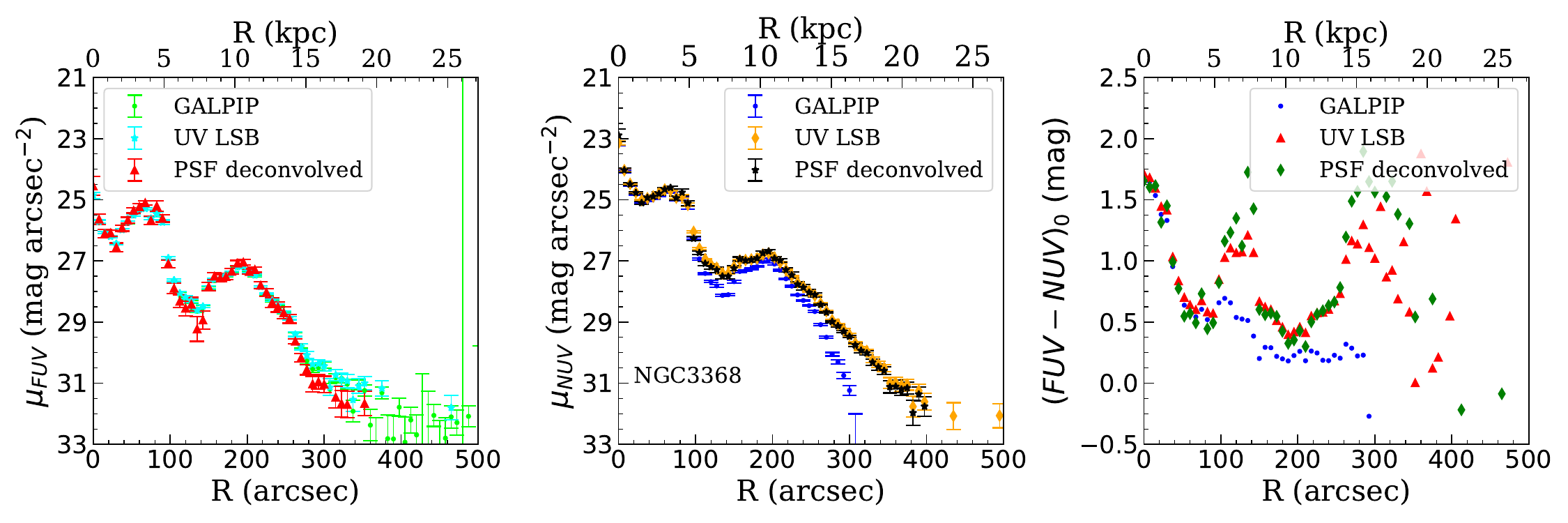}
        \includegraphics[width=0.8\textwidth]{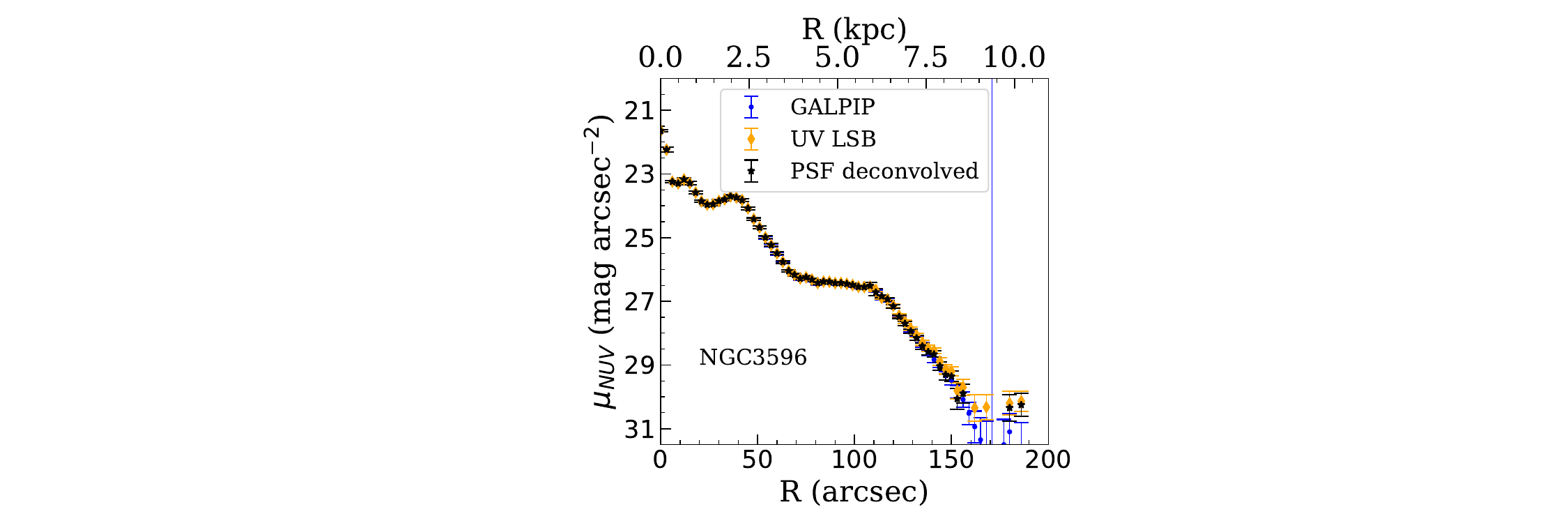}
        \caption{Cont.}
\end{figure*}
\begin{figure*}[ht]
    \centering
    \setcounter{figure}{0}
        \includegraphics[width=0.8\textwidth]{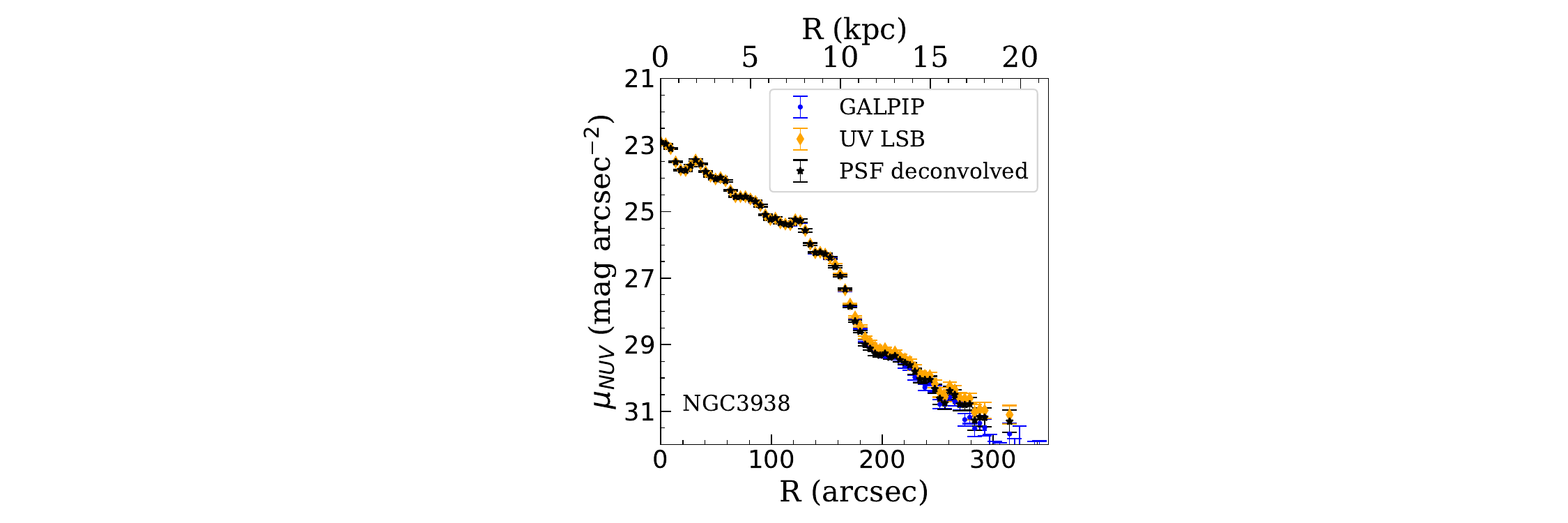}
        \includegraphics[width=0.8\textwidth]{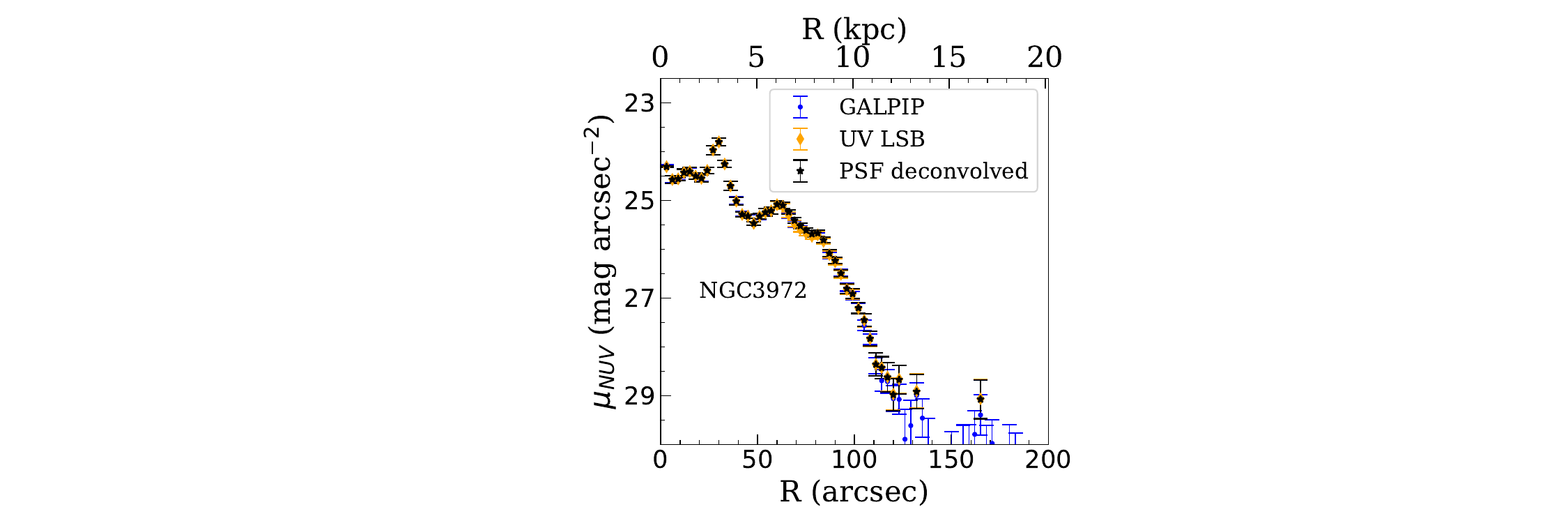}
        \includegraphics[width=0.8\textwidth]{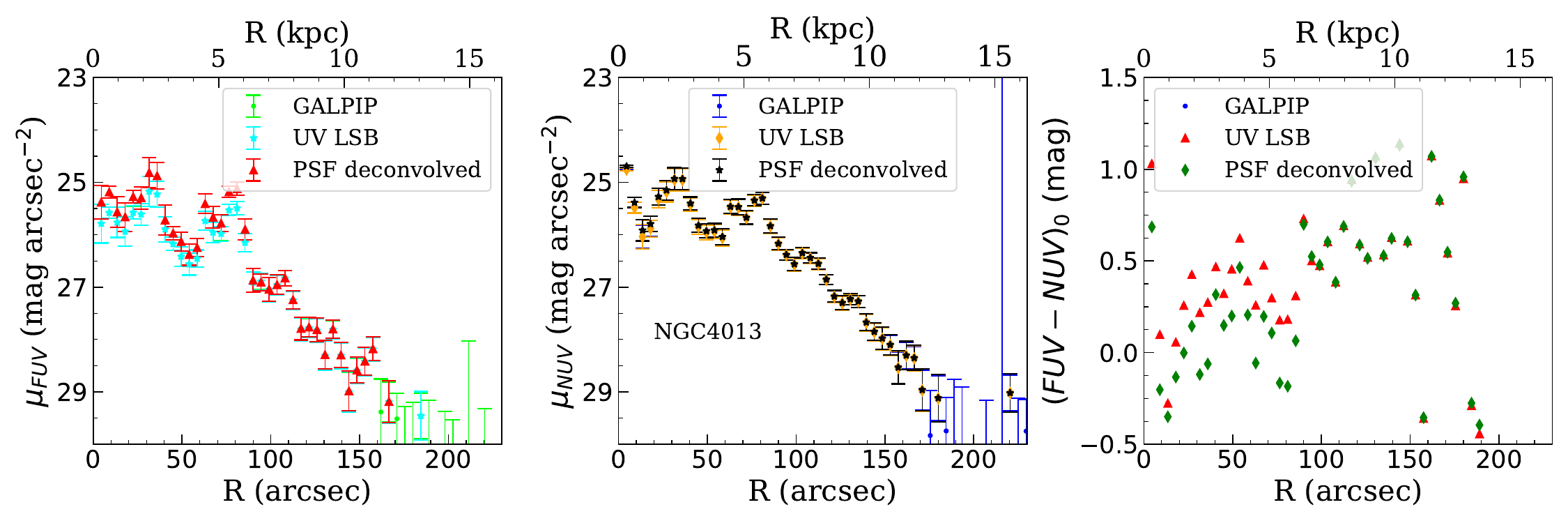}
        \includegraphics[width=0.8\textwidth]{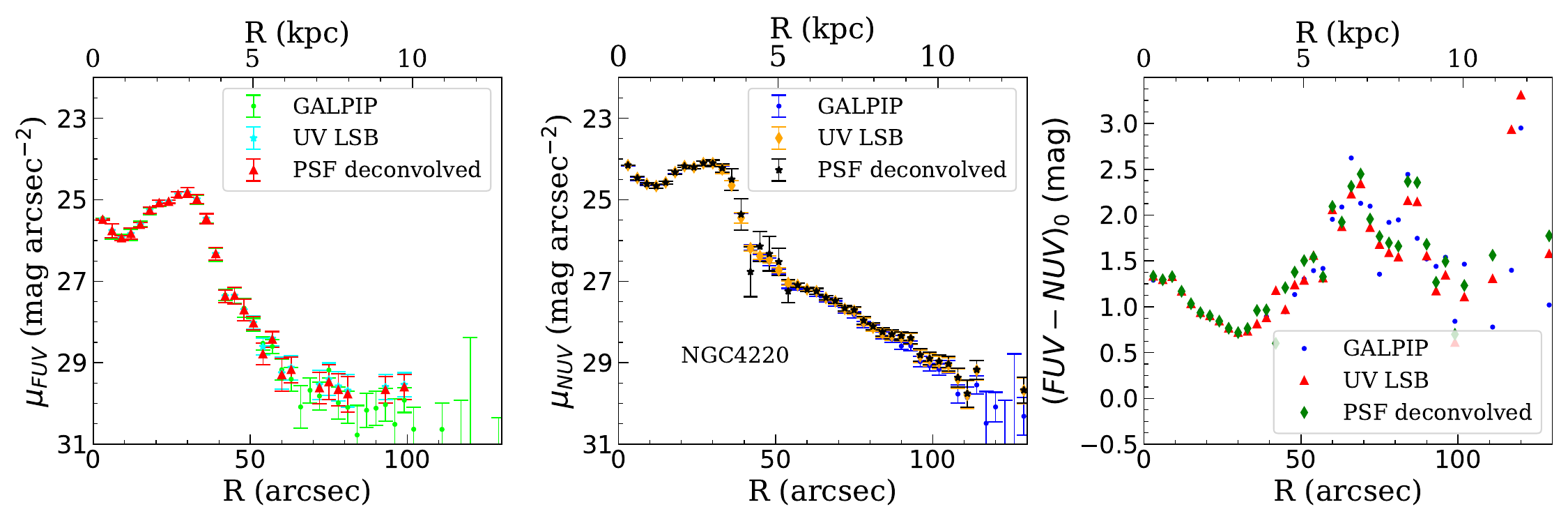}
        \caption{Cont.}
\end{figure*}
\begin{figure*}[h!]
    \centering
    \setcounter{figure}{0}
        \includegraphics[width=0.8\textwidth]{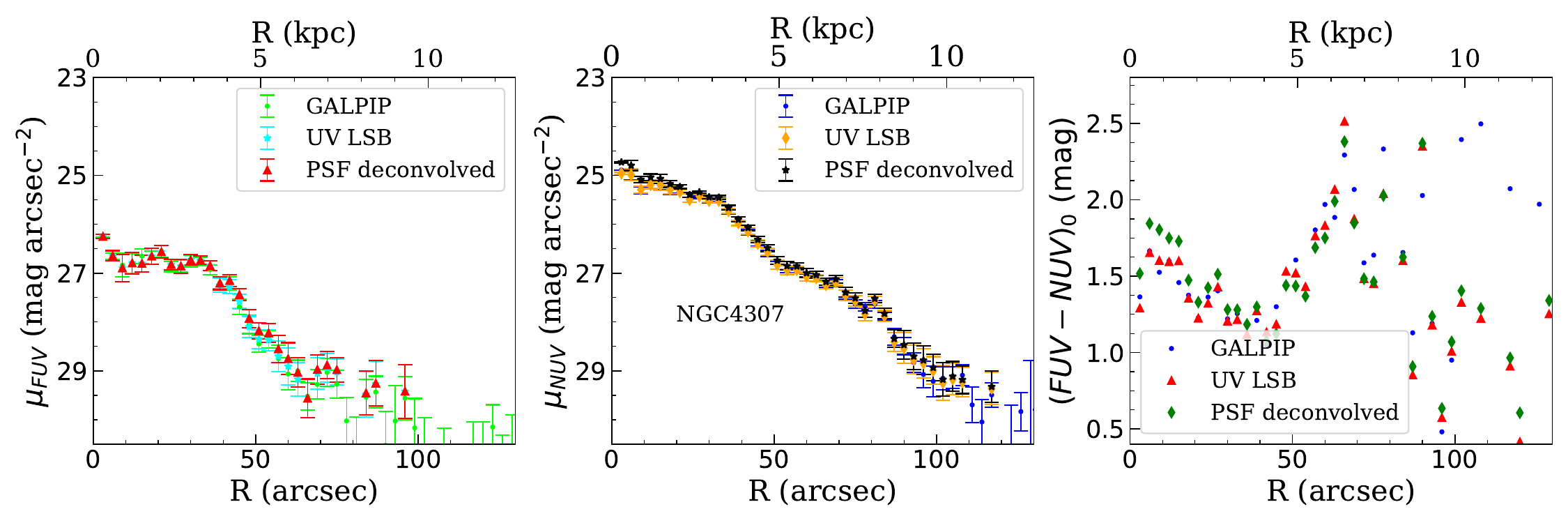}
        \includegraphics[width=0.8\textwidth]{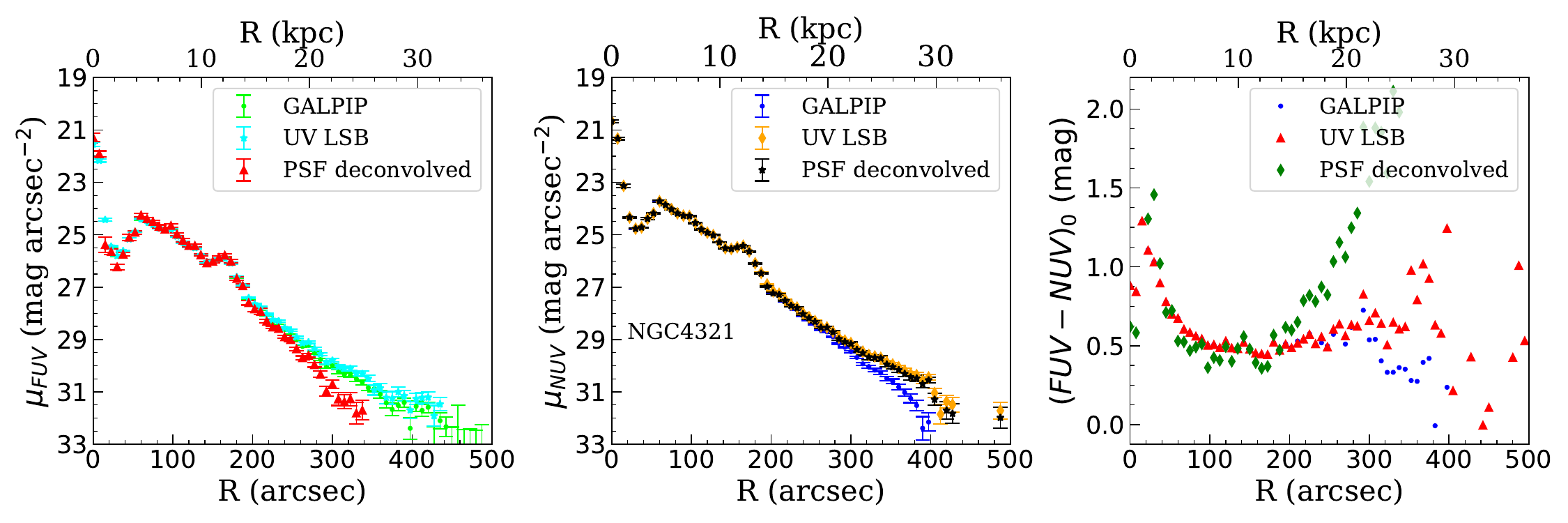}
        \includegraphics[width=0.8\textwidth]{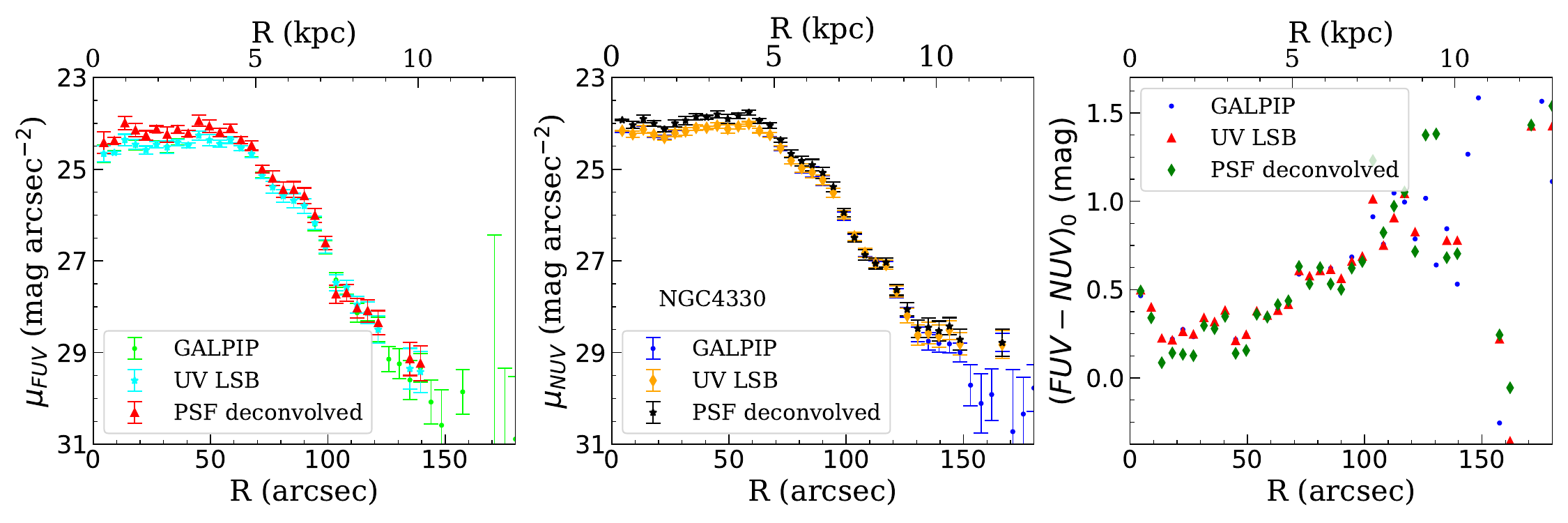}
        \includegraphics[width=0.8\textwidth]{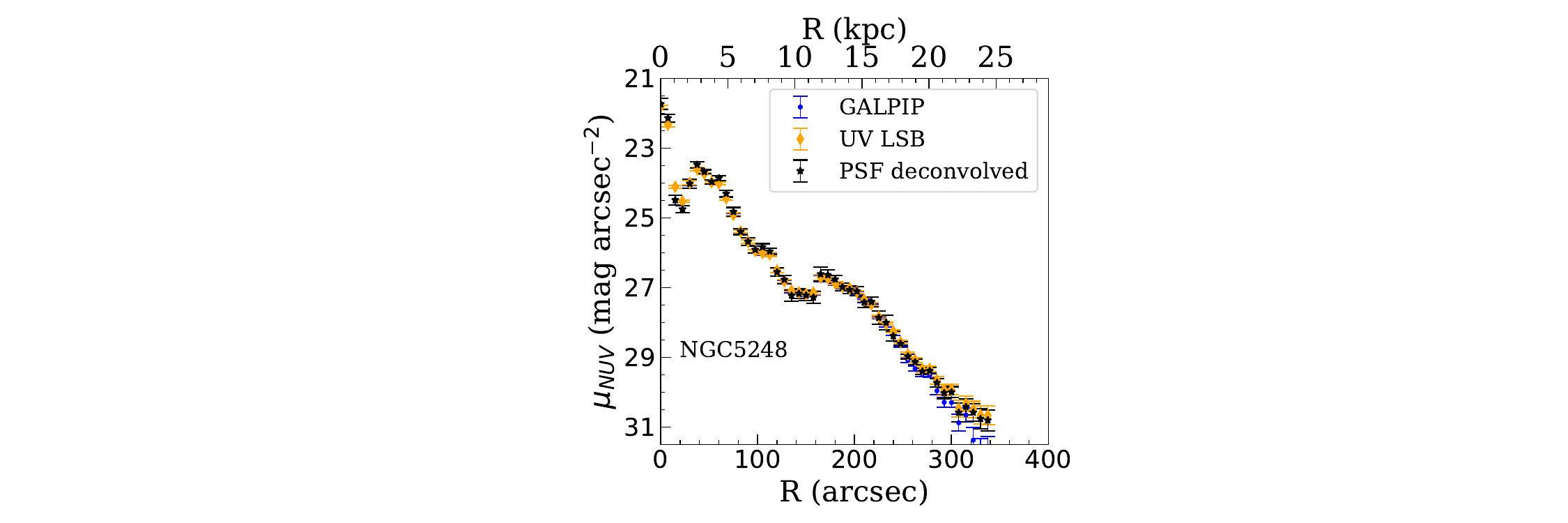}
        \caption{Cont.}
\end{figure*}
\begin{figure*}[h!]
    \centering
    \setcounter{figure}{0}
        \includegraphics[width=0.8\textwidth]{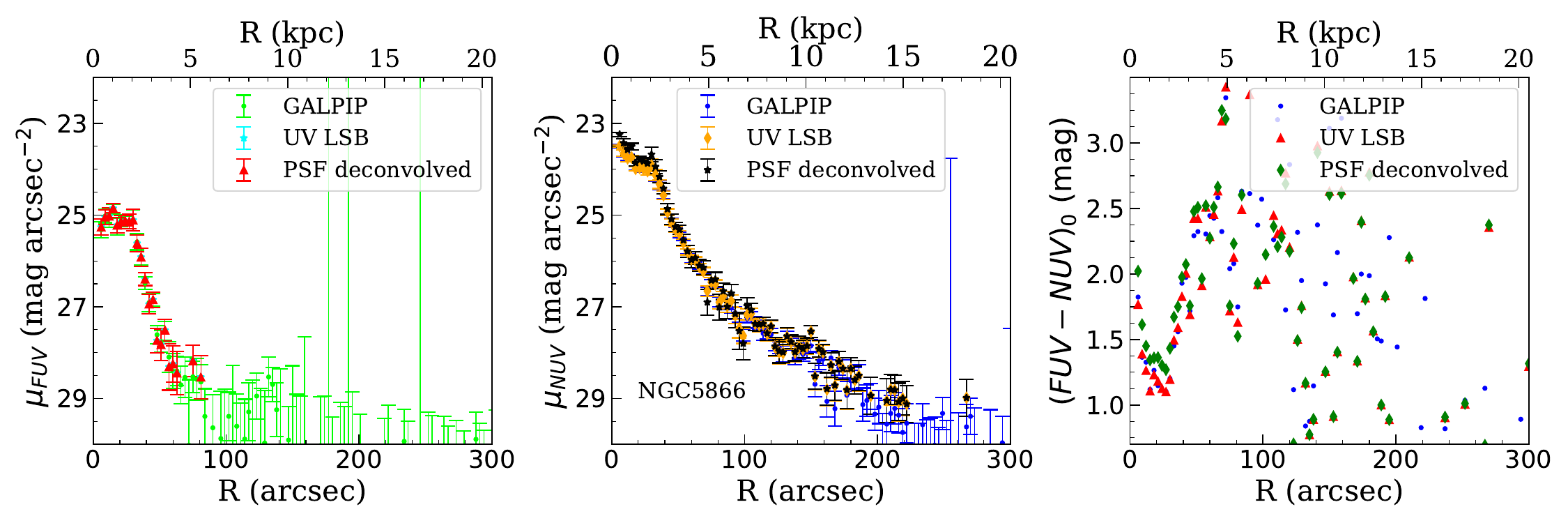}
        \includegraphics[width=0.8\textwidth]{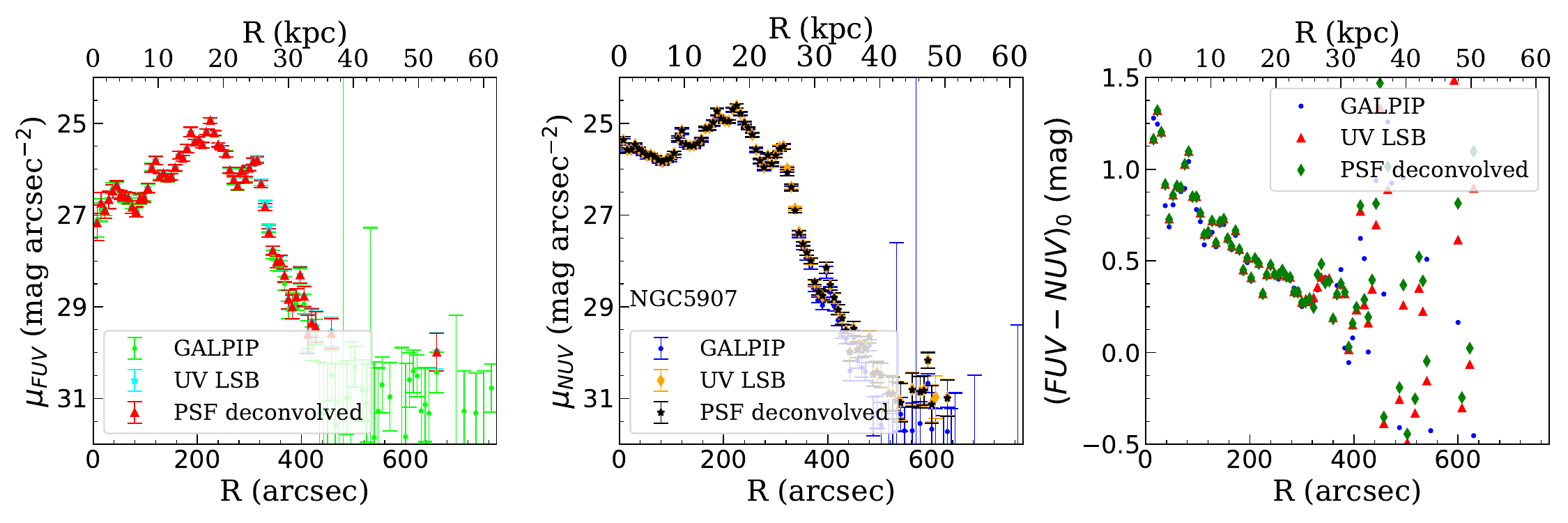}
        \includegraphics[width=0.8\textwidth]{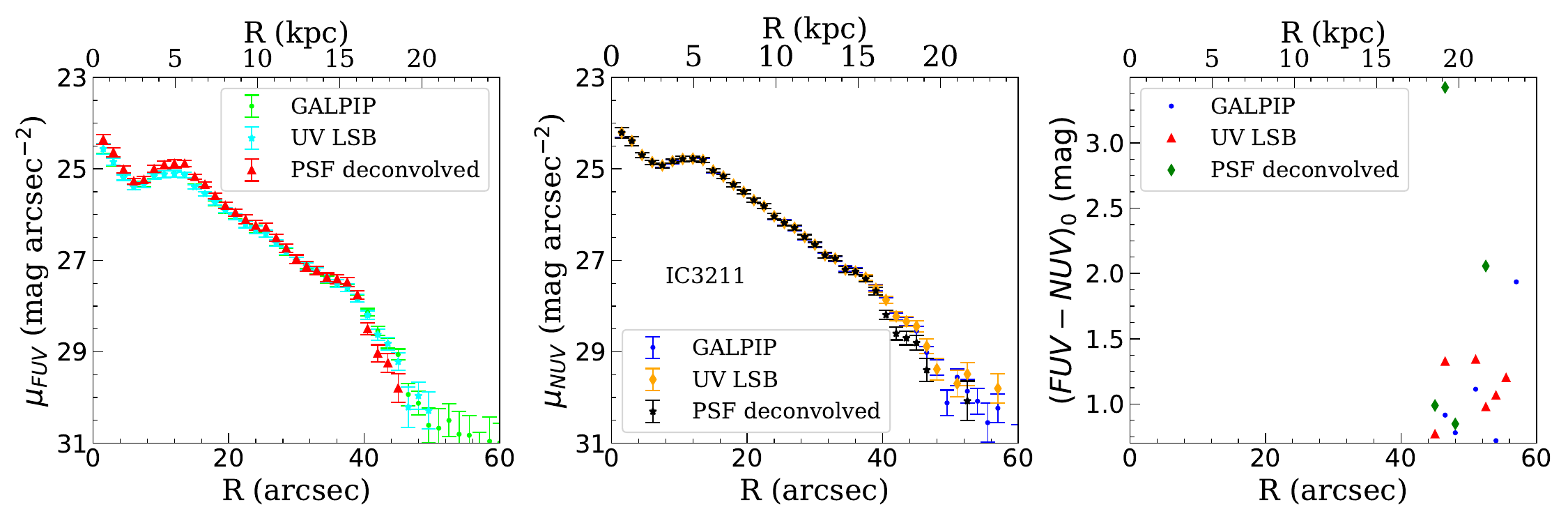}
        
        \caption{Cont.}
\end{figure*}
\FloatBarrier
\section{Comparison with previous works}\label{ap:comp_prev}

As mentioned in Sec. \ref{subsec:disc_bck}, this work is not the first to use a different background subtraction strategy from the GALEX pipeline. In \citet{GildePaz07} the background is also characterized as the mean of the local background surrounding the galaxy, a strategy also adopted by \citet{Bouquin18}. In this section we make a comparison between the galaxies in \citet{Bouquin18}'s sample and our dataset. We have 13 galaxies in common.

To make a proper comparison, in this ocassion the \textit{UV-LSB} profiles are reconstructed using the same centers, position angles, and axis ratios as in \citet{Bouquin18}, with elliptical apertures in steps of $4^{\prime\prime}$, even in those galaxies where wedge-shaped sectors were previously used. The comparison between the profiles is shown in Fig. \ref{fig:bouq_us}.

\begin{figure*}[h!]
    \centering

        \includegraphics[width=0.9\textwidth]{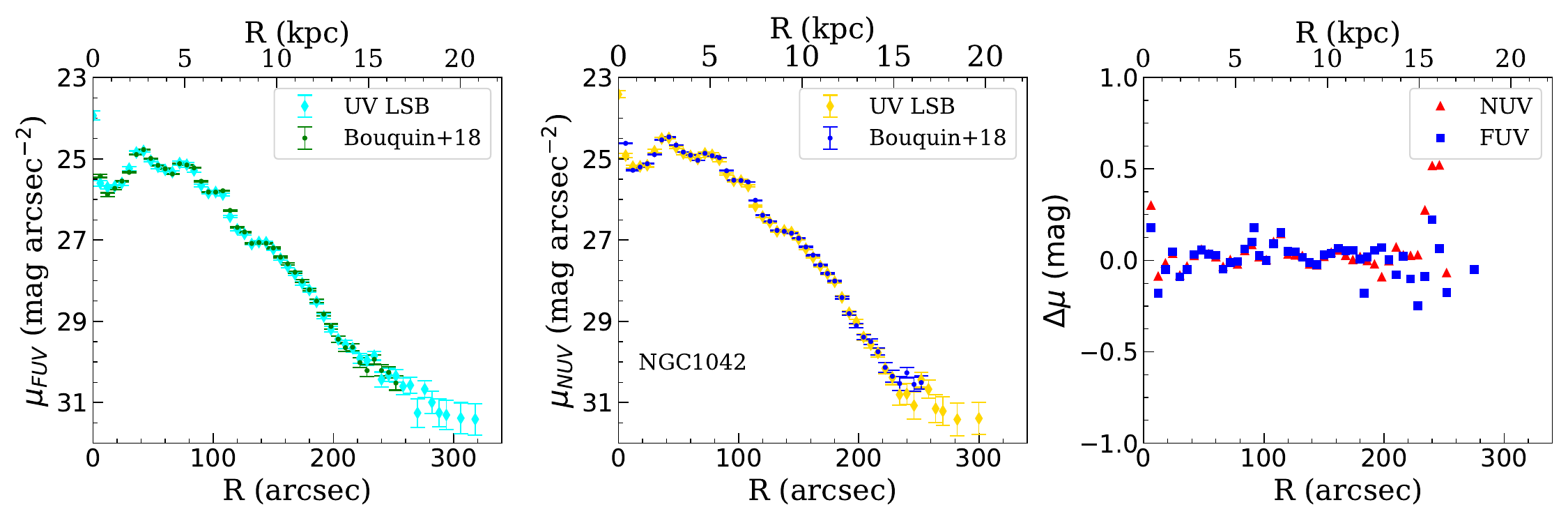}

        \includegraphics[width=0.9\textwidth]{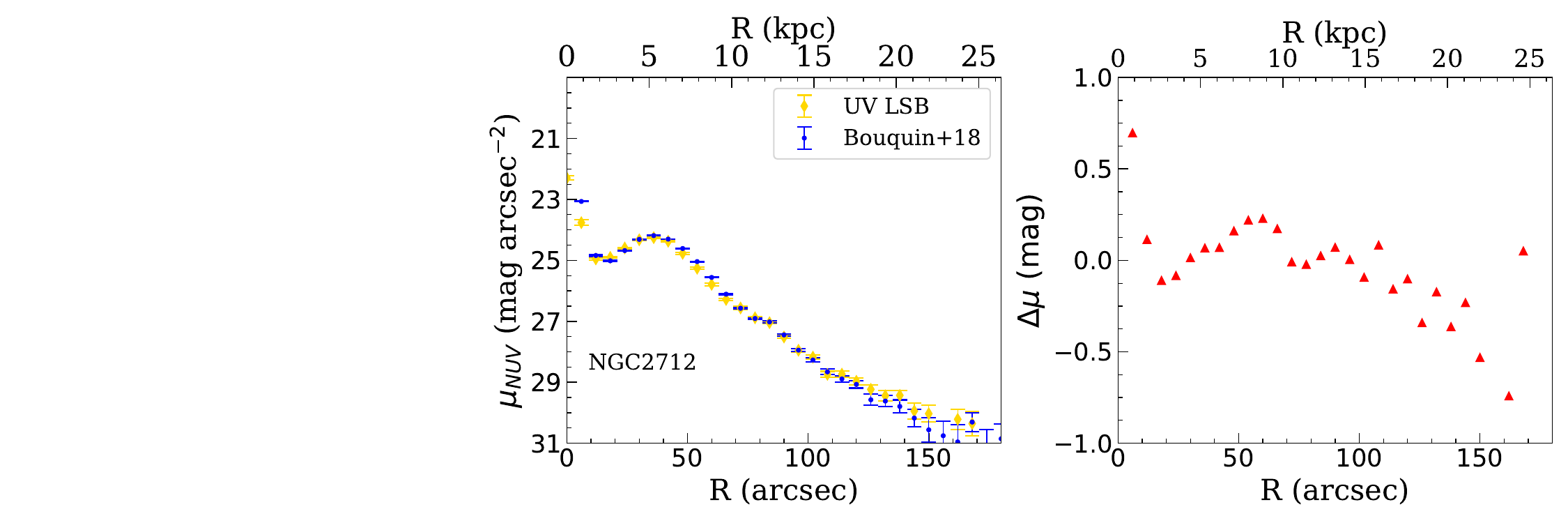}
        \includegraphics[width=0.9\textwidth]{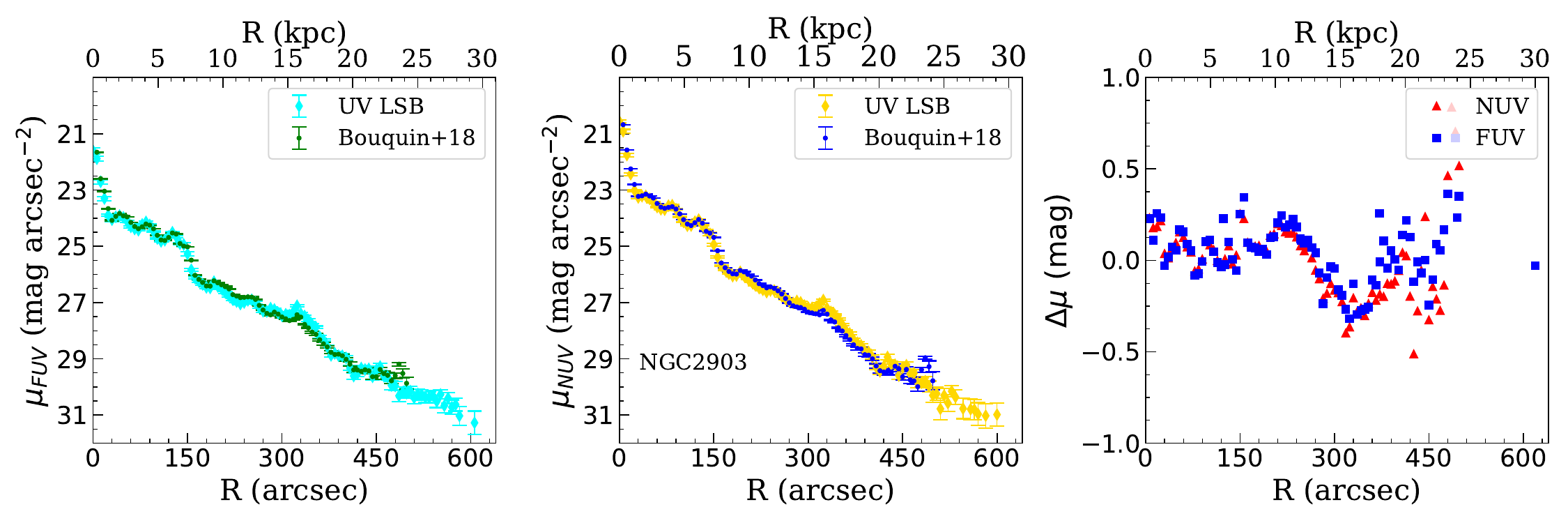}
        \includegraphics[width=0.9\textwidth]{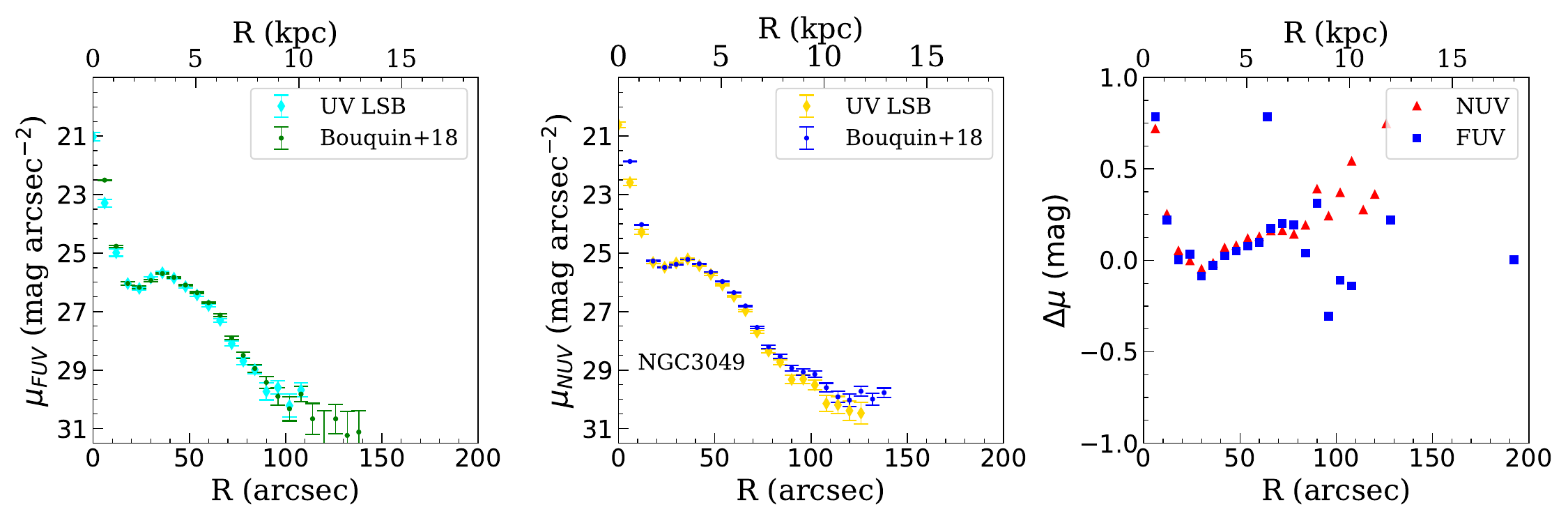}
        \caption{Comparison between \citet{Bouquin18} profiles and the \textit{UV-LSB} profiles obtained in this work after adapting the centers, position angles and axis ratios to the values used in \citet{Bouquin18}. Rightmost panels present the difference between the surface brightness profiles in the two works ($\Delta\mu=\mu_{UVLSB}-\mu_{B18}$).}
        \label{fig:bouq_us}
\end{figure*}
\begin{figure*}[ht]
    \centering
    \setcounter{figure}{0}
        \includegraphics[width=0.9\textwidth]{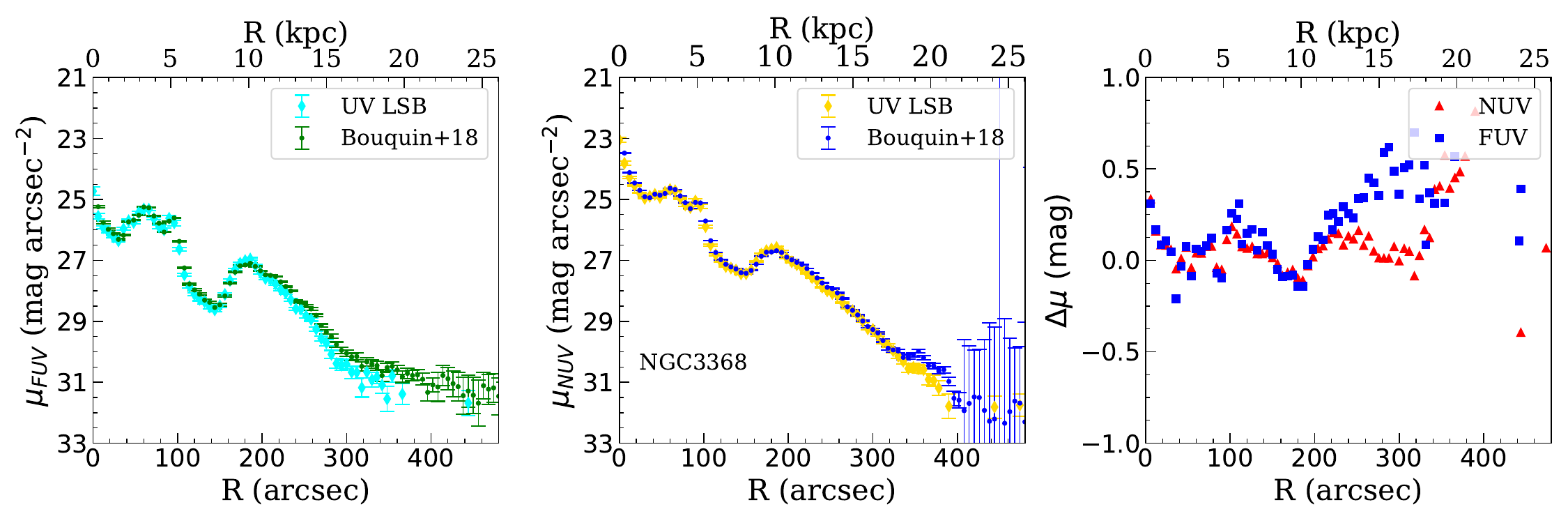}
        \includegraphics[width=0.9\textwidth]{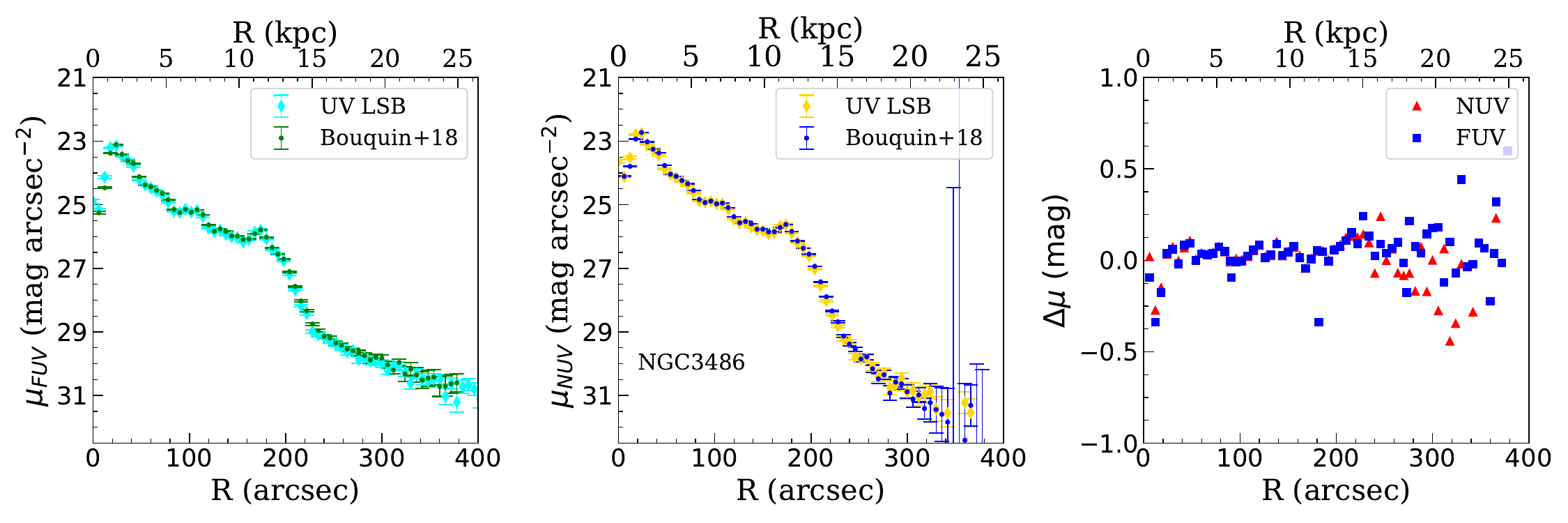}
        \includegraphics[width=0.9\textwidth]{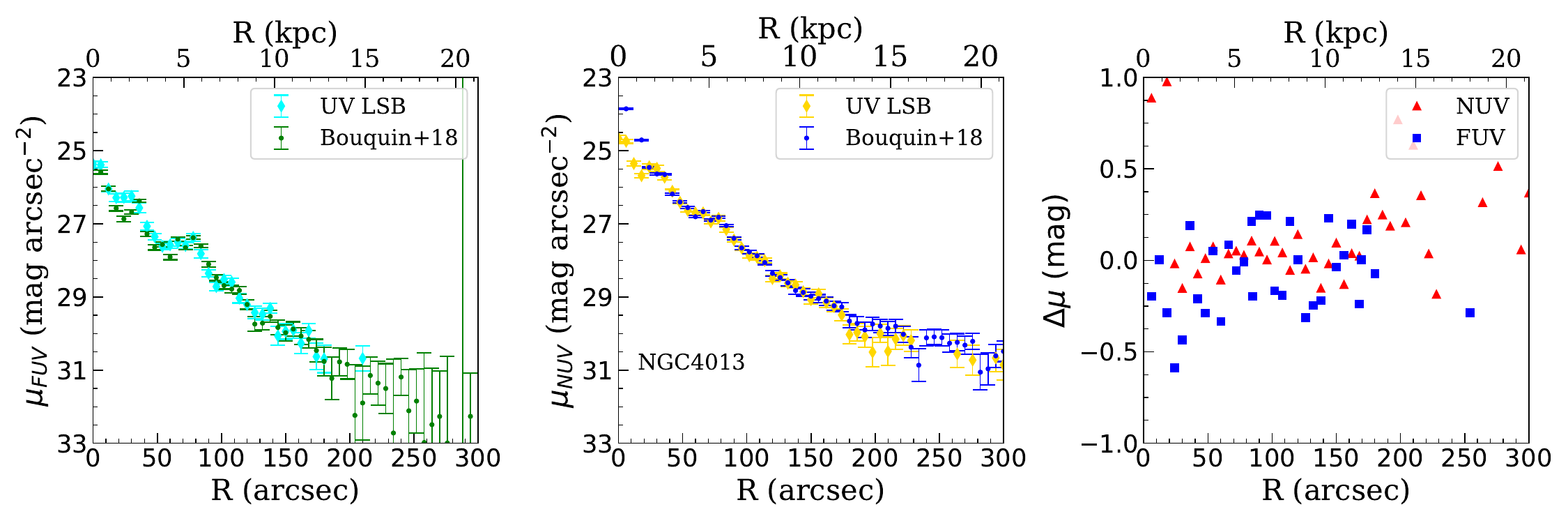}
        \includegraphics[width=0.9\textwidth]{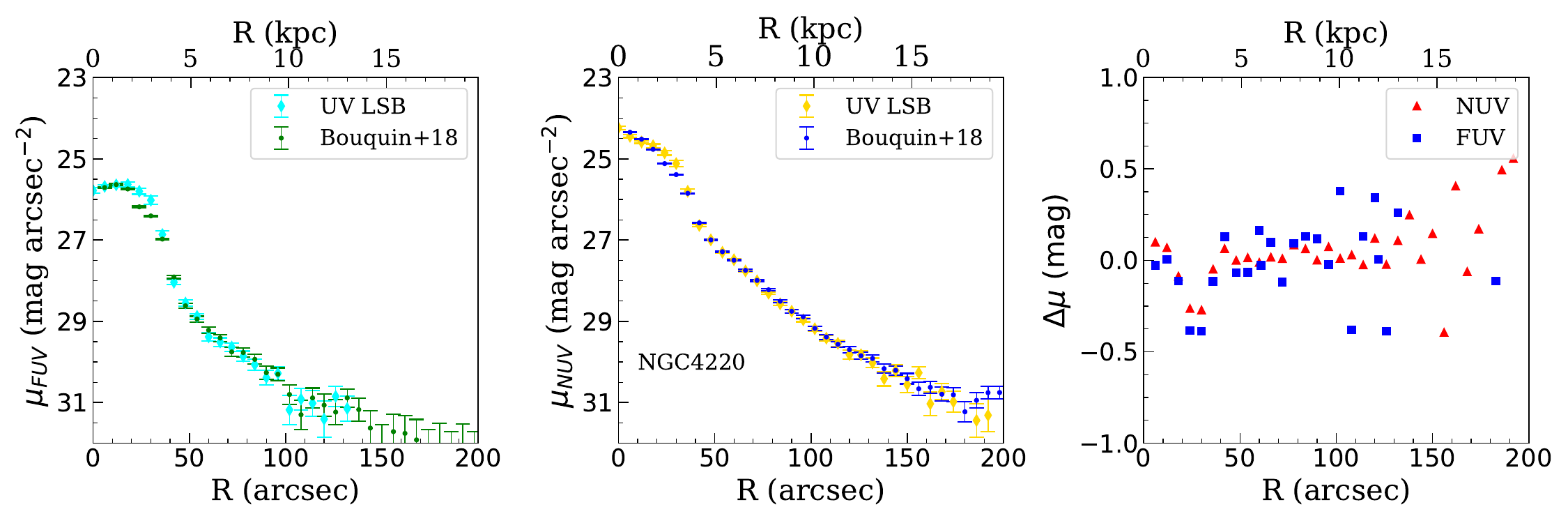}
        \caption{Cont.}
\end{figure*}
\begin{figure*}[ht]
    \centering
    \setcounter{figure}{0}
        \includegraphics[width=0.9\textwidth]{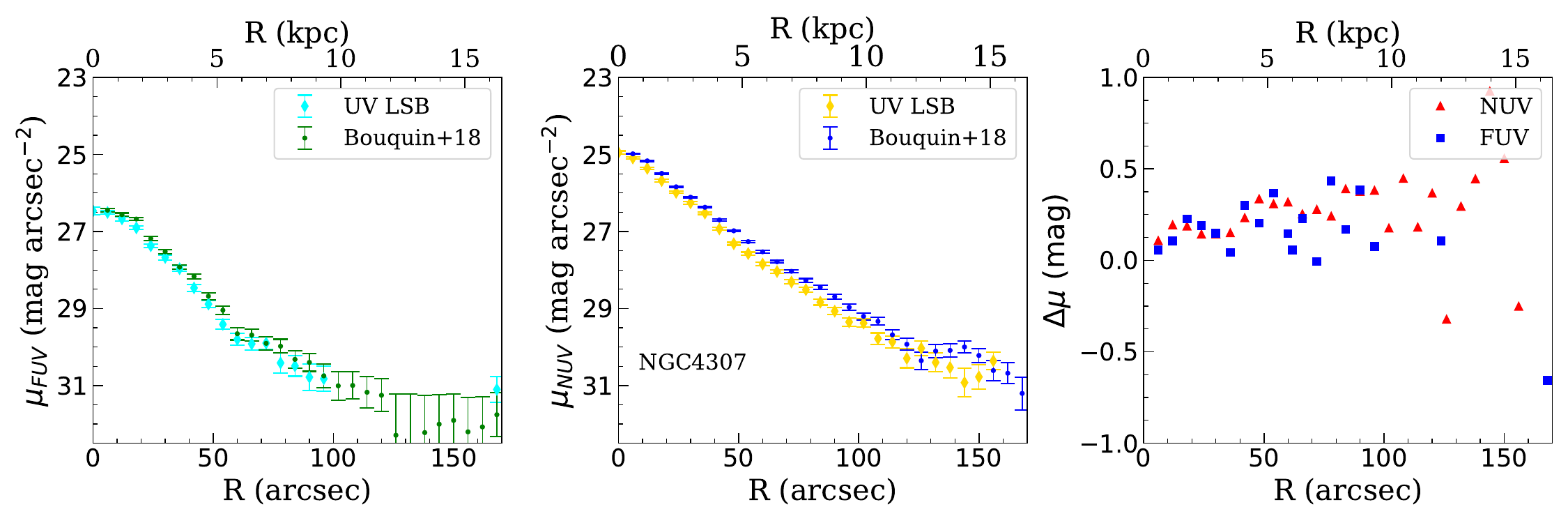}
        \includegraphics[width=0.9\textwidth]{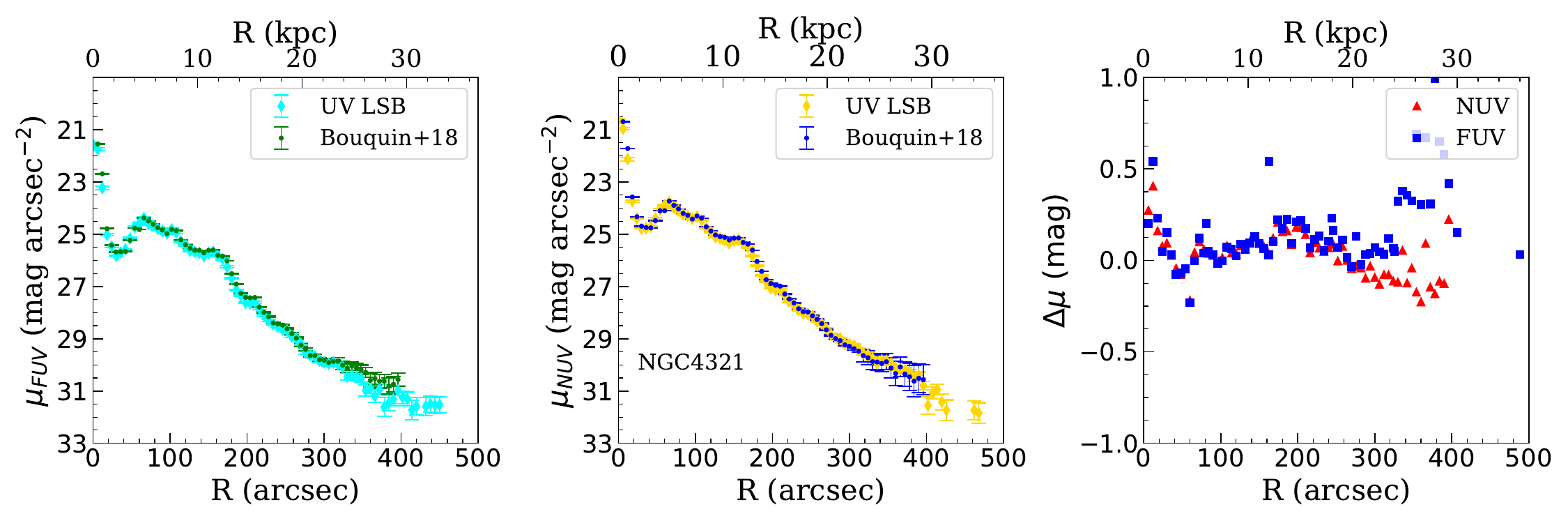}
        \includegraphics[width=0.9\textwidth]{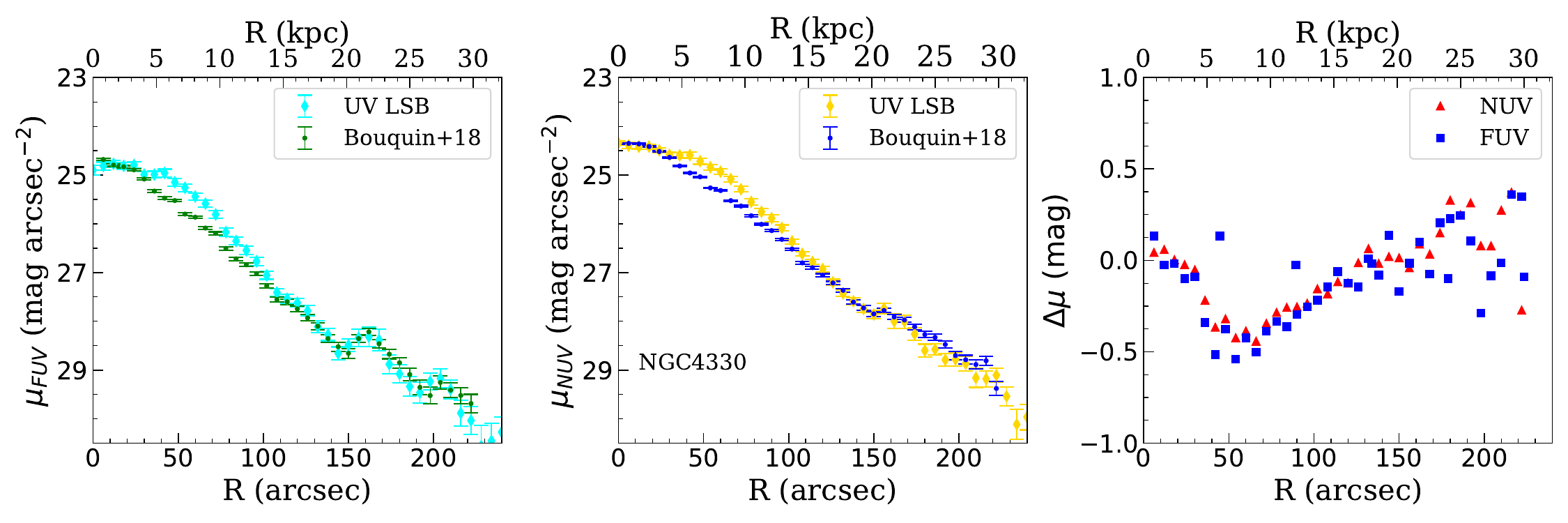}
        \includegraphics[width=0.9\textwidth]{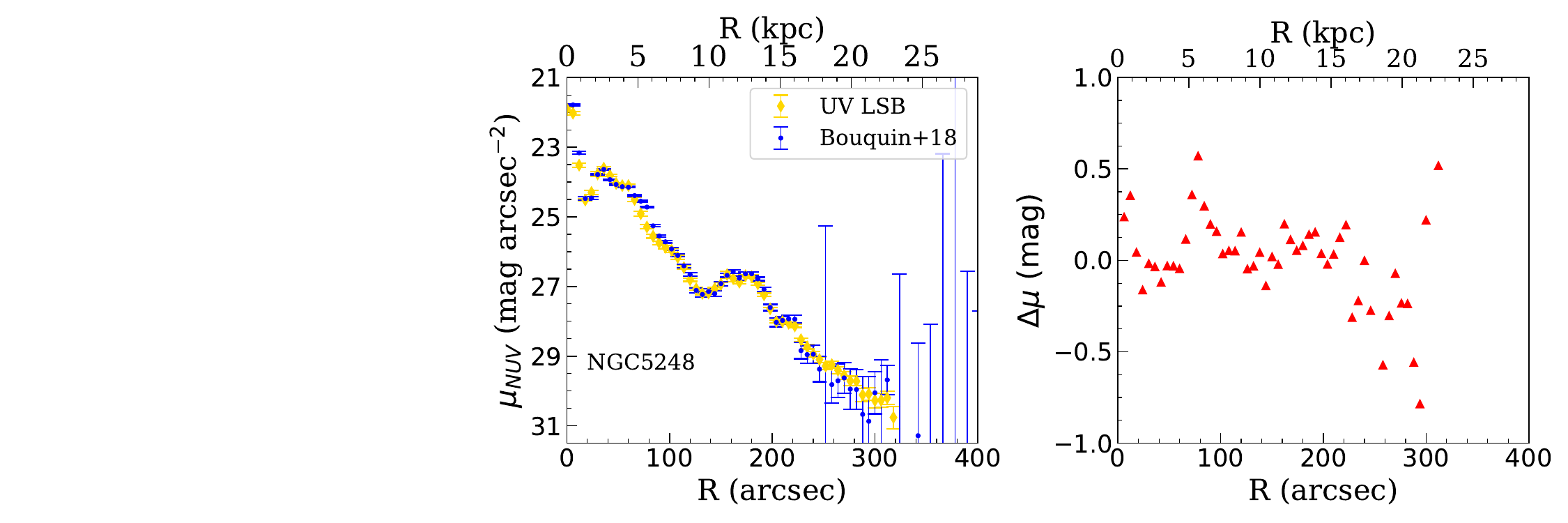}
        \caption{Cont.}
\end{figure*}
\begin{figure*}[ht]
    \centering
    \setcounter{figure}{0}
        \includegraphics[width=0.9\textwidth]{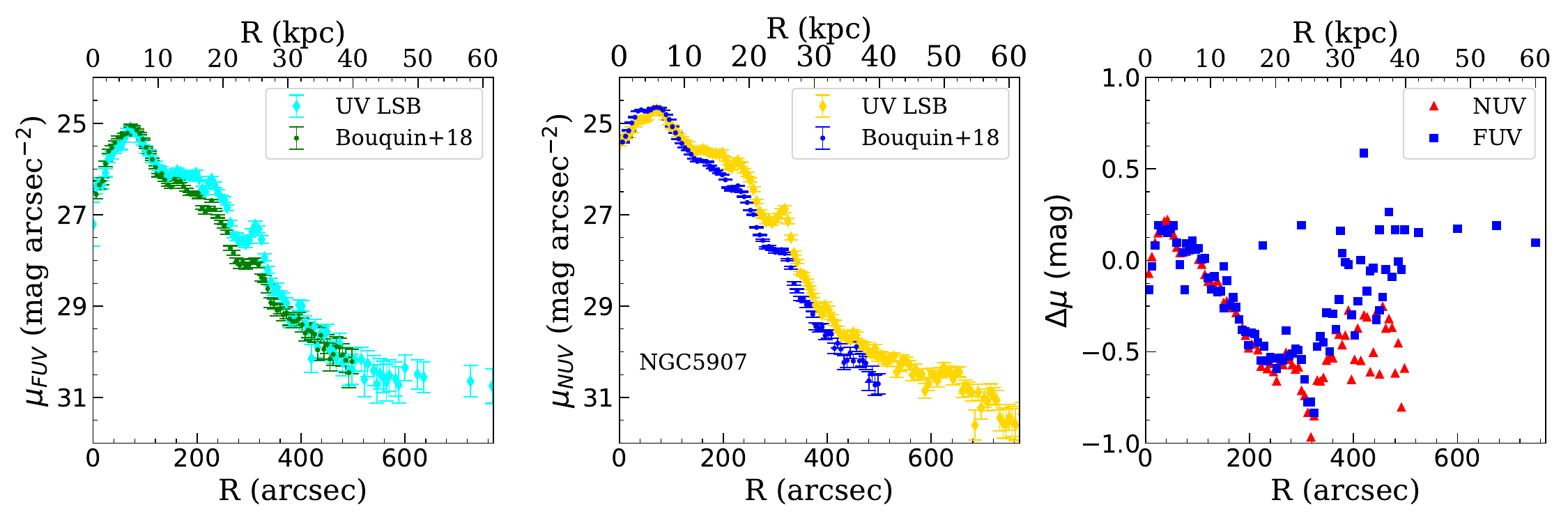}
        
        \caption{Cont.}
\end{figure*}

Since the background in both cases is characterized by the mean, the differences are related to the way the masks are constructed in the regions where the sky is estimated. The use of statistical background subtraction in our method facilitates the use of LSB-adapted detection algorithms. However, the mask construction used by \citet{Bouquin18} is a synthesis of GALPIP masks and (FUV-NUV) based detection of unresolved foreground sources \citep[see ][, Sec. 3.1]{Bouquin18}. Despite the different masking, the profiles agree very well for 10 out of 13 galaxies. In general, our surface brightness profiles extend a bit further than theirs. The 3 galaxies that show important differences in the profiles are the 3 highly inclined ones NGC4307, NGC4330, and NGC5907. Interestingly, the difference in surface brightness is in the inner regions of the galaxies and not in the outer parts. This leads us to suspect that the dust lane of these objects has been treated differently between the two analyses.
\section{Combination of different observing runs}\label{ap:coadd}

In this paper, we propose a methodology for analyzing individual GALEX observing runs, using those with the longest exposure times in both bands. It should be noted, however, that some of the galaxies in the present sample have been observed in different surveys. In this section we explore the potential improvement of our methodology and the reliability of our results by integrating multiple observing runs. To this end, the data presented in Table ~\ref{tab:data} have been combined with other archival data for six galaxies in our sample. The selection of these six galaxies was made to ensure that the new data had at least 0.3 times the original exposure time in one of the two bands, and that data were available in both GALEX bands. The details of the new data are presented in Table ~\ref{tab:new_data}.

\begin{table*}[h!]
    \caption{Galaxies selected for making a coadd.}
    \centering
    \begin{adjustbox}{width=\linewidth}

    \begin{tabular}{cccccccc}
    \hline
    \hline
       Name  & Survey & \multicolumn{2}{c}{$t_{exp}$} & \multicolumn{2}{c}{$f_{exp}$} & \multicolumn{2}{c}{$\mu_{lim}(3\sigma; 10\arcsec\times10\arcsec$)} \\
         & & \multicolumn{2}{c}{[s]} & & & \multicolumn{2}{c}{[$\rm{mag\, arcsec^{-2}}$]} \\
         & & FUV & NUV & FUV & NUV & FUV & NUV \\
    \hline
    NGC1042 & MIS & 1705 &  1705 & 0.578 & 0.452 & 29.12 & 29.34 \\
    NGC3368 & NGS & 1704 & 1704 & 0.376 & 0.149 & 29.55 & 29.77 \\
    NGC3486 & NGS & 541 & 1977.3 & 0.222 & 0.491 & 29.12 & 29.38 \\
    NGC4013 & GII & 1626.45 & 1626.5 & 0.991 & 0.991 &  29.46 & 29.15 \\
    NGC4321 & NGS & 1754.1 & 2932.2 & 0.365 & 0.453 & 29.47 & 29.66 \\
    NGC4330 & GII/GII\tablefootmark{a} & 1689.15/1241.05 & 3281.15/1241.05 & 0.437/0.321 & 0.850/0.321 & 29.31 & 29.60 \\ 
    \hline
    \hline     
    \end{tabular}
    \end{adjustbox}
    \vspace{0.1cm}
    \tablefoot{
    Galaxies are selected to have on the GALEX archive deep enough data (i.e., adding at least 30\% more time) from surveys other than those used in this paper.  We define $f_{exp}=\frac{t_{new}}{t_{orig}}$, where $t_{new}$ is the exposure time of the added data and $t_{orig}$ is the exposure time (see Table ~\ref{tab:data}) of the main data used in this paper. The limiting surface brightness of the final coadd (both surveys combined) is also given. \\
    \tablefoottext{a} {For NGC4330 there are two more observing runs that fulfill our selection criteria. We combine both.}}
    \label{tab:new_data}
\end{table*}

In order to combine the added data with the used in this paper, we have proceed as follows: 

\begin{enumerate}
    \item We apply the methodology outlined in Sec.~\ref{subsec:bcksub} to each of the observing runs separately in order to obtain the background-subtracted intensity maps. 
    \item We combine the  background-subtracted intensity maps of the added data with the original ones (the ones we called \textit{UV LSB} in Sec.~\ref{subsec:disc_bck}) using the weighted mean\footnote{\url{http://ned.ipac.caltech.edu/level5/Leo/frames.html}} \citep{leo+90}. To measure the weights, we run \texttt{NoiseChisel} on the intensity maps, mask all the sources, and we measure the standard deviation on the masked images. The squared standard deviation values are used as proxies for the weights. 
    \item The methodology described in Sec.~\ref{subsec:psf_sub} is applied to the final coadd to subtract the effects of the PSF. To this end, we employ the models delineated in Table ~\ref{tab:models}, but with the Wiener core constructed from the coadd. 
\end{enumerate}

\begin{figure*}[h!]
    \centering

        \includegraphics[width=0.9\textwidth]{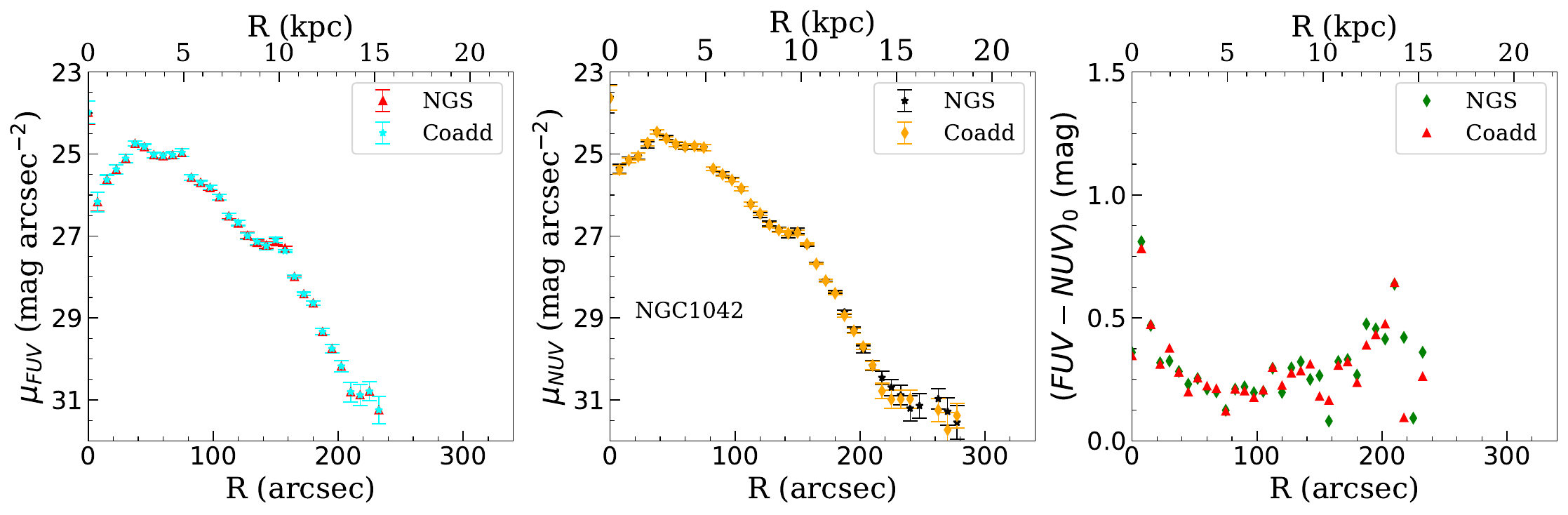}

        \includegraphics[width=0.9\textwidth]{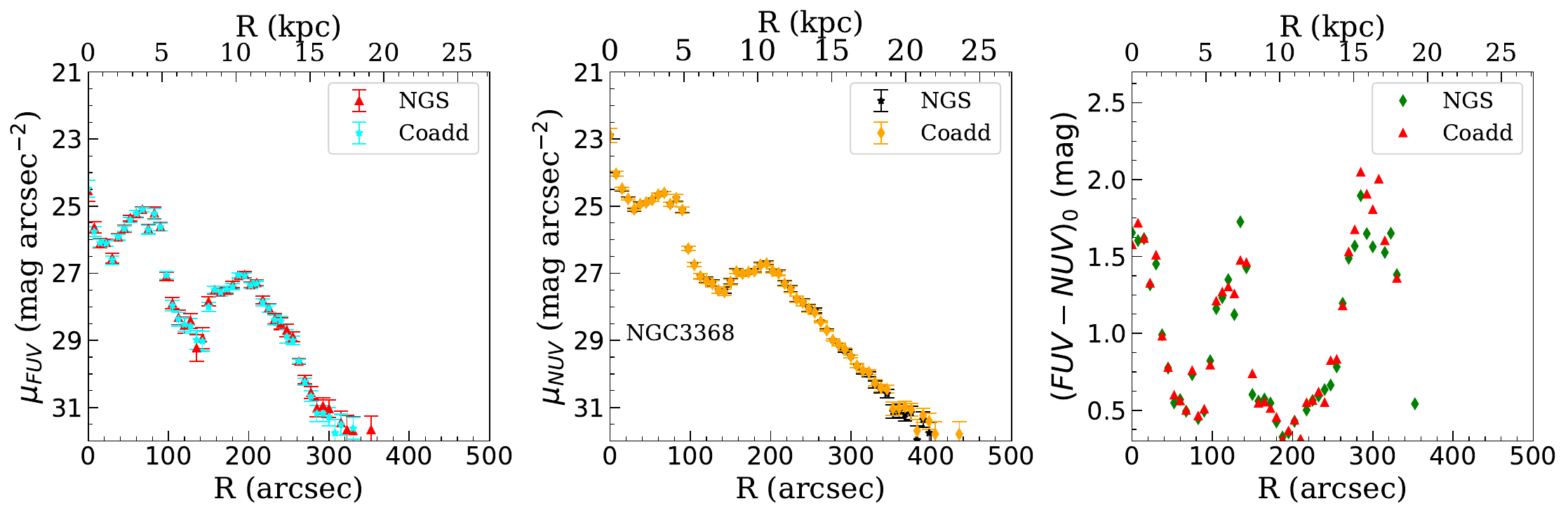}
        \includegraphics[width=0.9\textwidth]{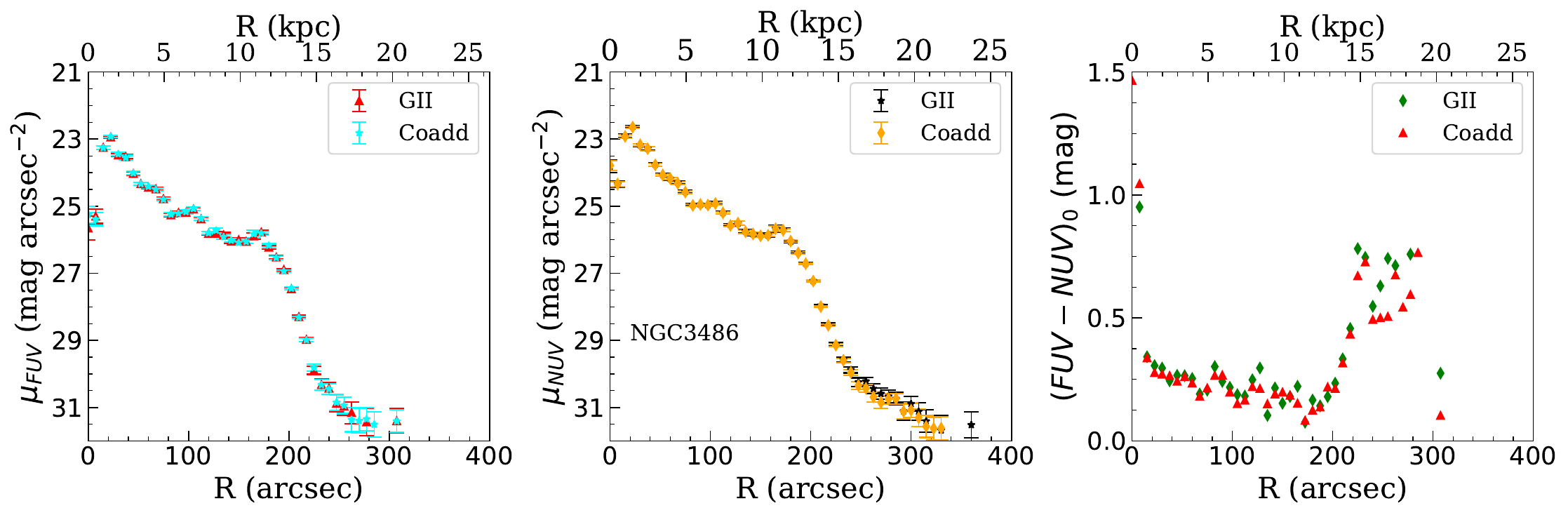}
        \includegraphics[width=0.9\textwidth]{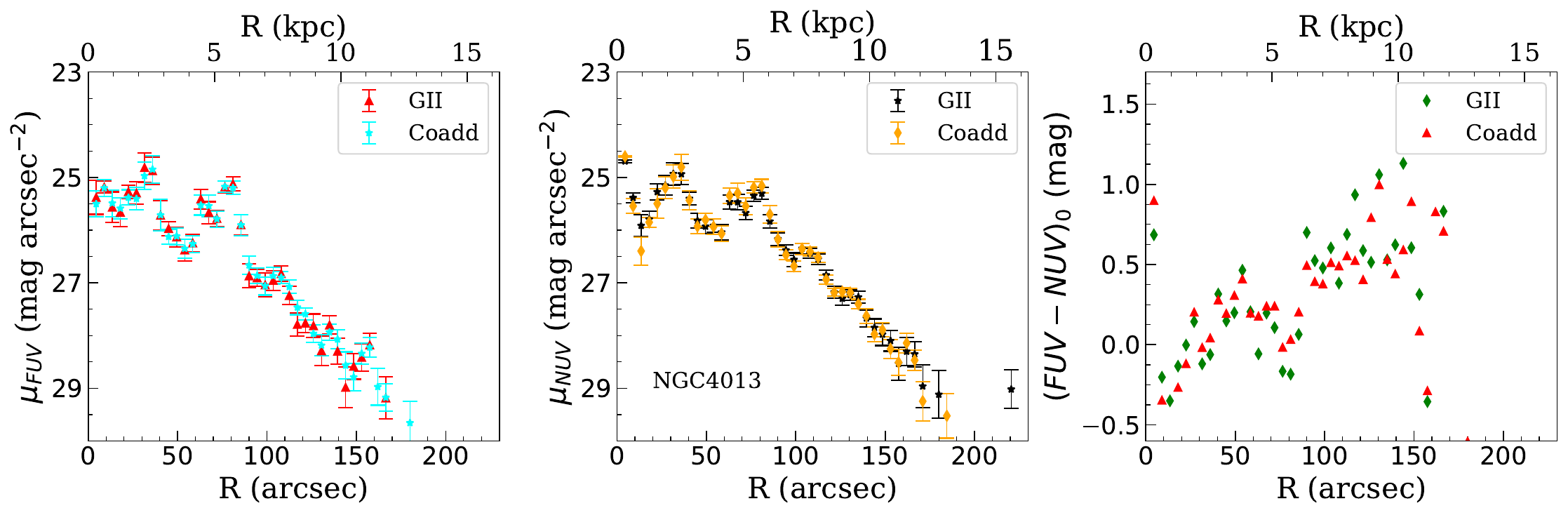}
        \caption{Comparison between the  surface brightness and color profiles of the galaxies using the combination of multiple GALEX runs against the profiles using a single survey (see Table ~\ref{tab:new_data}). See text for details.}
        \label{fig:coaddprofs}
\end{figure*}
\begin{figure*}[ht]
    \centering
    \setcounter{figure}{0}
        \includegraphics[width=0.9\textwidth]{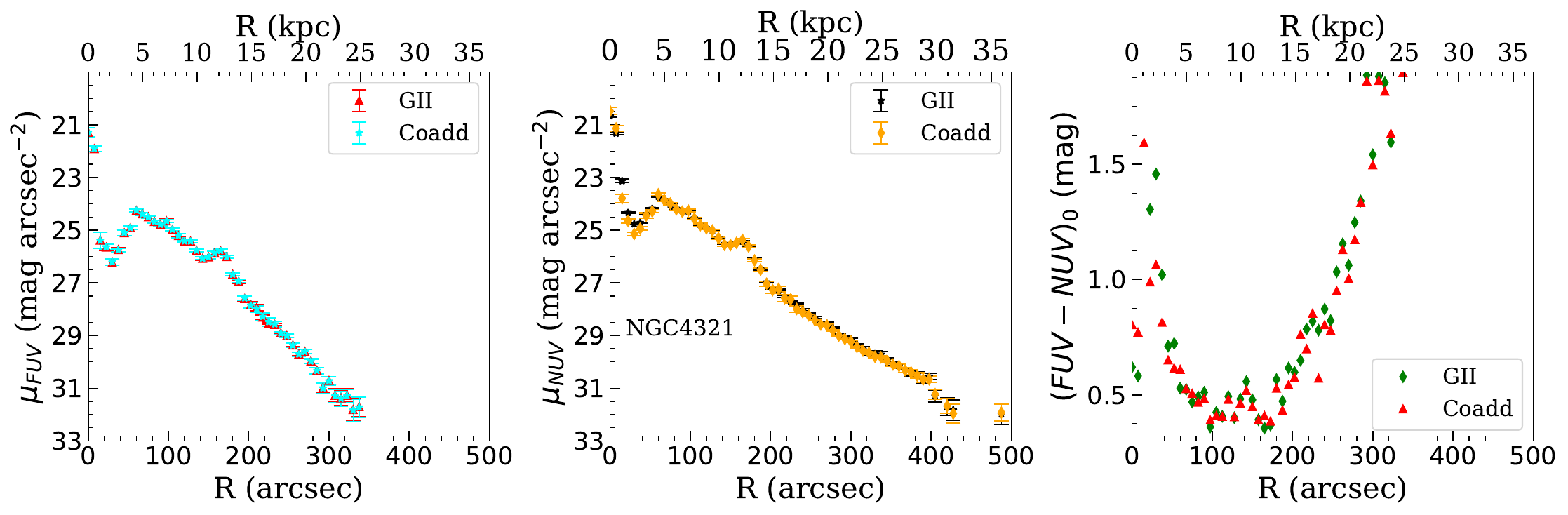}
        \includegraphics[width=0.9\textwidth]{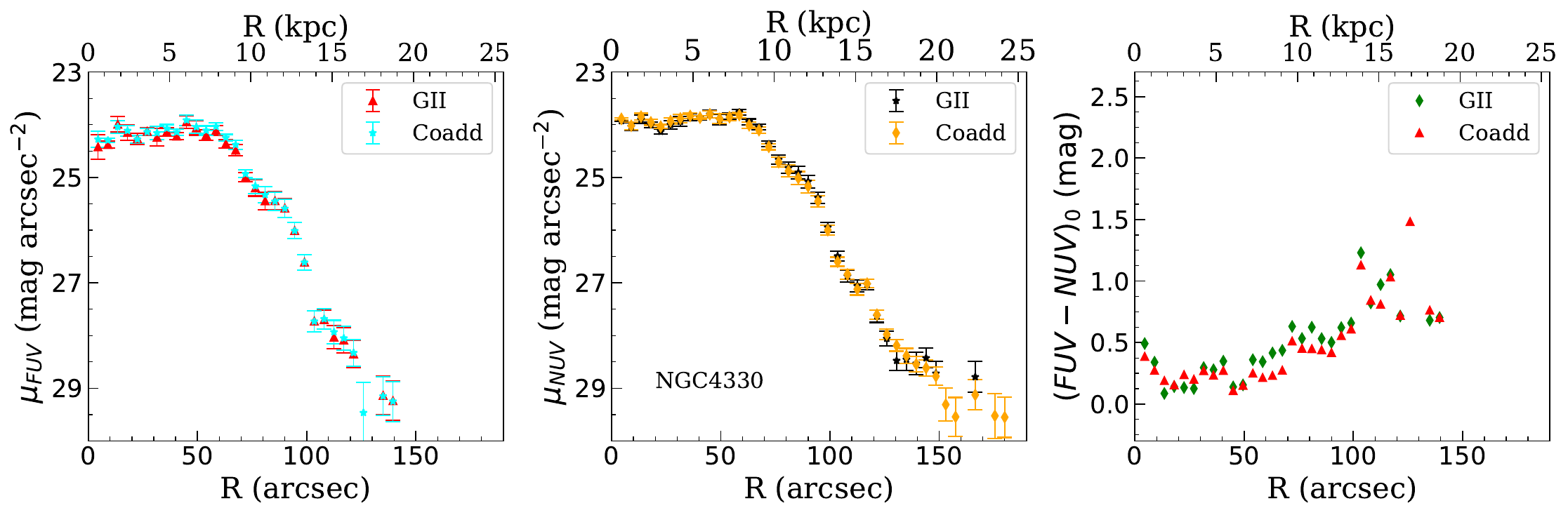}
        \caption{Cont.}
\end{figure*}

The profiles of the combined data are shown in Fig.~\ref{fig:coaddprofs}, in comparison with the profiles of Sec.~\ref{subsec:sbcp}. The comparison of the new (combined data) profiles and the estimation of the new surface brightness limits show that the depth gained by the combined data is in most cases very small. In the FUV the maximum gain is observed in NGC4013 with $\Delta\mu_{lim}=0.4\,\rm{mag\, arcsec^{-2}}$. In NUV, the maximum difference in depth is reached in NGC4330, with $\Delta\mu_{lim}=0.45\,\rm{mag\, arcsec^{-2}}$. In these two galaxies, coaddition of runs with comparable exposure times (NGC4013) or multiple runs (NGC4330) results in significantly deeper surface brightness profiles. \\

If the coaddition of multiple images does not significantly enhance the final depth, it should be used with caution. This process can introduce new sources of noise, such as the intrinsic scatter of the GALEX photometric calibration, which has an accuracy of 0.05 mag in FUV and 0.03 mag in NUV \citep[see ][section 4]{morrissey2007}. Other light sources not present in all images (such as reflections present in the new data of NGC1042 and NGC3368), or a different position of the source in the FoV of GALEX may also contribute to the final noise. However, the results obtained for these six galaxies suggest that our methodology can still be applied when combining multiple observing runs. While this may not have a significant impact on our sample, the integration of our methodology with the weighted mean coaddition of multiple images has the potential to improve the analysis of regions with extensive data.

\end{document}